\documentclass[preprint,review,3p]{elsarticle}

\usepackage{tikz}
\usetikzlibrary{spy}
\usepackage{pgfplots}
\usepackage{moresize}
\usepackage{multirow}
\usepackage{url}

\definecolor{mycolor1}{rgb}{0.00000,0.44700,0.74100}%
\definecolor{mycolor2}{rgb}{0.85000,0.32500,0.09800}%
\definecolor{mycolor3}{rgb}{0.92900,0.69400,0.12500}%
\definecolor{mycolor4}{rgb}{0.46667,0.67451,0.18824}%
\definecolor{mycolor5}{rgb}{0.49412,0.18431,0.30588}%
\definecolor{mycolor6}{rgb}{0.30196,0.69020,0.93333}%

\usepackage{graphicx}
\usepackage{amssymb}
\usepackage{amsthm,amsfonts,amsbsy,amsmath}
\DeclareMathAlphabet\mathbfcal{OMS}{cmsy}{b}{n}

\journal{Signal Processing}

\begin{document}

\begin{frontmatter}

\title{LADMM-Net: An Unrolled Deep Network For Spectral Image Fusion From Compressive Data}

\author[mymainaddress]{Juan Marcos Ramirez\corref{mycorrespondingauthor}}
\cortext[mycorrespondingauthor]{Corresponding author}
\ead{juanmarcos.ramirez@urjc.es}

\author[mymainaddress]{Jos\'e Ignacio Mart\'inez Torre}
\ead{joseignacio.martinez@urjc.es}

\author[mysecondaryaddress]{Henry Arguello}
\ead{henarfu@uis.edu.co}

\address[mymainaddress]{Computer Science Department, Universidad Rey Juan Carlos, M\'ostoles, Spain}
\address[mysecondaryaddress]{Computer Science Department, Universidad Industrial de Santander, Bucaramanga, Colombia}

\begin{abstract} 
Image fusion aims at estimating a high-resolution spectral image from a low-spatial-resolution hyperspectral image and a low-spectral-resolution multispectral image. In this regard, compressive spectral imaging (CSI) has emerged as an acquisition framework that captures the relevant information of spectral images using a reduced number of measurements. Recently, various image fusion methods from CSI measurements have been proposed. However, these methods exhibit high running times and face the challenging task of choosing sparsity-inducing bases. In this paper, a deep network under the algorithm unrolling approach is proposed for fusing spectral images from compressive measurements. This architecture, dubbed LADMM-Net, casts each iteration of a linearized version of the alternating direction method of multipliers into a processing layer whose concatenation deploys a deep network. The linearized approach enables obtaining fusion estimates without resorting to costly matrix inversions. Furthermore, this approach exploits the benefits of learnable transforms to estimate the image details included in both the auxiliary variable and the Lagrange multiplier. Finally, the performance of the proposed technique is evaluated on two spectral image databases and one dataset captured at the laboratory. Extensive simulations show that the proposed method outperforms the state-of-the-art approaches that fuse spectral images from compressive measurements.
\end{abstract}


\end{frontmatter}

\section{Introduction}
A hyperspectral (HS) image can be considered as a three-dimensional (3-D) data set that contains the light reflective responses of a two-dimensional (2-D) scene across tens or hundreds of spectral bands ranging from the visible spectrum (VIS, $400-700$ nm) to the shortwave infrared region (SWIR, $700-2400$ nm) \cite{Ghamisi2017Advances,camps2011remote}. HS images provide detailed spectral information of the scene, enabling the identification of distinct materials included in the covered region. These images have been considered for various applications such as precision agriculture, environment monitoring, and medical diagnosis \cite{martin2018applications, Khan2018Modern}. HS imaging sensors commonly capture low-spatial-resolution data sets with the goal of achieving a high signal-to-noise ratio (SNR) in the acquired observations. To overcome this drawback, data fusion has emerged as a signal processing task that focuses on merging the information in HS images with the information provided by the high-spatial-resolution but low-spectral-resolution multispectral (MS) images \cite{ghassemian2016review, Yokoya2017Hyperspectral, dian2020recent}.

Fusion methods are typically applied to HS and MS images captured by scanning sensors whose functioning is based on the Nyquist-Shannon sampling theorem \cite{Wei2015Fast, YokoyaCoupled2012}. Nevertheless, scanning sensors require a huge amount of observations to capture the relevant information in the covered scenes. In this regard, compressive spectral imaging (CSI) has emerged as an alternative sensing framework that captures the relevant information embedded in spectral images using a reduced set of projections \cite{CaoComputational2016}. The coded aperture snapshot spectral imaging (CASSI) is the most representative CSI system whose measurements basically are projections of encoded versions of the input spectral field \cite{WagadarikarSingle2008}. Multiple variants of the CASSI system have been developed including the three-dimensional CASSI (3D-CASSI) \cite{CaoComputational2016}, the colored CASSI (C-CASSI) \cite{ArguelloColored2014} and the snapshot colored CSI (SCCSI) \cite{CorreaImaging2015}, among others. Various image fusion methods from multi-sensor compressive measurements have been recently proposed. More precisely, these methods determine the acquisition model that describes the multi-sensor optical system and formulate an optimization problem. Then, an iterative algorithm is developed to estimate the high-resolution spectral image from compressive samples \cite{Vargas2019Spectral, Vargas2019Spectral2, Gelve2020Nonlocal}. These methods assume that high-resolution spectral image can be sparsely represented on a predefined transform basis and they also exhibit high running times. On the other hand, various deep learning approaches have been recently developed providing state-of-the-art performances for different imaging applications including hyperspectral image analysis \cite{petersson2016hyperspectral} and target detection in SAR imagery \cite{Pu2021Deep, Pu2021Shuffkle}. However, deep learning approaches typically suffer from a lack of explainability of the inverse mapping that recovers the image of interest from degraded measurements \cite{Jin2017Deep, Monga2021Algorithm}.

In this regard, deep algorithm unrolling is an approach for solving inverse problems whose performance and interpretability have attracted the attention of the signal and image processing community in the past decade. This approach takes advantage of both the interpretability of the model-based iterative algorithms and the remarkable performance of deep neural networks \cite{Monga2021Algorithm}. Specifically, this approach casts each update step of an iterative algorithm into a network-based structure whose concatenation describes a deep network. The first work in deep algorithm unrolling was proposed by Gregor and LeCun in the context of sparse coding \cite{gregor2010learning}. Basically, this method maps each update of the iterative soft-thresholding algorithm (ISTA) to a network-based structure that consists of learning the inversion induced by two operators to compute the output estimate. This technique is referred to as Learned ISTA (LISTA). A second approach based on the ISTA algorithm is the ISTA-Net architecture \cite{zhang2018ISTA} that has been designed to recovering images captured under the compressive sensing (CS) scheme. This approach exploits the representation power of the convolutional neural networks (CNN) to build a sparsity-inducing nonlinear transform for representing the target image. Furthermore, authors introduce an enhanced version, called ISTA-Net+, that relies on the fact that the image residuals can be better described in a sparsity-promoting dictionary. Recently, the ADMM-CSNet has been reported in \cite{yang2016deep, Yang2020ADMM} for reconstructing images from CS samples. In essence, this work proposes two deep learning architectures that unroll two versions of the ADMM algorithm, where the matrix inversions are computed using techniques based on the fast Fourier transform (FFT). Other unrolling approaches for solving inverse problems has been also introduced such as the model-based deep learning (MoDL) architecture \cite{Aggarwal2019MoDL}, the FBPConvNet \cite{Jin2017Deep}, and the analytic compressive iterative deep (ACID) framework \cite{wu2020stabilizing}.

This paper focuses on developing a network architecture based on the linearized version of the ADMM algorithm \cite{esser2010general} to solve the image fusion problem from HS and MS compressive measurements. This algorithm unfolding-based network, called LADMM-Net, is an interpretable architecture that exploits the sparsity-inducing nonlinear transform to estimate the image high-frequency content embedded in both auxiliary and dual variables of the ADMM algorithm \cite{boyd2011distributed}. In essence, the proposed architecture casts each iteration of the LADMM algorithm into a network-based processing layer whose cascading deploys a deep network. Every layer consists of an approximation unit (AU) followed by a network-based refinement unit (NRU). The AU considers the information embedded in the coded aperture patterns as well as the image high-frequency content computed by the previous layer to obtain an estimate of the fused image. In contrast, the NRU is a network-based structure that can be seen as a nonlinear transform function relied on a CNN that estimate the high-frequency information of the target spectral image. The main contributions of this work are synthesized as follows.

\begin{enumerate}
    \item The proposed approach dubbed LADMM-Net basically is a deep learning architecture that solves the data fusion problem from HS and MS compressive measurements. In particular, this approach casts each iteration of a linearized version of the ADMM algorithm into a CNN structure whose cascading describes a deep network. Moreover, the linearized version of the ADMM allows to estimate approximations of the variable of interest without resorting to costly matrix inversions. In addition, the information embedded in both the auxiliary variable and the Lagrange multiplier is obtained by the network-based structure to remarkably improve the fusion estimation.
    \item Secondly, the performance of the proposed architecture is evaluated for two spectral image databases and one real data set captured in the laboratory. Furthermore, the proposed deep network is tested for different processing layer numbers and distinct compression ratios.
    \item Extensive simulations show that that the proposed deep learning architecture outperforms other state-of-the-art methods that obtain high-resolution images from HS and MS compressive measurements. Since data fusion can be considered a particular inverse problem, the proposed architecture is adapted for reconstructing images from compressive random projections.
\end{enumerate}

\subsection{Related work}

Recently, various algorithm unrolling methods have been developed for solving inverse problems \cite{Monga2021Algorithm}. In this regard, LISTA is considered the first deep algorithm unrolling technique that maps each ISTA iteration into a fully connected neural network. This method aims at learning the inverse mapping performed by two matrices used in the ISTA algorithm \cite{gregor2010learning}. It is worth noting that this approach is computationally unfeasible for recovery large-size images such as high-resolution spectral images. On the other hand, ISTA-Net casts each ISTA iteration into a network structure that learns the soft-thresholding parameter, the regularization parameter, and a nonlinear transform \cite{zhang2018ISTA}. Although the ISTA-Net structure is computationally efficient, this method is affected by the limited capacity of convolutional neural networks to learn the entire information embedded in the target image. Compared to ISTA-Net, the proposed approach unfolds a linearized version of the ADMM algorithm that alternately solves simple subproblems. Furthermore, our method attempts to learn a nonlinear transform that describes image details embedded in both the auxiliary variable and the Lagrange multiplier improving thus, the reconstruction performance while exploiting the computational efficiency of convolutional networks.

ADMM-CSNet unfolds the ADMM algorithm in the context of imaging compressive sensing \cite{yang2016deep, Yang2020ADMM}. In particular, this technique develops two matrix inversion procedures to estimate the target variable update, one of which is tailored to the measurement matrices used by compressive sensing magnetic resonance imaging (CS-MRI). In addition, ADMM-CSNet deploys a set of linear filters that attempts to learn the algorithm parameters, a linear transform, and the regularization function. Compared to the ADMM-CSNet, the proposed approach avoids resorting to matrix inversions by updating the target variable performing a computationally efficient proximal descent step. This minimization step can use any efficient measurement matrix such as random matrices and CSI sampling matrices. Furthermore, LADMM-Net aims at solving an $\ell_1$-regularized inverse problem whose proximal mapping reduces to the soft-thresholding operator on a shifted version of the auxiliary variable. Finally, our method optimizes an invertible nonlinear transform based on convolutional neural networks to sparsely represent image details.

This paper is organized as follows. Section \ref{sec:model} describes the dual-arm architecture based on the 3D-CASSI systems, and Section \ref{sec:ladmm_net} introduces the proposed approach for solving the image fusion problem from compressive projections. The results of extensive simulations are shown in Section \ref{sec:results}, and some concluding remarks are exposed in Section \ref{sec:conclusions}.

\section{Observational model}\label{sec:model}

In this work, the proposed algorithm unfolding method is used to fuse spectral images from compressive measurements captured by a dual-arm acquisition system based on the 3D-CASSI architecture. To this end, we first introduce the foundations of the 3D-CASSI optical architecture, and then, we present the acquisition model of the dual-arm system.

\subsection{3D-CASSI system}

\begin{figure}
    \centering
    \includegraphics[width=0.50\linewidth]{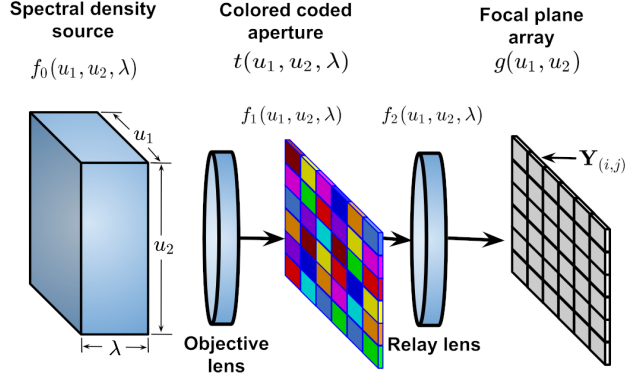}
    \caption{Schematic of the 3D-CASSI optical architecture.}
    \label{fig:cassi_anlg}
\end{figure} 

In general, the 3D-CASSI architecture is a compressive spectral imaging (CSI) system that has been designed to obtain the relevant information of spectral images by capturing a reduced number of camera snapshots \cite{CaoComputational2016}. This compressive acquisition system has been recently used in different applications such as clustering \cite{HinojosaCoded2018}, feature fusion \cite{RamirezMultiresolution2019, ramirez2021feature}, spectral image classification \cite{HinojosaSpectral2019}, and spectral image fusion \cite{Vargas2019Spectral, Vargas2019Spectral2}. In essence, the 3D-CASSI system projects an encoded version of the input spectral image onto a focal plane array (FPA). A schematic of the 3D-CASSI optical architecture is illustrated in Fig. \ref{fig:cassi_anlg}. As can be seen in this figure, $f_0(u_1, u_2, \lambda)$ stands for the spectral density source to be sensed, where ($u_1$, $u_2$) represents the spatial location and $\lambda$ stands the wavelength axis \cite{WagadarikarSingle2008}. The input spectral field $f_0(u_1, u_2, \lambda)$ is firstly affected by the optical system as follows
\begin{equation}
    f_1(u_1, u_2, \lambda) = \iiint f_0(u_1^{\prime}, u_2^{\prime}, \lambda^{\prime}) h(u_1^{\prime}-u_1, u_2^{\prime}-u_2, \lambda^{\prime}-\lambda) d u_1^{\prime} d u_2^{\prime}, d\lambda^{\prime}.
\end{equation}
where $h(u_1, u_2, \lambda)$ is commonly modeled as a shift-invariant impulse response that describes the propagation losses induced by the imaging optics. For the sake of simplicity, this impulse response has been typically modeled as $h(u_1,u_2,\lambda) = \delta(u_1,u_2,\lambda)$ \cite{WagadarikarSingle2008}. Subsequently, $f_1(u_1, u_2, \lambda)$ is modulated by a colored coded aperture that consists of a 2D array of optical filters, where each optical filter at a specific spatial location modulates the incoming light according to a particular spectral response. It is worth noting that the encoding function depends on the pattern of optical filters forming the colored coded aperture \cite{ArguelloColored2014}. More precisely, the encoding operation can be described as
\begin{equation}
    f_2(u_1, u_2, \lambda) = f_1(u_1, u_2, \lambda)t(u_1, u_2, \lambda),
\end{equation}
where $t(u_1, u_2, \lambda)$ is the spatial-spectral encoding function performed by the colored coded aperture. Then, the encoded spectral field is integrated across the system spectral sensitivity $\Lambda$ onto the FPA. In this case, the projected plane is given by
\begin{equation}
    g(u_1,u_2) = \int_{\Lambda} f_2(u_1, u_2, \lambda) d\lambda.
\end{equation}

\begin{figure}
    \centering
    \includegraphics[width=0.65\linewidth]{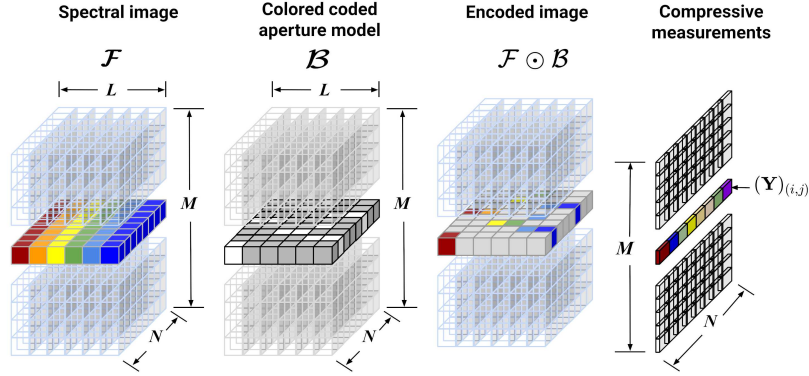}
    \caption{A discrete model of the effects induced by the 3D-CASSI architecture to obtain the compressive measurements. The input spectral image $\mathbfcal{F}$ is encoded by the colored coded aperture model $\mathbfcal{B}$ according to $\mathbfcal{F} \odot \mathbfcal{B}$. Then, the encoded spectral image is projected onto a camera detector to capture the compressive measurements.}
    \label{fig:cassi_sch}
\end{figure}

\vspace{-20pt}
\subsubsection{3D-CASSI discrete model}

To describe the intensity captured by an FPA element, consider $\Delta$ the pixel width and $\mathrm{rect}\left(\frac{u_1}{\Delta} - i, \frac{u_2}{\Delta} -j\right)$ the surface covered by the FPA element at the discrete spatial location$(i,j)$ \cite{CorreaImaging2015}. Hence, the pixel intensity captured by the detector at the spatial coordinate $(i,j)$ can be expressed as
\begin{equation}
    (\mathbf{Y})_{_{(i,j)}} = \int_{\Delta} \int_{\Delta}  g(u_1,u_2) \mathrm{rect}\left(\frac{u_1}{\Delta} - i, \frac{u_2}{\Delta} -j\right) du_1 du_2,
\end{equation}
for $i=0,\ldots,M-1$ and $j=0,\ldots, N-1$, therefore, the detector has dimensions of $M \times N$ pixels.

A discrete model is frequently used to describe the 3D-CASSI compressive samples. To this end, consider the discrete data cube $\mathbfcal{F} \in \mathbb{R}^{M \times N \times L}$ that describes the input spectral image whose elements at the spatial coordinate ($i$,$j$) and the $\ell$-th spectral band are denoted as $(\mathcal{F})_{(i,j,\ell)}$  for $i = 0,\ldots,M-1$, $j = 0,\ldots,N-1$, and $\ell = 0,\ldots,L-1$. Fig. \ref{fig:cassi_sch} shows the discrete model of the optical phenomenon induced by the 3D-CASSI system to capture the compressive measurements. As can be observed in this figure, the encoding operation performed by the colored coded aperture is modeled as a binary data cube $\mathbfcal{B} \in \{0,1\}^{M \times N \times L}$ with entries $(\mathcal{B})_{(i,j,\ell)}$ whose values depend on the pattern of optical filters \cite{HinojosaCoded2018}. Afterward, the encoded image is obtained as the Hadamard product between the input spectral image $\mathbfcal{F}$ and the colored coded aperture model $\mathbfcal{B}$, i.e. $\mathbfcal{F} \odot \mathbfcal{B}$ with $\odot$ denoting the element-wise multiplication \cite{ArguelloColored2014}. This encoded version is subsequently projected along the spectral axis onto a detector plane, and the measurement captured at the spatial location ($i$, $j$) can be obtained as
\begin{equation}
    (\mathbf{Y})_{_{(i,j)}} = \left[\sum_{\ell=0}^{L-1} \left(\mathbfcal{F} \odot \mathbfcal{B}\right)_{_{(i,j,\ell)}} \right] + \left(\mathbf{N}\right)_{_{(i,j)}}
\end{equation}
where $\mathbf{N} \in \mathbb{R}^{M \times N}$ is the noise matrix that affects the detector measurements whose entries $(\mathbf{N})_{_{(i,j)}}$ are commonly assumed as independent and identically distributed (iid) random samples following a Gaussian distribution. In general, multiple snapshots are required to recover a reliable version of the spectral image from compressive measurements. In this regard, the 3D-CASSI sample obtained at the spatial coordinate ($i$, $j$) and the $w$-th snapshot can be defined as
\begin{equation}
    (\mathbf{Y}^{(w)})_{_{(i,j)}} = \left[\sum_{\ell=0}^{L-1} \left(\mathbfcal{F} \odot \mathbfcal{B}^{(w)}\right)_{_{(i,j,\ell)}} \right] + \left(\mathbf{N}^{(w)}\right)_{_{(i,j)}}
\end{equation}
for $w = 0, \ldots, W-1$, where $\mathbfcal{B}^{(w)}$ is the model of the colored coded aperture used at the $w$-th snapshot, $W$ the number of snapshots captured by the multi-frame 3D-CASSI system, and $\mathbf{N}^{(w)}$ is the noise matrix affecting the $w$-th snapshot. Various methods that design the coded aperture patterns in the context of CASSI acquisition architectures have been developed in order to improve spectral image reconstructions \cite{ArguelloColored2014, ArguelloRank2013}. In particular, we use the algorithm to design multi-frame colored-coded aperture patterns introduced in \cite{RamirezMultiresolution2019} that captures the entire spatial-spectral information of the input image without redundancy. Notice also that each 3D-CASSI snapshot captures an image with dimensions $M \times N$ pixels, therefore, the compression ratio between the captured data and the spectral image of interest is defined as $\eta = \frac{MNW}{MNL} = \frac{W}{L}$. Hence, a particular compression ratio is achieved by controlling the number of snapshots.

\subsection{Dual-arm acquisition system}

\begin{figure}
    \centering
    \includegraphics[width=0.65\linewidth]{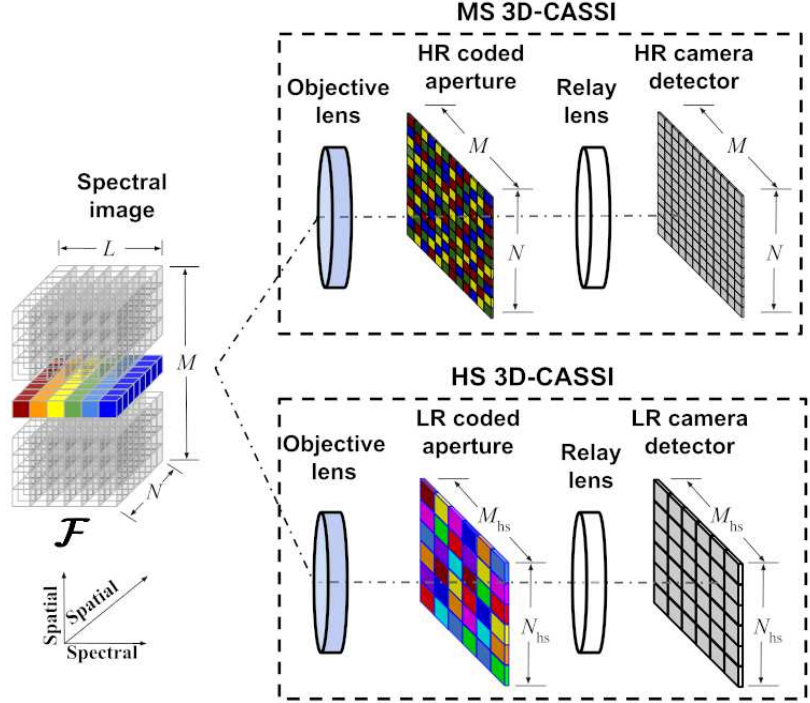}
    \caption{Dual-arm optical architecture to capture hyperspectral and multispectral 3D-CASSI compressive measurements. \vspace{-10pt}}
    \label{fig:dual_arm}
\end{figure} 

Figure \ref{fig:dual_arm} illustrates a schematic of the dual-arm architecture that captures the multi-sensor compressive measurements. As can be seen in this figure, each arm consists of a 3D-CASSI optical system that projects an encoded version of the input spectral image onto the respective camera detector \cite{CaoComputational2016, rueda2020compressive}. In particular, the upper part of the dual-arm system called MS 3D-CASSI includes a high-resolution coded aperture and a high-resolution camera detector. This arm captures poor spectral information due to the small sizes of the optical filters that form the colored coded aperture. More precisely, this arm aims at recovering a low-spectral-resolution multispectral image $\mathbfcal{F}_{\mathrm{ms}} = \boldsymbol{\zeta}(\mathbfcal{F})$ with size $M \times N \times L_{\mathrm{ms}}$, where $\boldsymbol{\zeta}(\cdot)$ represents a spectral degradation function, and $L_{\mathrm{ms}} = L/q$ denotes the number of bands of the multispectral image with $q$ as the spectral decimation factor. Moreover, the colored coded aperture model used at the $w$-th snapshot can be described as $\mathbfcal{B}_{\mathrm{ms}}^{(w)} \in \{0,1\}^{M \times N \times L_{\mathrm{ms}}}$ with entries $(\mathcal{B}^{(w)}_{\mathrm{ms}})_{_{(i,j,\ell)}}$ for $i = 0,\ldots,M-1$, $j = 0,\ldots,N-1$, $\ell = 0,\ldots,L_{\mathrm{ms}} -1$, and $w = 0,\ldots,W_{\mathrm{ms}}-1$, where $W_{\mathrm{ms}}$ denotes the number of snapshots captured by the MS 3D-CASSI system. In consequence, a single sample captured by the MS 3D-CASSI system at the spatial coordinate ($i$,$j$) and $w$-th snapshot can be expressed as
\begin{eqnarray}
   (\mathbf{Y}_{\mathrm{ms}}^{(w)})_{_{(i,j)}} & = & \left[\sum_{\ell=0}^{L_{\mathrm{ms}}-1} \left(\mathbfcal{F}_{\mathrm{ms}} \odot \mathbfcal{B}_{\mathrm{ms}}^{(w)}\right)_{_{(i,j,\ell)}} \right] + \left(\mathbf{N}_{\mathrm{ms}}^{(w)}\right)_{_{(i,j)}} \nonumber \\
   & = & \left[ \sum_{\ell=0}^{L_{\mathrm{ms}}-1} \left(\boldsymbol{\zeta}(\mathbfcal{F}) \odot \mathbfcal{B}_{\mathrm{ms}}^{(w)} \right)_{_{(i,j,\ell)}} \right] + \left(\mathbf{N}_{\mathrm{ms}}^{(w)}\right)_{_{(i,j)}},
\end{eqnarray}
where $\mathbf{N}_{\mathrm{ms}}^{(w)} \in \mathbb{R}^{M \times N}$ is the noise matrix affecting the MS 3D-CASSI detector at the $w$-th snapshot.

On the other hand, the bottom part of the dual system called HS 3D-CASSI consists of a low-spatial-resolution coded aperture and a low-spatial-resolution camera detector. In contrast to the MS 3D-CASSI sensor, this arm captures rich spectral information of the input image. In this case, the HS 3D-CASSI sensor attempts to estimate the low-spatial-resolution image $\mathbfcal{F}_{\mathrm{hs}} = \boldsymbol{\xi}(\mathbfcal{F})$ with dimensions $M_{\mathrm{hs}} \times N_{\mathrm{hs}} \times L$, where $\boldsymbol{\xi}(\cdot)$ represents the spatial downsampling function, and $M_{\mathrm{hs}} = M /p$; $N_{\mathrm{hs}} = N /p$ with $p$ as the spatial decimation factor. Furthermore, the coded aperture used at the $w$-th snapshot is modeled as $\mathbfcal{B}_{\mathrm{hs}}^{(w)} \in \{0,1\}^{M_{\mathrm{hs}} \times N_{\mathrm{hs}} \times L}$ with entries $(\mathcal{B}^{(w)}_{\mathrm{hs}})_{(i,j,\ell)}$ for $i = 0,\ldots,M_{\mathrm{hs}} -1$, $j = 0,\ldots,N_{\mathrm{hs}}-1$, and $\ell = 0,\ldots,L-1$, and $w=0,\ldots, W_{\mathrm{hs}}-1$, where $W_{\mathrm{hs}}$ is the number of snapshots captured by the HS 3D-CASSI system. Therefore, a single measurement captured by the HS 3D-CASSI sensor at the spatial coordinate ($i$, $j$) and $w$-th snapshot can be defined as
\begin{equation}
    (\mathbf{Y}_{\mathrm{hs}}^{(w)})_{_{(i,j)}} = \left[ \sum_{l=0}^{L-1} \left(\boldsymbol{\xi}(\mathbfcal{F}) \odot \mathbfcal{B}_{\mathrm{hs}}^{(w)}\right)_{_{(i,j,\ell)}} \right] + \left(\mathbf{N}_{\mathrm{hs}}^{(w)}\right)_{_{(i,j)}}.
\end{equation}

In summary, the entire set of compressive measurements obtained by dual-arm system can be succinctly described as
\begin{eqnarray}
    \pmb{y}_{\mathrm{ms}} &  = & \pmb{H}_{\mathrm{ms}} \pmb{\textit{f}} + \pmb{n}_{\mathrm{ms}}, \\
    \pmb{y}_{\mathrm{hs}} & = & \pmb{H}_{\mathrm{hs}} \pmb{f} + \pmb{n}_{\mathrm{hs}},
\end{eqnarray}
where $\pmb{y}_{\mathrm{ms}} \in \mathbb{R}^{MNW_{\mathrm{ms}}}$ and $\pmb{y}_{\mathrm{hs}} \in \mathbb{R}^{M_{\mathrm{hs}}N_{\mathrm{hs}}W_{\mathrm{hs}}}$ are the vectors that contain the MS and HS compressive samples, respectively; $\pmb{H}_{\mathrm{ms}} \in \mathbb{R}^{MNW_{\mathrm{ms}} \times MNL}$ and $\pmb{H}_{\mathrm{hs}} \in \mathbb{R}^{M_{\mathrm{hs}}N_{\mathrm{hs}}W_{\mathrm{hs}} \times MNL}$ are the measurement matrices describing MS 3D-CASSI and HS 3D-CASSI acquisition processes, respectively; $\pmb{f}$ is the spectral image of interest in vector form; and $\pmb{n}_{\mathrm{ms}}$ and $\pmb{n}_{\mathrm{hs}}$ are noise vectors.

\begin{figure}
    \begin{center}
    \begin{tabular}{c}
        \includegraphics[width=0.50\linewidth]{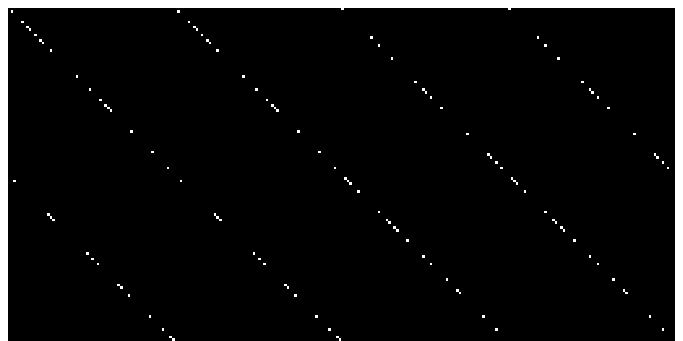}
         \\
         \footnotesize(a)
         \\
    \includegraphics[width=0.50\linewidth]{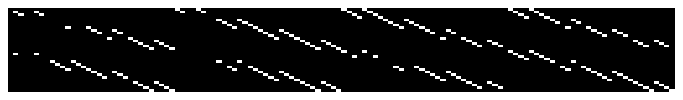}
         \\
         \footnotesize(b)
    \end{tabular}
    \end{center}
    \vspace{-10pt}
    \caption{(a) Representation of the measurement matrix $\pmb{H}_{\mathrm{ms}}$ of the MS 3D-CASSI system for $M = 8$, $N = 8$, $L=4$, $q=2$, and $W_{\mathrm{ms}} = 2$. Instance of the measurement matrix $\pmb{H}_{\mathrm{hs}}$ of the HS 3D-CASSI system for $M = 8$, $N = 8$, $L=4$, $p=2$, and $W_{\mathrm{hs}} = 2$.}
    \label{fig:sampling_matrix}
\end{figure}

In particular, the measurement matrix of the MS 3D-CASSI sensor includes both the spectral decimation function $\boldsymbol{\zeta}(\cdot)$ and the effect of the colored coded aperture. In this case, the measurement matrix can be defined as

\vspace{-5pt}
\small
\begin{equation}
    (\pmb{H}_{\mathrm{ms}})_{(u,v)} = \left\lbrace 
    \begin{tabular}{c l}
          $\frac{1}{q}(\mathcal{B}^{(w)}_{\mathrm{ms}})_{_{(i,j,\ell)}}$, & if $u = i + jM + wMN$ and \\
          & $v = i + jM + (\ell q + z_m)MN$ \\
         0, & otherwise 
    \end{tabular}
    \right.
\end{equation}\normalsize
for $z_m=0,\ldots,q$. A representation of the measurement matrix $\pmb{H}_{\mathrm{ms}}$ is illustrated in Fig. \ref{fig:sampling_matrix}(a) for $M = 8$, $N = 8$, $L=4$, $q=2$, and $W_{\mathrm{ms}} = 2$. On the other hand, the measurement matrix of the HS 3D-CASSI system should consider the spatial downsampling function $\boldsymbol{\xi}(\cdot)$ and the coded aperture pattern. More precisely, this measurement matrix can be expressed as 

\vspace{-5pt}
\small
\begin{equation}
    (\pmb{H}_{\mathrm{hs}})_{(u,v)} = \left\lbrace 
    \begin{tabular}{c l}
          $\frac{1}{p^2}(\mathcal{B}^{(w)}_{\mathrm{hs}})_{_{(i,j,\ell)}}$, & if $u = i + jM_{\mathrm{hs}} + wM_{\mathrm{hs}}N_{\mathrm{hs}}$ \\ 
          & and $v = ip + jp^2M_{\mathrm{hs}} + \ell p^2 M_{\mathrm{hs}}N_{\mathrm{hs}}$ \\
          & $+ z_{h_1} + pM_{\mathrm{hs}}z_{h_2}$\\
         0, & otherwise 
    \end{tabular}
    \right.
\end{equation}\normalsize
for $z_{h_1}=0,\ldots,p$ and $z_{h_2}=0,\ldots,p$. An example of the measurement matrix $\pmb{H}_{\mathrm{hs}}$ is displayed in Fig. \ref{fig:sampling_matrix}(b) for $M = 8$, $N = 8$, $L=4$, $p=2$, and $W_{\mathrm{hs}} = 2$.

Assuming that the noise vector entries are random samples obeying to a Gaussian statistical model, the image fusion problem reduces to minimize of the sum of the squared errors. However, this problem is ill-posed and its respective solution leads to severely degraded images. Hence, a regularization term is often included to exploit previous knowledge about the target spectral image. A sparsity-inducing term in a given transform transform is commonly included to solve this problem. Thus, the spectral image fusion problem from compressive measurements is formulated as
\begin{equation}\label{eq:imagefusionproblem}
    \hat{\pmb{f}} = \arg\min_{\pmb{f}} \frac{1}{2}\|\pmb{y}_{\mathrm{hs}} - \pmb{H}_{\mathrm{hs}}\pmb{f} \|_2^2 + \frac{\lambda_1}{2}\|\pmb{y}_{\mathrm{ms}} - \pmb{H}_{\mathrm{ms}}\pmb{f} \|_2^2 + \lambda_2 \|\pmb{\Psi}\pmb{f}\|_1,
\end{equation}
where $\lambda_1$ and $\lambda_2$ are regularization parameters.

\section{Proposed architecture}\label{sec:ladmm_net}


\subsection{Linearized ADMM}

The alternating direction method of multipliers (ADMM) is a variable splitting optimization approach that has been widely used to solve inverse problems involving large-size datasets. This approach builds an augmented Lagrangian from a constrained optimization. Furthermore, the ADMM algorithm estimates the target variable by alternately solving small subproblems \cite{boyd2011distributed}. This optimization framework has successfully applied to solve compressive sensing problems with proven convergence properties \cite{yang2011alternating, wang2019global}. In contrast to ISTA-based algorithms, ADMM-based approaches have shown in compressive sensing applications faster convergence rates and lower reconstruction errors \cite{yang2011alternating}. In addition, in comparison with AMP-based methods, ADMM algorithms do not require tailoring a denoising engine to the application at hand. In the context of alternating direction methods, the spectral image fusion problem (\ref{eq:imagefusionproblem}) can be reformulated as a constrained optimization given by
\begin{equation} \label{eq:augLag}
\begin{split}
\hat{\pmb{f}} & = \arg\min_{\pmb{f}} \frac{1}{2}\|\pmb{y}_{\mathrm{hs}} - \pmb{H}_{\mathrm{hs}}\pmb{f} \|_2^2 + \frac{\lambda_1}{2}\|\pmb{y}_{\mathrm{ms}} - \pmb{H}_{\mathrm{ms}}\pmb{f} \|_2^2 + \lambda_2 \|\pmb{b}\|_1  \\
 & \mathrm{s.t.}~\pmb{\Psi}\pmb{f} - \pmb{b} = 0,
\end{split}
\end{equation}
where $\pmb{b}$ is an auxiliary variable under the ADMM framework. In this sense, the augmented Lagrangian associated with the constrained optimization (\ref{eq:augLag}) can be written as
\begin{equation}\label{eq:augm}
    \mathbfcal{L}(\pmb{f},\pmb{b},\pmb{d}) = \frac{1}{2}\|\pmb{y}_{\mathrm{hs}} - \pmb{H}_{\mathrm{hs}}\pmb{f} \|_2^2 + \frac{\lambda_1}{2}\|\pmb{y}_{\mathrm{ms}} - \pmb{H}_{\mathrm{ms}}\pmb{f} \|_2^2  + \lambda_2 \|\pmb{b}\|_1 + \frac{\rho}{2} \|\pmb{\Psi}\pmb{f} - \pmb{b} + \pmb{d} \|_2^2,
\end{equation}
where $\rho > 0$ is a penalty parameter and $\pmb{d}$ plays the role of the Lagrangian multiplier vector. The ADMM approach attempts to optimize the augmented Lagrangian by iteratively updating the target variable $\pmb{f}$, the auxiliary variable $\pmb{b}$, and the Lagrange multiplier vector $\pmb{d}$ \cite{boyd2011distributed}. In this regard, the updating of the target variable at the iteration $k$ is estimated by solving the minimization
\begin{equation}
\pmb{f}^{(k)}  = \arg\min_{\pmb{f}}
\frac{1}{2}\|\pmb{y}_{\mathrm{hs}} - \pmb{H}_{\mathrm{hs}}\pmb{f} \|_2^2 + \frac{\lambda_1}{2}\|\pmb{y}_{\mathrm{ms}} - \pmb{H}_{\mathrm{ms}}\pmb{f} \|_2^2 + \frac{\rho}{2} \|\pmb{\Psi}\pmb{f} - \pmb{b} + \pmb{d} \|_2^2,  \label{eq:xmin_admm}
\end{equation}
whose solution leads to a closed-form expression. However, notice that the estimation of $\pmb{f}^{(k)}$ involves computationally expensive matrix operations. To overcome this drawback, a linearized version of the ADMM algorithm obtains the first-order approximation of the cost function around $\pmb{f}^{(k-1)}$ \cite{esser2010general} shown as follows

\vspace{-15pt}
\small
\begin{eqnarray}\label{eq:approx}
\frac{1}{2}\|\pmb{y}_{\mathrm{hs}} - \pmb{H}_{\mathrm{hs}}\pmb{f} \|_2^2 + \frac{\lambda_1}{2}\|\pmb{y}_{\mathrm{ms}} - \pmb{H}_{\mathrm{ms}}\pmb{f} \|_2^2 + \frac{\rho}{2} \|\pmb{\Psi}\pmb{f} - \pmb{b} + \pmb{d} \|_2^2 \hspace{-5pt} & \approx & \hspace{-5pt} \frac{1}{2}\|\pmb{y}_{\mathrm{hs}} - \pmb{H}_{\mathrm{hs}}\pmb{f}^{(k-1)} \|_2^2 + \frac{\lambda_1}{2}\|\pmb{y}_{\mathrm{ms}} - \pmb{H}_{\mathrm{ms}}\pmb{f}^{(k-1)} \|_2^2  \nonumber \\ & \cdots + &  \frac{\rho}{2} \|\pmb{\Psi}\pmb{f}^{(k-1)} - \pmb{b}^{(k-1)} + \pmb{d}^{(k-1)} \|_2^2 + \nonumber\\ & \cdots + & \left\langle \pmb{f} - \pmb{f}^{(k-1)}, \nabla(\pmb{f}^{(k-1)})  \right\rangle + \frac{\alpha}{2}\|\pmb{f} - \pmb{f}^{(k-1)}\|_2^2  \nonumber \\ & \approx & \frac{1}{2}\|\pmb{y}_{\mathrm{hs}} - \pmb{H}_{\mathrm{hs}}\pmb{f}^{(k-1)} \|_2^2 + \frac{\lambda_1}{2}\|\pmb{y}_{\mathrm{ms}} - \pmb{H}_{\mathrm{ms}}\pmb{f}^{(k-1)} \|_2^2 \nonumber \\ & \cdots + & \frac{\rho}{2} \|\pmb{\Psi}\pmb{f}^{(k-1)} - \pmb{b}^{(k-1)} + \pmb{d}^{(k-1)} \|_2^2  \nonumber\\
& \cdots + & \frac{\alpha}{2} \|\pmb{f} - \pmb{f}^{(k-1)} + \frac{1}{\alpha} \nabla(\pmb{f}^{(k-1)}) \|_2^2.   \label{eq:xmin_admm}
\end{eqnarray}\normalsize
where $\nabla(\pmb{f}^{(k-1)})$ denotes the gradient of the cost function around $\pmb{f}^{(k-1)}$ given by 
\begin{equation}\label{eq:gradient}
    \nabla (\pmb{f}^{(k-1)}) =  \pmb{H}_{\mathrm{hs}}^{\top}\left(\pmb{H}_{\mathrm{hs}}\pmb{f}^{(k-1)} - \pmb{y}_{\mathrm{hs}}\right) + \lambda_1 \pmb{H}_{\mathrm{ms}}^{\top}\left(\pmb{H}_{\mathrm{ms}}\pmb{f}^{(k-1)} - \pmb{y}_{\mathrm{ms}}\right) + \rho \pmb{\Psi}^{\top} \left(\pmb{\Psi}\pmb{f}^{(k-1)} - \pmb{b}^{(k-1)} + \pmb{d}^{(k-1)} \right).
\end{equation}
By substituting (\ref{eq:gradient}) in (\ref{eq:approx}), and minimizing the resulting cost function, the update $\pmb{f}^{(k)}$ can be obtained as
\begin{equation}
    \pmb{f}^{(k)}  =  \pmb{f}^{(k-1)} - \frac{1}{\alpha} \left[\pmb{H}_{\mathrm{hs}}^{\top}\left(\pmb{H}_{\mathrm{hs}}\pmb{f}^{(k-1)} - \pmb{y}_{\mathrm{hs}}\right)  + \lambda_1 \pmb{H}_{\mathrm{ms}}^{\top}\left(\pmb{H}_{\mathrm{ms}}\pmb{f}^{(k-1)} - \pmb{y}_{\mathrm{ms}}\right) + \rho  \pmb{\Psi}^{\top} \left(\pmb{\Psi}\pmb{f}^{(k-1)} - \pmb{b}^{(k-1)} + \pmb{d}^{(k-1)} \right)  \right]. \label{eq:updatex1}
\end{equation}


In order to update the dual variable at the $k$-th iteration, the algorithm aims at solving the optimization
\begin{equation}
\hat{\pmb{b}}^{(k)}  = \arg\min_{\pmb{b}} \left\lbrace
\frac{\rho}{2} \|\boldsymbol{\Psi}\pmb{f}^{(k)} - \pmb{b} + \pmb{d}^{(k-1)} \|_2^2 +  \lambda_2 \|\pmb{b} \|_1 
\right\rbrace ,
\end{equation}
that describes the proximal mapping when the regularizer is given by $\| \pmb{b} \|_1$. Notice that the solution to this optimization problem is reduced to the soft-thresholding operation when the signal of interest is sparsely represented in a transform domain, i.e.
\begin{equation}
    \pmb{b}^{(k)}  =  \mathcal{S}_{\tilde{\lambda}}\left(\boldsymbol{\Psi} \pmb{f}^{(k)} + \pmb{d}^{(k-1)} \right), \label{eq:lmbd1}
\end{equation}
where $\tilde{\lambda} = \lambda_2 / \rho$ and $\mathcal{S}_{\tilde{\lambda}}(x) = \mathrm{sign}(x) \max\{x-\tilde{\lambda}, 0\}$. Finally, the Lagrange multiplier vector is updated at the $k$-th iteration as follows
\begin{equation}\label{eq:lagrange_upd}
    \hat{\pmb{d}}^{(k)} =  \pmb{d}^{(k-1)} + \pmb{\Psi}\pmb{f} - \pmb{b}^{(k)}.
\end{equation}

\subsection{LADMM-Net}

\begin{figure*}
    \centering
    \includegraphics[width=1.00\linewidth]{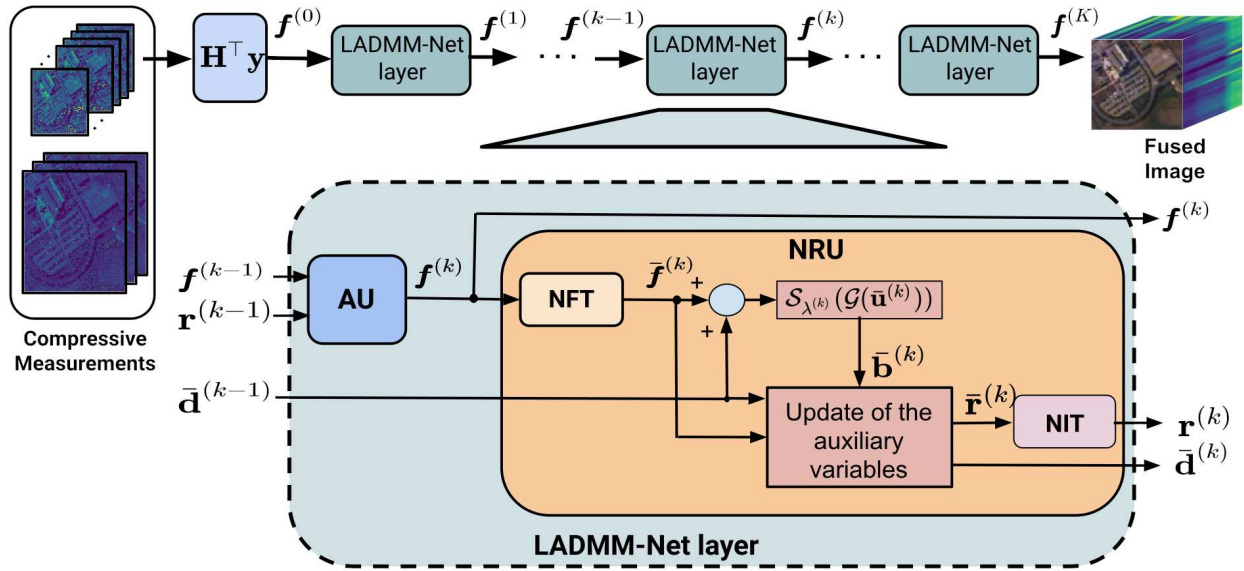}
    \caption{Schematic of the proposed LADMM-Net architecture to solve the spectral image fusion problem from multi-sensor compressive measurements.}
    \label{fig:deep_arch}
\end{figure*}

In this section, we introduce a deep learning architecture called LADMM-Net that exploits the advantages of both the LADMM optimization algorithm and the unrolling approach to solve the spectral image fusion problem from HS and MS compressive measurements. More precisely, the basic idea is to cast each iteration of the LADMM algorithm into a processing block with a convolutional neural network (CNN). Therefore, the overall architecture consists of multiple processing layers deploying a deep network. Fig. \ref{fig:deep_arch} shows a schematic of the proposed architecture for solving the spectral image fusion problem from HS and MS compressive measurements.

Notice that the data fusion problem attempts to describe the measurements that minimize the sparse image representation in a given transform domain. In general, combinations of orthogonal transforms such as the discrete cosine transform (DCT) or the orthogonal discrete wavelet transform (DWT) lead to degraded reconstructions. Therefore, an approach that includes nonlinear transform functions whose parameters can be learned from available data can be developed to improve the image recovery performance. In this work, we follow a similar approach to that developed for the ISTA-Net method \cite{zhang2018ISTA}. Basically, this approach substitutes the hand-crafted transform by a network structure that can be considered as a nonlinear transform function $\mathbfcal{G}(\cdot)$ that induces sparsity to the image representation. In contrast to the ISTA-Net approach, the transform relied on CNN is optimized to estimate both the auxiliary variable and the Lagrange multiplier that capture the high-frequency information of the spectral image of interest. Notice that the Lagrangian term encloses the information embedded in both the auxiliary variable and the Lagrange multiplier, therefore, the proposed network architecture attempts to optimize the image prior information and convexity-inducing penalty to improve the algorithm convergence. As shown below, the proposed approach exhibits a superior performance without resorting to the target variable mapping onto the residual domain.

\begin{figure}
    \centering
    \includegraphics[width=0.40\linewidth]{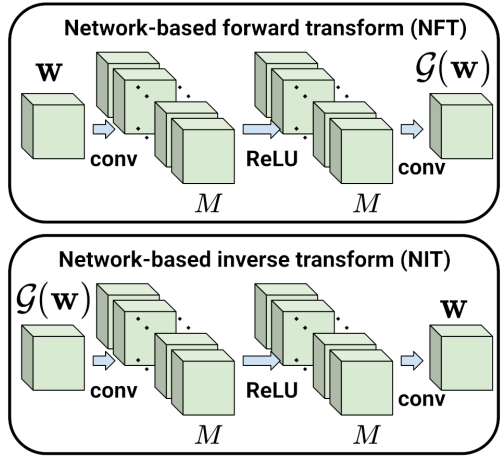}
    \caption{Schematic of the network-based forward transform (NFT) and the network-based inverse transform (NIT) presented for 3-D images.}
    \label{fig:deep_Psi}
\end{figure}

Figure \ref{fig:deep_Psi} shows the structure of the network-based forward transform (NFT). As can be seen, this structure consists of a rectified linear unit (ReLU) between two convolutional operators \cite{zhang2018ISTA}. Notice the 3-D shape of the data processed by the network. Specifically, the first convolutional operator consists of $M$ filters with size $3 \times 3 \times L$ while the second convolutional network contains $L$ filters with size $3 \times 3 \times M$. In addition, the proposed architecture also includes a network structure that recovers the image in its conventional domain. This structure, called network-based inverse transform (NIT), is also shown in Fig. \ref{fig:deep_Psi}. As can be seen in this figure, the network exhibits the same structure as the NFT network. It is worth noting that the NIT should be able to invert the effect of the forward transform, i.e. $\tilde{\mathbfcal{G}}(\mathbfcal{G}(\mathbf{x})) \approx \mathbf{x}$. This leads to architectures with end-to-end training schemes such that the whole network structure recovers the target image from compressive measurements. Furthermore, the symmetry of the network-based transform functions is considered during the training stage by including an invertibility error term in the loss function \cite{zhang2018ISTA}. In this context, the spectral image fusion problem from HS and MS compressive measurements can be described as
\begin{equation}\label{eq:new_problem}
    \hat{\pmb{f}} = \arg\min_{\pmb{f}} \frac{1}{2}\|\pmb{y}_{\mathrm{hs}} - \pmb{H}_{\mathrm{hs}}\pmb{f} \|_2^2 + \frac{\lambda_1}{2}\|\pmb{y}_{\mathrm{ms}} - \pmb{H}_{\mathrm{ms}}\pmb{f} \|_2^2 + \lambda_2 \|\mathbfcal{G}( \pmb{f})\|_1.
\end{equation}

An unrolled algorithm approach is developed to solve this problem that casts each iteration of the LADMM algorithm to a processing module dubbed LADMM-Net layer. This layer attempts to efficiently solve the updates (\ref{eq:updatex1}), (\ref{eq:lmbd1}), and (\ref{eq:lagrange_upd}) by exploiting the flexibility of the network-based transform functions. In essence, each LADMM-Net layer consists of an approximation unit (AU) and a network-based refinement unit (NRU) that are described as follows. 

\subsection{Approximation unit (AU)}

This module estimate the target image according to (\ref{eq:updatex1}). This unit is also shown in Fig. \ref{fig:deep_arch}. Therefore, taking into account that $\pmb{f}^{(k-1)}$ and $\pmb{r}^{(k-1)}$ have been yielded by the previous layer, and considering the information embedded in the measurement matrices $\pmb{H}_{\mathrm{ms}}$ and $\pmb{H}_{\mathrm{hs}}$, the output of this processing unit is expressed as
\begin{equation}\label{eq:au_eq}
    \pmb{f}^{(k)} = \pmb{f}^{(k-1)} - \frac{1}{\alpha^{(k)}} \left[ \pmb{H}_{\mathrm{hs}}^{\top}\left(\pmb{H}_{\mathrm{hs}}\pmb{f}^{(k-1)} - \pmb{y}_{\mathrm{hs}}\right) +  \lambda_1^{(k)} \pmb{H}_{\mathrm{ms}}^{\top}\left(\pmb{H}_{\mathrm{ms}}\pmb{f}^{(k-1)} - \pmb{y}_{\mathrm{ms}}\right) + \rho^{(k)} \pmb{r}^{(k-1)} \right]
\end{equation}
where $\alpha^{(k)}$ is a learnable step-size at the $k$-the phase; $(\lambda_1^{(k)}, \rho^{(k)})$ are the learnable regularization parameters at the $k$-the phase; and  $\pmb{r}^{(k-1)}$ is the augmented Lagragian term computed by the NRU of the previous processing layer. Notice that, the learnable parameters ($\alpha^{(k)}$,$\lambda_1^{(k)}$,$\rho^{(k)}$) are different across the layers. Furthermore, under the LADMM approach, the estimate of the target variable $\pmb{f}^{(k)}$ can be interpreted as a gradient of the cost function quadratic terms around $\pmb{f}^{(k-1)}$. This update does not resort to expensive matrix inversions.

\subsection{Network-based refinement unit (NRU)}

An schematic of this module is also illustrated in Fig. \ref{fig:deep_arch}. In particular, the NRU aims at obtaining the update of the dual variable $\pmb{b}^{(k)}$ and the Lagrange multiplier $\pmb{d}^{(k)}$. Taking into account the fusion problem described in (\ref{eq:new_problem}), the dual variable estimate aims to solve the optimization problem given by 
\begin{equation}\label{eq:new_dual_opt}
\hat{\pmb{b}}^{(k)}  = \arg\min_{\pmb{b}} \left\lbrace
\frac{\rho}{2} \|\mathbfcal{G}(\pmb{f}^{(k)}) - \pmb{b} + \pmb{d}^{(k-1)} \|_2^2 +  \lambda_2 \|\pmb{b} \|_1 
\right\rbrace ,
\end{equation}
whose solution is obtained as 
\begin{equation}\label{eq:new_dual_upd}
    \bar{\pmb{b}}^{(k)} = \mathcal{S}_{\tilde{\lambda}^{(k)}}(\bar{\pmb{u}}^{(k)}) = \mathcal{S}_{\tilde{\lambda}^{(k)}}(\mathbfcal{G}({\pmb{f}}^{(k)}) + \pmb{d}^{(k-1)}),
\end{equation}
where $\tilde{\lambda}^{(k)} = \lambda_2 /\rho$ is the soft-thresholding parameter that is learned at the $k$-th processing stage. To obtain the dual variable update $\pmb{b}^{(k)}$, the current estimate of the target variable $\pmb{f}^{(k)}$ should be sparsely represented by the nonlinear transform function $\mathbfcal{G}(\cdot)$ recreated by the convolutional neural network. This transform performed by the NFT block can be described as $\bar{\pmb{f}}^{(k)} = \mathbfcal{G}(\pmb{f}^{(k)})$. Then, the dual variable update is obtained according to (\ref{eq:new_dual_upd}). In other words, the soft-thresholding is applied to a shifted version of the image transform. The Lagrange multiplier vector estimated by the previous layer $\pmb{d}^{(k-1)}$ should be available.

Subsequently, the Lagrange multiplier update is obtained as 
\begin{equation}
    \pmb{d}^{(k)} = \pmb{d}^{(k-1)} + \mathbfcal{G}(\pmb{f}^{(k)}) - \pmb{b}^{(k)},
\end{equation}
that also requires $\pmb{d}^{(k-1)}$. Therefore, as can be seen in Fig. \ref{fig:deep_arch}, the dual variable update $\pmb{d}^{(k)}$ is defined as a layer output.

Upon a closer look of (\ref{eq:updatex1}), the term $\boldsymbol{\Psi}^{\top}(\boldsymbol{\Psi} \pmb{f}^{(k)} + \pmb{d}^{(k)} - \pmb{b}^{(k)})$ can be seen as the contribution of the augmented Lagragian term. In particular, this vector contains the information of the coefficients removed by the soft-thresholding function that commonly contains both the high-frequency components of the image of interest and coefficients associated with the measurement noise. As can be seen in (\ref{eq:au_eq}), the terms to compute the target variable update are available. Therefore, the residual is computed in the transformed domain and then the network-based inverse transform (NIT) is applied to this subtraction, in other words,  
\begin{eqnarray}
    \pmb{r}^{(k)} & = &
    \tilde{\mathbfcal{G}}\left(\mathbfcal{G}(\pmb{f}^{(k)}) + \pmb{d}^{(k)} - {\pmb{b}}^{(k)}\right),
\end{eqnarray}
where $\pmb{r}^{(k)}$ is also defined as an output of the processing layer. The information of the dual variable is implicitly included in this residual vector. Hence, the proposed architecture considers the high-frequency information of the spectral image without remarkably increasing the memory requirements.

\subsection{LADMM-Net parameters}

Basically, the proposed architecture attempts to learn a set of parameters denoted by $\boldsymbol{\Theta}$ whose elements are described as follows. First, the step-size $\alpha^{(k)}$ and the regularization parameters ($\rho^{(k)}$,$\lambda_1^{(k)}$) are updated in every approximation unit. Furthermore, the threshold $\tilde{\lambda}^{(k)}$ and the parameters of the network-based nonlinear transforms $\mathbfcal{G}^{(k)}$ and $\tilde{\mathbfcal{G}}^{(k)}$ are learned in the NRU. In summary, the proposed deep network can be described by the set of parameters $\boldsymbol{\Theta} = \{\alpha^{(k)},\rho^{(k)}, \lambda_1^{(k)}, \tilde{\lambda}^{(k)},\mathbfcal{G}^{(k)}, \tilde{\mathbfcal{G}}^{(k)}\}_{k=1}^{K}$, where $K$ is the number of the LADMM-Net layers. The number of learnable parameters that contains the proposed architecture is $K(4 + 36ML)$.

\subsubsection{Initialization}

An initial estimate $\pmb{f}^{(0)}$ is required by the proposed network-based architecture. In this sense, since the information of the measurement matrices ($\pmb{H}_{\mathrm{ms}}$, $\pmb{H}_{\mathrm{hs}}$) is available, an initial approximation of the target variable can be computed as $\pmb{f}^{(0)} = \frac{1}{2}\pmb{H}_{\mathrm{ms}}^{\top}\pmb{y}_{\mathrm{ms}} + \frac{1}{2}\pmb{H}_{\mathrm{hs}}^{\top}\pmb{y}_{\mathrm{hs}}$. Moreover, the convolutional neural networks implementing the invertible nonlinear transforms ($\mathbfcal{G}^{(k)}$, $\tilde{\mathbfcal{G}}^{(k)}$) are initialized with random values generated by the Xavier method. Finally, algorithm parameters along the multiples layers are initialized as $\alpha^{(k)} = 0.5$, $\rho^{(k)} = 0.1$, $\lambda_{1}^{(k)} = 1$, $\tilde{\lambda}^{(k)} = 0.01$, for $k =1, \ldots, K$.

\subsection{LADMM-Net training}

In the context of the spectral image fusion from HS and MS compressive measurements, the ground truth image is denoted by $\pmb{f}_v$. On the other hand, the corresponding HS and MS compressive projections $(\pmb{y}_{\mathrm{ms}},\pmb{y}_{\mathrm{hs}})_{v}$ can be described as the input dataset. Therefore, the training set can be built as $\Gamma = \{\pmb{f}_v,(\pmb{y}_{\mathrm{ms}},\pmb{y}_{\mathrm{hs}})_{v} \}_{b=1}^{B}$. In this work, we use a similar loss function than that reported in \cite{zhang2018ISTA} for the ISTA-Net. Specifically, the training stage attempts to minimize the mean square error $\mathcal{H}(\boldsymbol{\Theta})_{data}$ while an invertibility error $\mathcal{H}(\boldsymbol{\Theta})_{inv}$ that estimates the symmetry degree of network-based nonlinear transform functions is also considered. Thus, given the training set $\Gamma$, the loss function is defined as
\begin{equation}
 \mathcal{H}(\boldsymbol{\Theta}) = \mathcal{H}(\boldsymbol{\Theta})_{data} + \gamma \mathcal{H}(\boldsymbol{\Theta})_{inv},
\end{equation}
with
\begin{eqnarray}
    \mathcal{H}(\boldsymbol{\Theta})_{data} & = & \frac{1}{B} \sum_{b=1}^B \|\pmb{f}_v(\boldsymbol{\Theta},(\pmb{y}_{\mathrm{ms}},\pmb{y}_{\mathrm{hs}})_{v}) - \pmb{f}_v\|_2^2, \\
    \mathcal{H}(\boldsymbol{\Theta})_{inv}& = &\frac{1}{BK} \sum_{b=1}^B \sum_{k=1}^K \|\tilde{\mathbfcal{G}}(\mathbfcal{G}(\pmb{f}^{(k)})) - \pmb{f}^{(k)}  \|_2^2,
\end{eqnarray}
where $\pmb{f}_v(\boldsymbol{\Theta},\Gamma)$ is the fused image recovered by the proposed architecture using the set of learnable parameters $\boldsymbol{\Theta}$ and the compressive measurements $(\pmb{y}_{\mathrm{ms}},\pmb{y}_{\mathrm{hs}})_{v}$. Note that $K$ is the number of LADMM-Net layers and $B$ is the batch size. In this work, $\gamma$ is set to 0.1.

As can be seen, the proposed approach unfolds the linearized ADMM algorithm whose convergence analysis is developed in \cite{yang2011alternating, wang2019global}. In contrast to the linearized ADMM algorithm that uses a predefined transform, the proposed approach attempts to learn a nonlinear transform describing the image detail information embedded in both the dual variable and the Lagrange multiplier. In this regard, the convergence of the proposed approach shall also depend on the parameter setting at the training stage, including the initialization of the learnable weights and the learning rate. Therefore, since the nonlinear nature of the learned transform and the huge amount of parameters involved at the learning stage, the convergence analysis is considered an open research problem.

\section{Results}\label{sec:results}

To evaluate the proposed spectral image fusion approach, we use two available databases: the Harvard database and the CAVE database. Furthermore, we test the proposed approach to fuse a spectral image from HS and MS compressive measurements captured at the laboratory. Finally, the proposed algorithm unrolling technique is extended to recover grayscale images from compressive random projections.  

\subsection{Harvard database}

\begin{figure*}
    \begin{center}
    \begin{tabular}{c c c c c c}
    \hspace{-15pt}
    \includegraphics[width=0.165\linewidth]{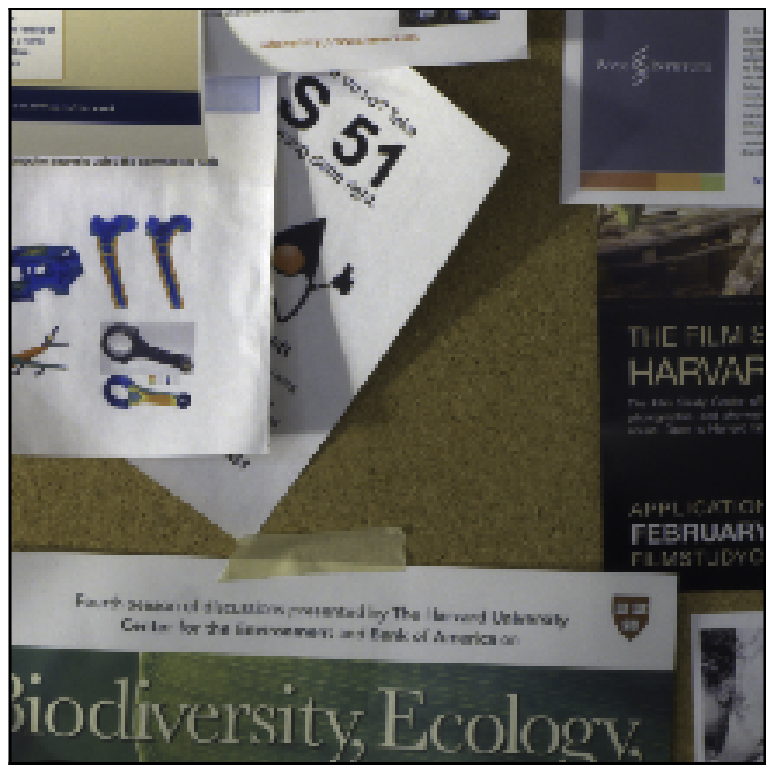}
    &
    \hspace{-15pt}
    \includegraphics[width=0.165\linewidth]{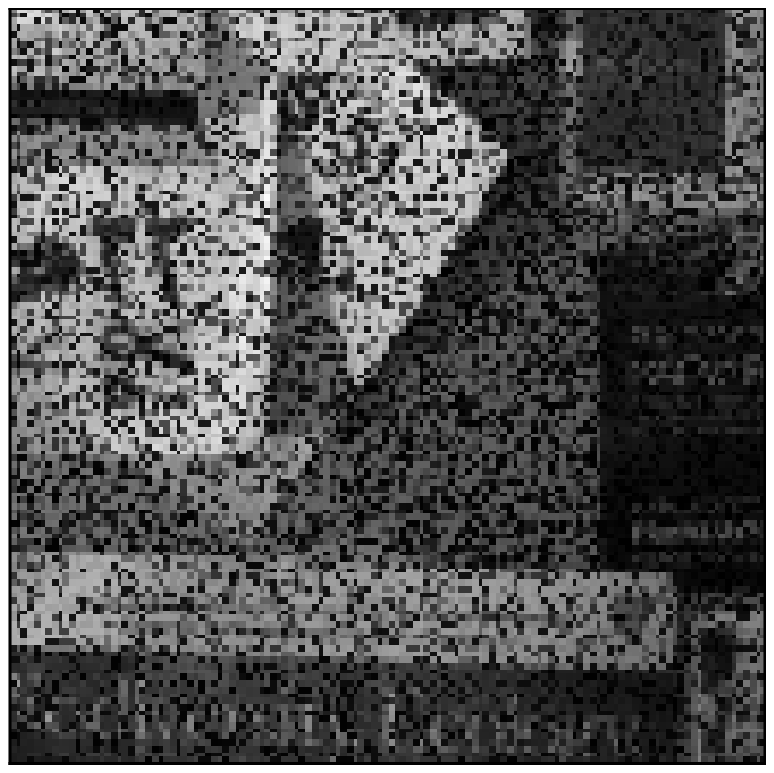}
    &
    \hspace{-15pt}
    \includegraphics[width=0.165\linewidth]{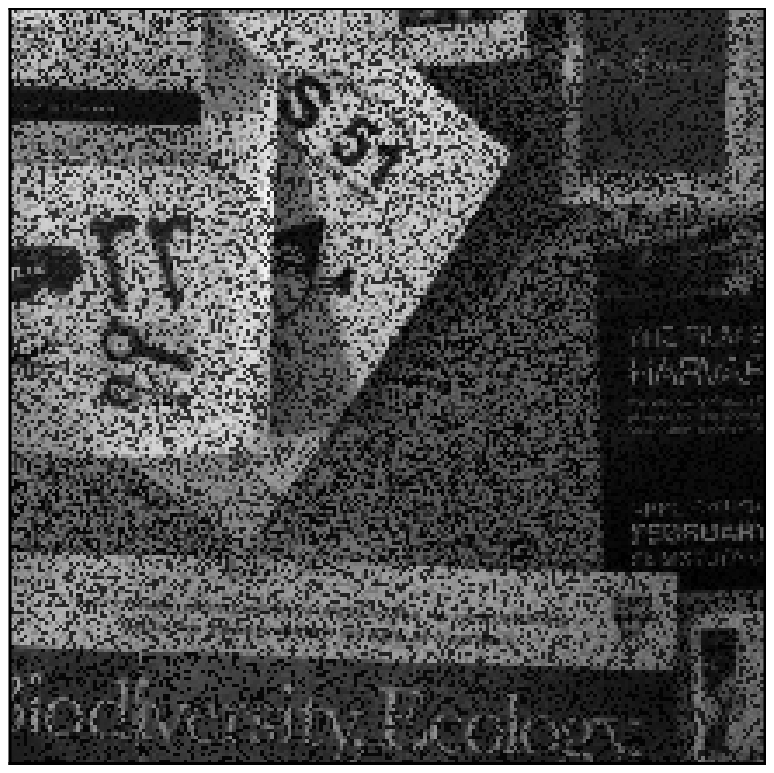}
    &
    \hspace{-15pt}
    \includegraphics[width=0.165\linewidth]{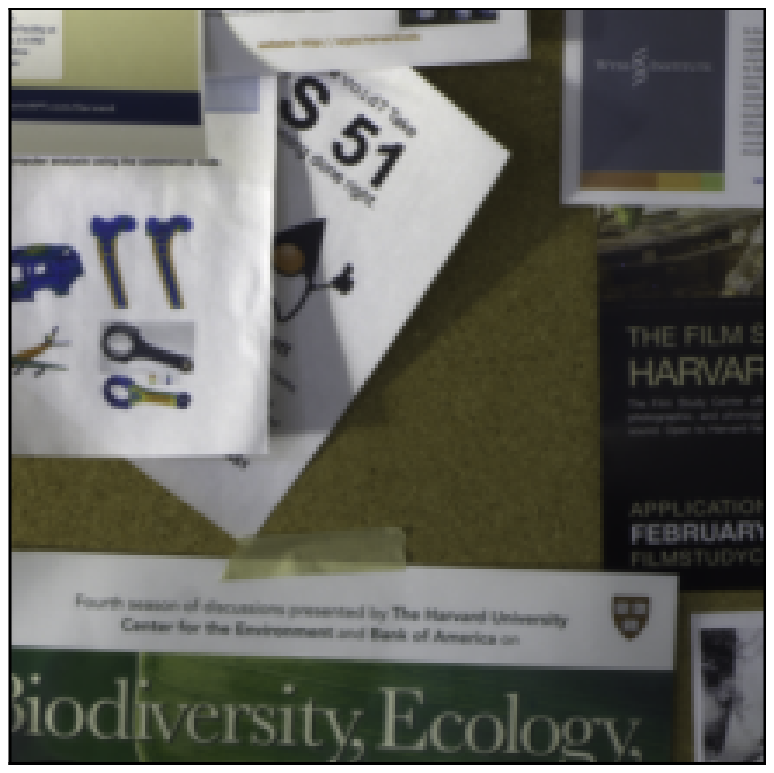}
    &
    \hspace{-15pt}
    \includegraphics[width=0.165\linewidth]{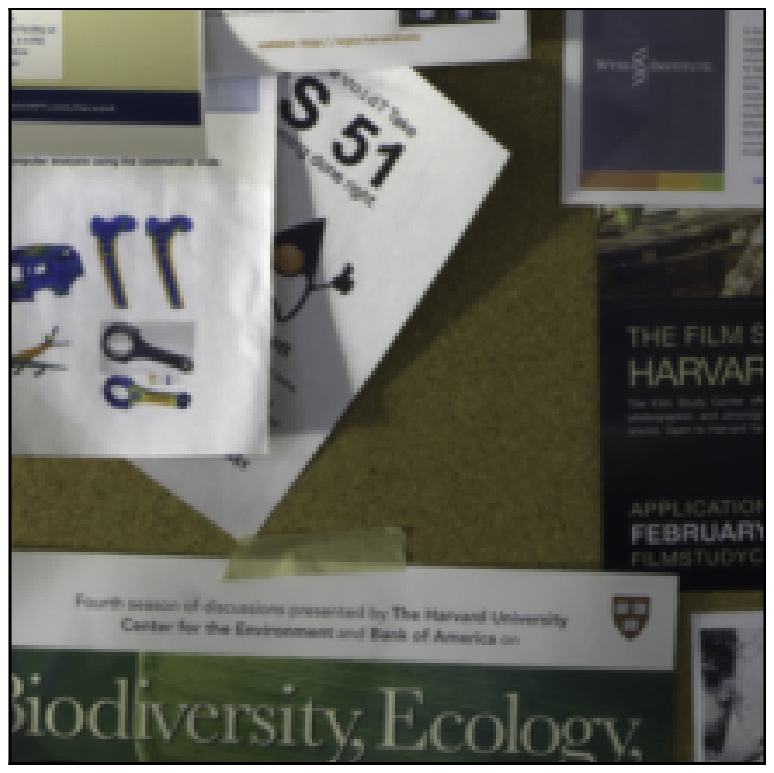}
    &
    \hspace{-15pt}
    \includegraphics[width=0.165\linewidth]{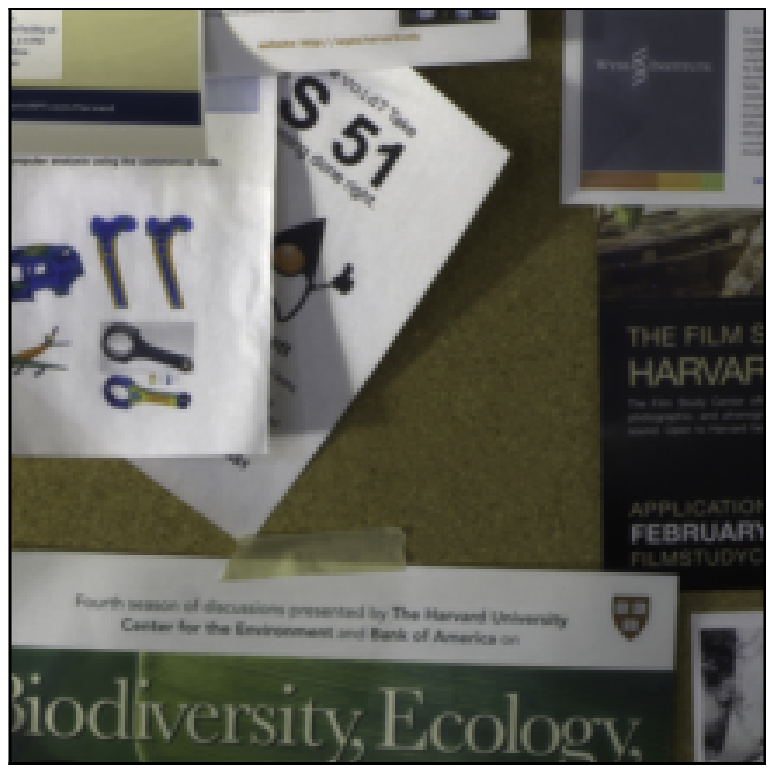}
        \\
    \scriptsize
     \hspace{-15pt}
         (a) Original Image
     & \scriptsize
     \hspace{-15pt}
         (b) HS CASSI shot
     & \scriptsize
     \hspace{-15pt}
         (c) MS CASSI shot
     & \scriptsize
     \hspace{-15pt}
         (d) 5 layers
     & \scriptsize
     \hspace{-15pt}
         (e) 7 layers
     & \scriptsize
     \hspace{-15pt}
         (f) 10 layers
    \end{tabular}
    \end{center}
    \caption{Harvard database. (a) RGB composite of the high-resolution spectral image; (b) 3D-CASSI shot; (c) 3D-CASSI shot. RGB composite of the spectral image fused by the proposed method using (d) 5 layers with PSNR = 33.18 dB and SSIM = 0.927, (E) 7 layers with PSNR = 35.53 dB and SSIM = 0.940, (f) 10 layers with PSNR = 36.38 dB and SSIM = 0.945.}
    \label{fig:layers}
\end{figure*}

This database consists of fifty high-resolution images that exhibit indoor and outdoor scenes under natural illumination \cite{chakrabarti2011statistics}. Since the proposed approach is compared to other model-based methods that sparsely represent the spectral image in a predefined transform domain, we extracted $72$ spectral images with dimensions $512 \times 512$ pixels and $31$ spectral bands in the wavelength range from $420$ nm to $720$ nm. Figure \ref{fig:layers}(a) displays the RGB composite of a high-resolution spectral image.

A dual-arm optical architecture is simulated to generate the multi-sensor compressive measurements. In this case, the HS images are obtained as degraded versions of the high-resolution images across the spatial coordinates with a $16:1$ decimation factor, i.e. $p=4$. More precisely, every HS image signature is yielded by averaging neighboring spectral pixels of the corresponding high-resolution image. Therefore, every HS image comprises a datacube with dimensions of $128 \times 128$ pixels and $31$ spectral bands. Afterward, the HS compressive measurements are obtained by simulating a multi-frame HS 3D-CASSI system, where the number of acquired snapshots depends on the desired compression ratio. Fig. \ref{fig:layers}(b) displays an HS 3D-CASSI projection for a compression ratio of $25\%$. On the other hand, the MS images are spectrally degraded versions of the high-resolution images with a $2:1$ decimation factor ($q=2$). In particular, each band of an MS image is obtained by averaging contiguous spectral bands of the corresponding high-resolution image. Thus, an MS image exhibits a size of $512 \times 512 \times 16$. Subsequently, the MS compressive samples are obtained by simulating a multi-frame MS 3D-CASSI optical architecture, where the number of captured snapshots also depends on the compression ratio. Fig. \ref{fig:layers}(c) shows a MS 3D-CASSI snapshot for a compression ratio of $25\%$.

Specifically, the proposed feature fusion method is implemented in Pytorch\footnote{The source code of the proposed method can be downloaded from this link:\url{https://github.com/JuanMarcosRamirez/LADMM_Net_Pytorch}}. The LADMM-Net models are trained for different layer numbers $N_{\ell}$ and different compression ratios $\{25.00\%, 37.50\%, 50.00\%\}$. In addition, we use 48 high-resolution spectral images and their respective compressive projections to train the proposed unrolled algorithm. Since sampling matrices demand high storage costs, we exploit the structure of the 3D-CASSI system to efficiently implement the matrix products that involve $\pmb{H}_{\mathrm{hs}}$ and $\pmb{H}_{\mathrm{ms}}$ and their transposes. Moreover, the Adam optimization algorithm was used to train the unfolded network with a learning rate $\zeta = 0.0005$, and $256$ epochs. To avoid memory overloading, the batch size was fixed to $1$. The experiments were performed on a desktop computer with Intel Core i7 CPU, $3.00$ GHz, $64$ GB RAM, and GTX-1050 GPU.

Fig. \ref{fig:layers}(d)-(f) show the RGB composites of the spectral images fused by the proposed algorithm unrolling approach as the number of layers increases. Notice that the peak signal-to-noise ratio (PSNR) and the structural similarity (SSIM) index yielded by fused images are included in the caption of this figure. Furthermore, to quantitatively evaluate the performance of the proposed fusion approach, Table \ref{tab:layers} shows the averages of various performance metrics over 24 testing images as the number of layers increases. More precisely, we include the ensemble average of the PSNR, SSIM, the spectral angle mapper (SAM), the running time, and the training time. As can be observed in this table, the performance of the proposed approach improves as the number of processing layers increases. It can be also seen that the computation times are larger as the number of layers increases. Nevertheless, the proposed method yields a remarkable performance for a small number of processing blocks. It is worth noting that the spectral image fusion involves large data sizes, and multiple variables with the target image size should be stored by each processing layer. Therefore, the maximum number of processing layers is constrained by the GPU memory size.

\begin{table}\footnotesize
    \caption{Comparison of the LADMM-Net for different number of layers}
    \centering
    \begin{tabular}{c||c c c}
    \hline
    \hline
         & 5 layers & 7 layers  & 10 layers \\
    \hline
    \hline
        PSNR[dB] & 34.47 $\pm$ 4.16  & 35.39 $\pm$ 3.65 & \textbf{36.15} $\pm$ \textbf{3.84} \\
        SSIM & 0.942 $\pm$ 0.036  & 0.952 $\pm$ 0.027 & \textbf{0.955} $\pm$ \textbf{0.029} \\
        SAM & 0.111 $\pm$ 0.058 & 0.100 $\pm$ 0.046 & \textbf{0.095} $\pm$ \textbf{0.051}\\
        Running time [s] & \textbf{0.15} $\pm$ \textbf{0.01} & 0.22 $\pm$ 0.01 & 0.30 $\pm$ \textbf 0.02\\
        Training time & \textbf{1h29min} & 2h2min & 2h54min \\
    \hline
    \hline
    \end{tabular}
    \label{tab:layers}
\end{table}

For comparison purposes, different methods that fuse spectral images from compressive measurements are implemented. The first approach aims at recovering both the HS image and the MS image from their respective compressive measurements, and then, the fast fusion based on Sylvester equation (FUSE) \cite{Wei2015Fast} is applied to obtain the high-resolution spectral image. This approach is referred to as Rec + FUSE. In this case, an ADMM-based algorithm that solves an $\ell_1$-regularized inverse problem is used to reconstruct both the HS image and the MS image. Secondly, the spectral image fusion from compressive measurements (SIFCM) method \cite{Vargas2019Spectral} is implemented. Additionally, ISTA-Net \cite{zhang2018ISTA} and OPINE-Net methods \cite{zhang2020optimization} are adapted to the spectral image fusion problem from compressive measurements. In this experiment, LADMM-Net, ISTA-Net, and OPINE-Net are implemented using 10 processing layers and 48 training samples. For these deep learning architectures, we use the Adam optimization algorithm and a learning rate $\zeta=0.0005$. Fig. \ref{fig:image_comparison} displays the RGB composite of the reconstructed images using a compression ratio of $25\%$. Notice that the PSNR values are included in the bottom part of every recovered image. In addition, Fig. \ref{fig:signatures} shows the spectral signatures of selected pixels belonging to the recovered images. As can be observed in these figures, the proposed architecture achieves the best performance with respect to the other approaches.

\newcommand\setwideb{0.165}
\begin{figure*}
\begin{center}
\begin{tabular}{c c c c c c}
    \hspace{-15pt}
    \begin{minipage}{\setwideb\linewidth}
    \includegraphics[width=\linewidth]{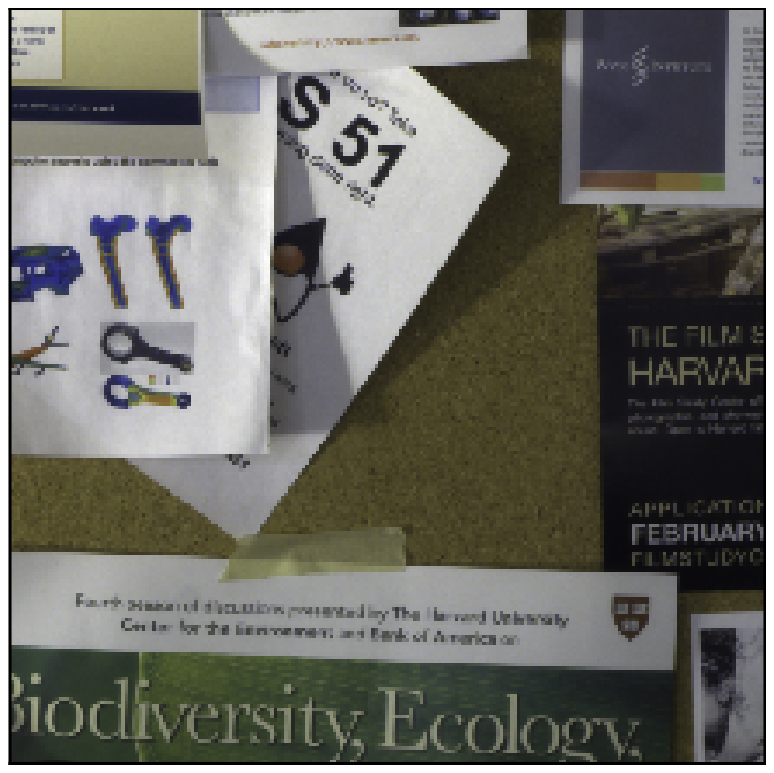}
    \end{minipage}
    &
    \hspace{-15pt}
    \begin{minipage}{\setwideb\linewidth}
    \includegraphics[width=\linewidth]{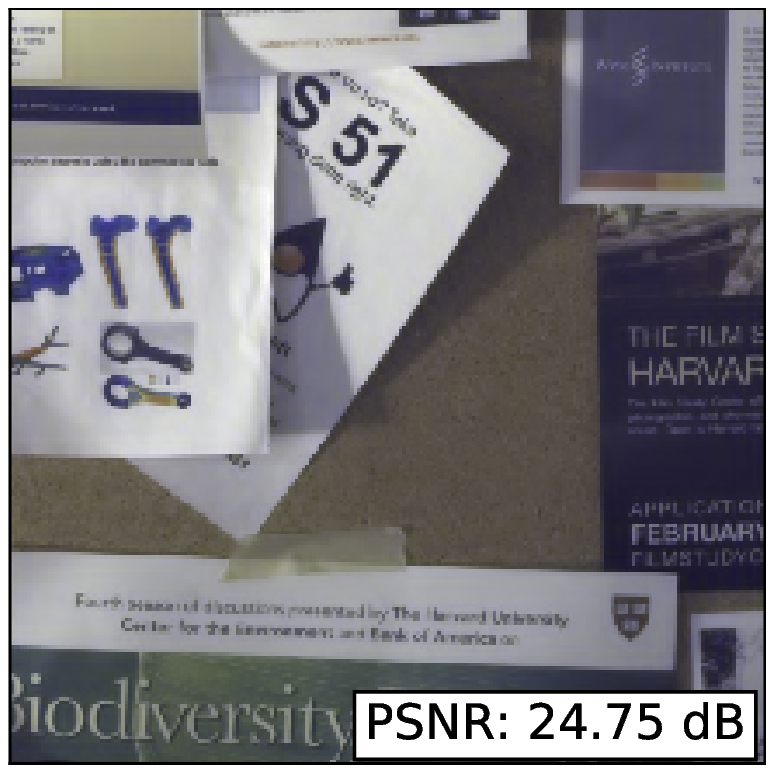}
    \end{minipage}
    &
    \hspace{-15pt}
    \begin{minipage}{\setwideb\linewidth}
    \includegraphics[width=\linewidth]{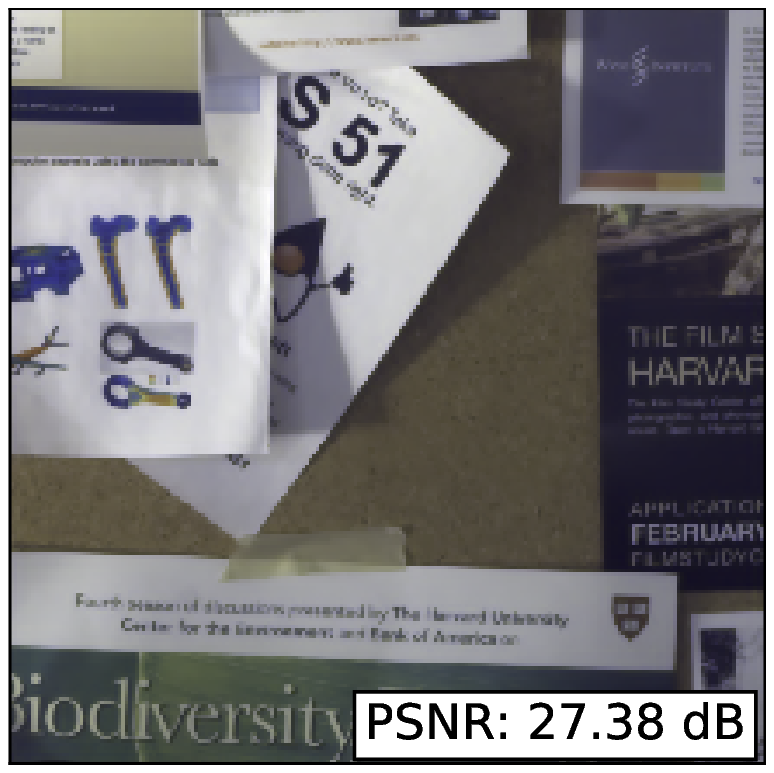}
    \end{minipage}
    &
    \hspace{-15pt}
    \begin{minipage}{\setwideb\linewidth}
    \includegraphics[width=\linewidth]{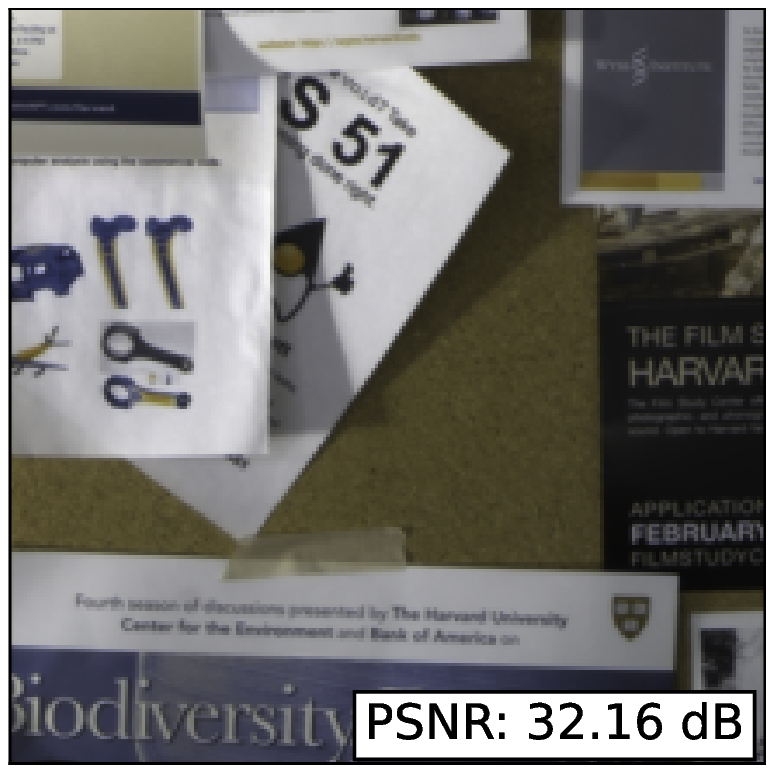}
    \end{minipage}
    &
    \hspace{-15pt}
    \begin{minipage}{\setwideb\linewidth}
    \includegraphics[width=\linewidth]{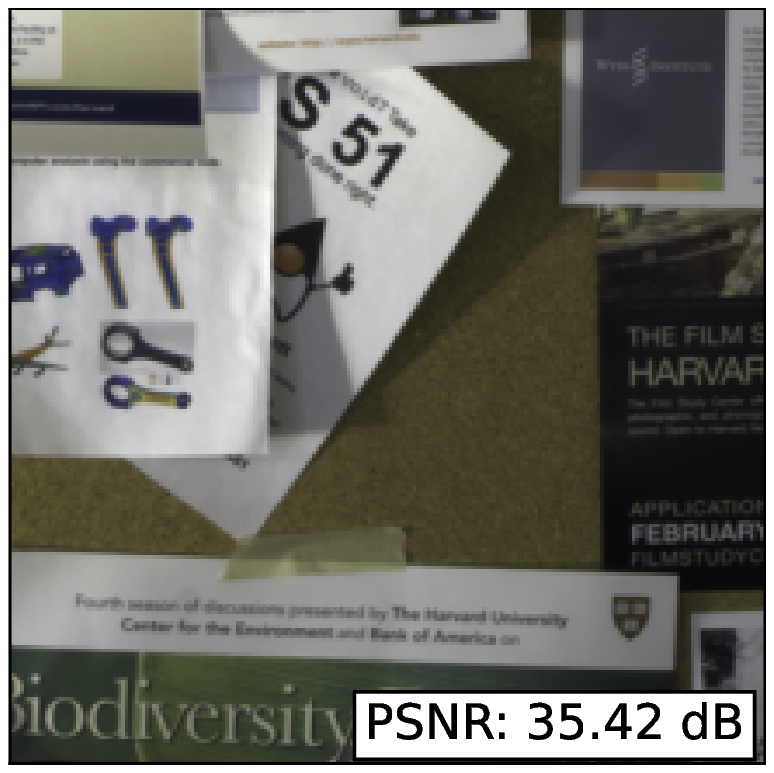}
    \end{minipage}
    &
    \hspace{-15pt}
    \begin{minipage}{\setwideb\linewidth}
    \includegraphics[width=\linewidth]{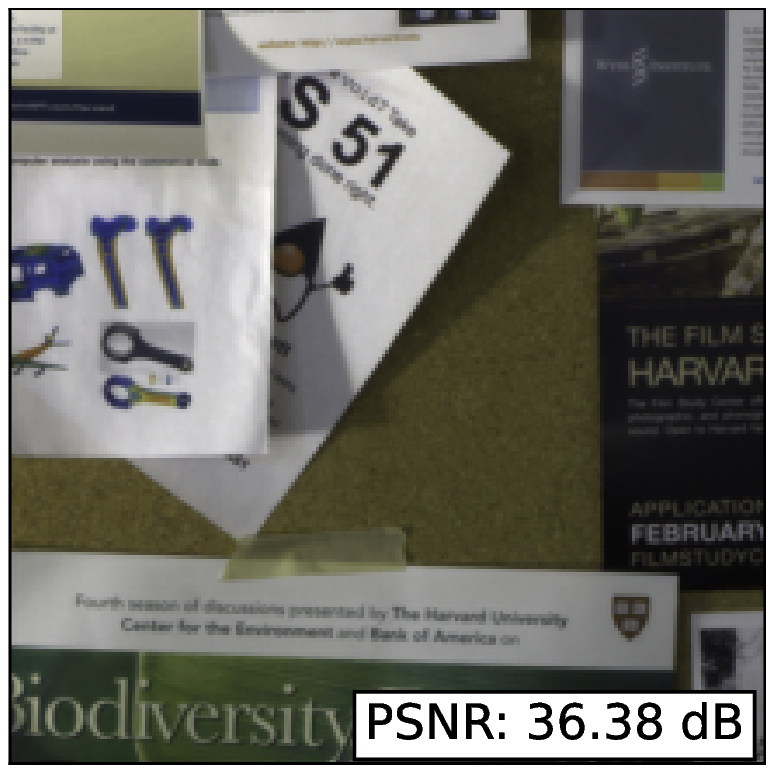}
    \end{minipage}
    \\
    \hspace{-15pt}
    \begin{minipage}{\setwideb\linewidth}
    \includegraphics[width=\linewidth]{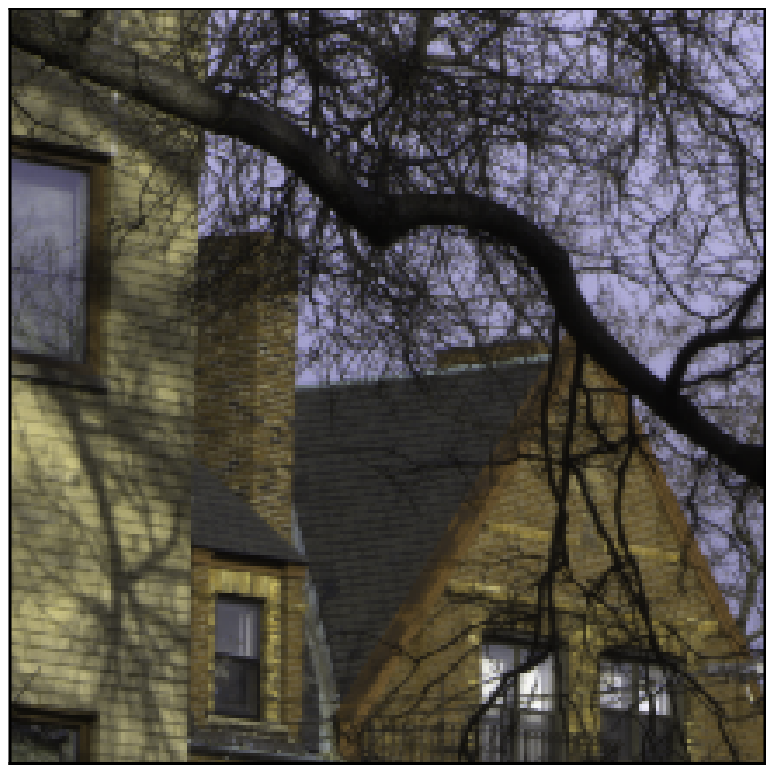}
    \end{minipage}
    &
    \hspace{-15pt}
    \begin{minipage}{\setwideb\linewidth}
    \includegraphics[width=\linewidth]{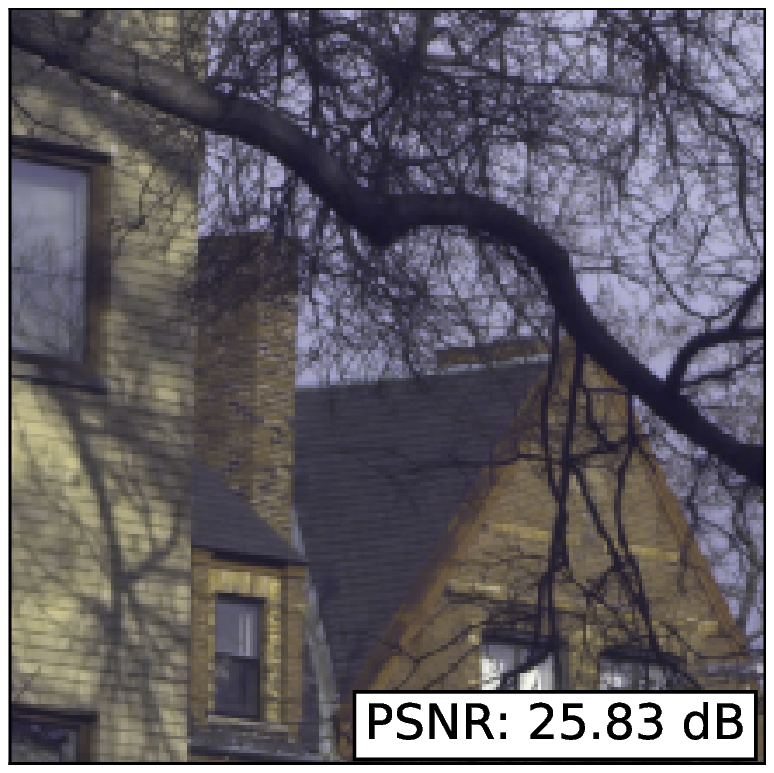}
    \end{minipage}
	&
    \hspace{-15pt}
    \begin{minipage}{\setwideb\linewidth}
    \includegraphics[width=\linewidth]{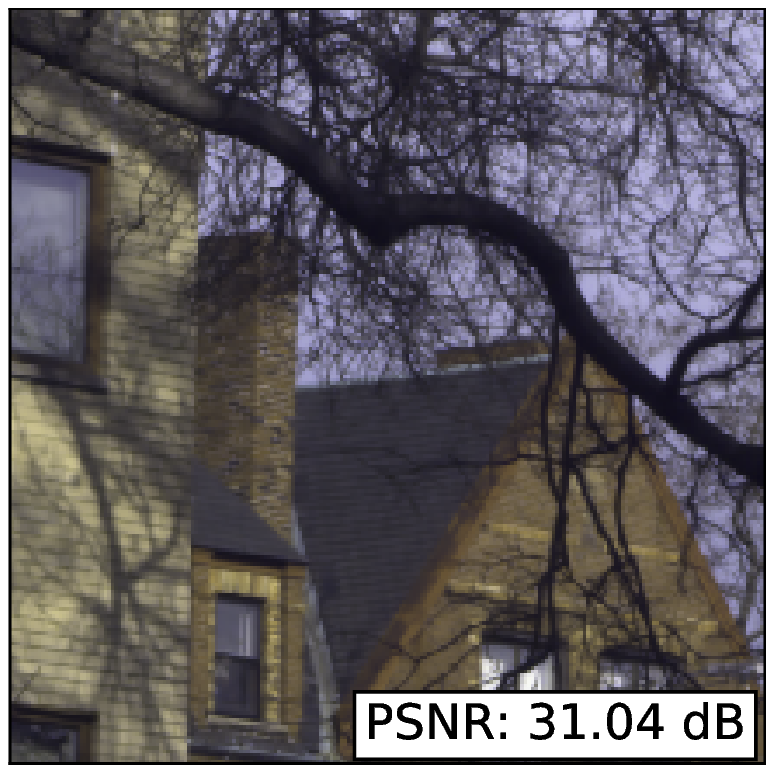}
    \end{minipage}
	&
    \hspace{-15pt}
    \begin{minipage}{\setwideb\linewidth}
    \includegraphics[width=\linewidth]{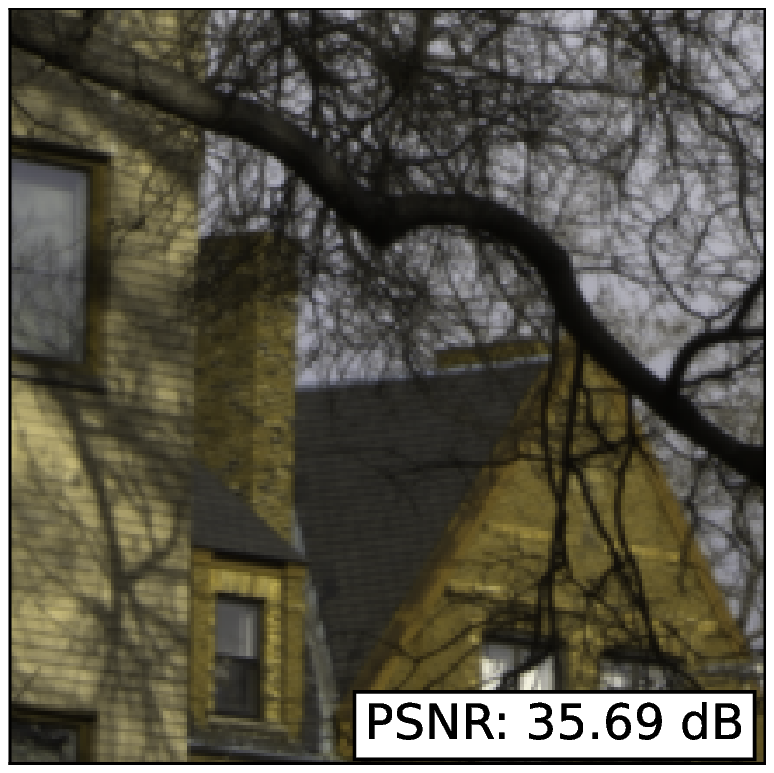}
    \end{minipage}
	&
    \hspace{-15pt}
    \begin{minipage}{\setwideb\linewidth}
    \includegraphics[width=\linewidth]{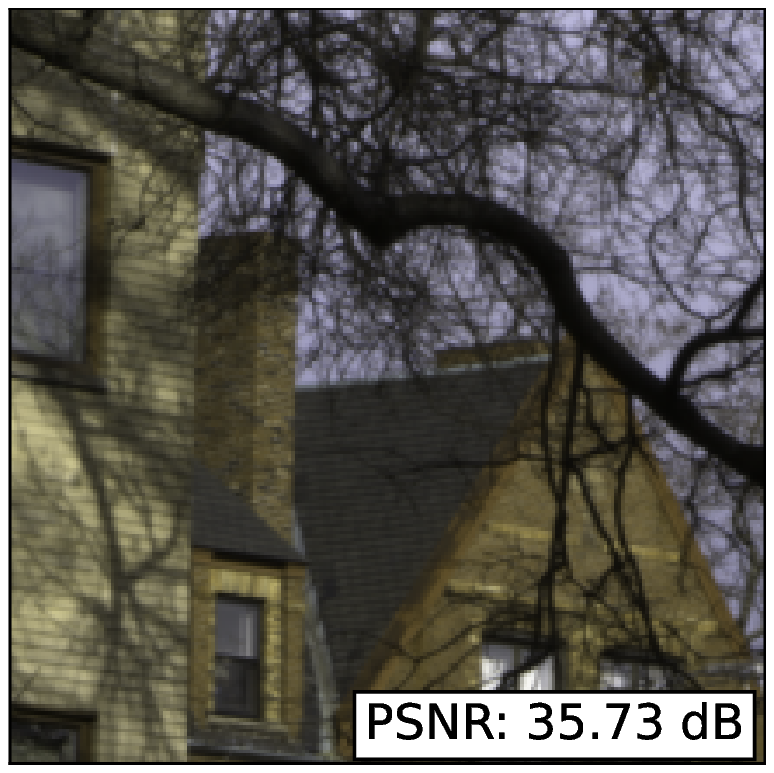}
    \end{minipage}
	&
    \hspace{-15pt}
    \begin{minipage}{\setwideb\linewidth}
    \includegraphics[width=\linewidth]{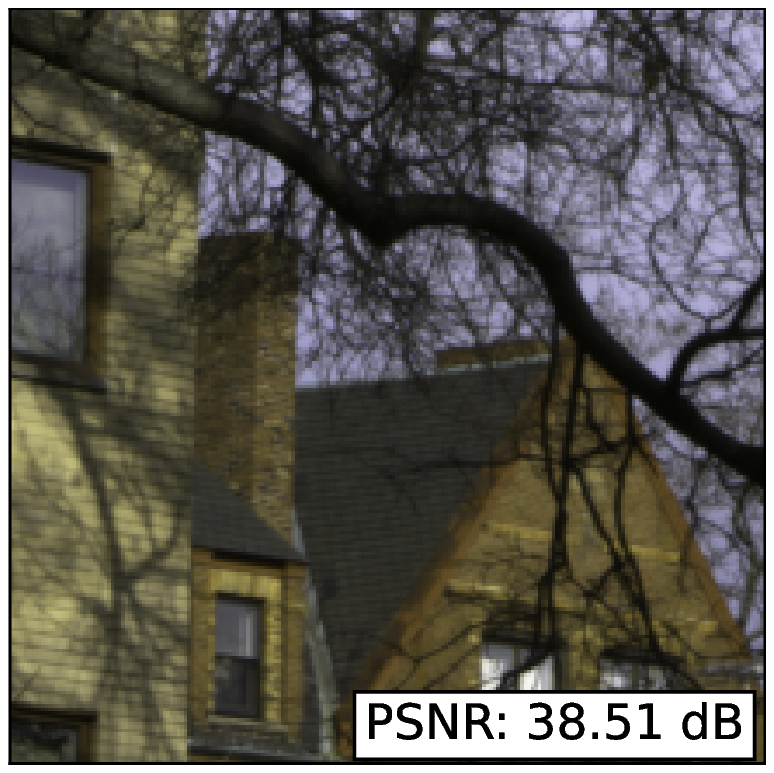}
    \end{minipage}
	\\
    \hspace{-15pt}
    \begin{minipage}{\setwideb\linewidth}
    \includegraphics[width=\linewidth]{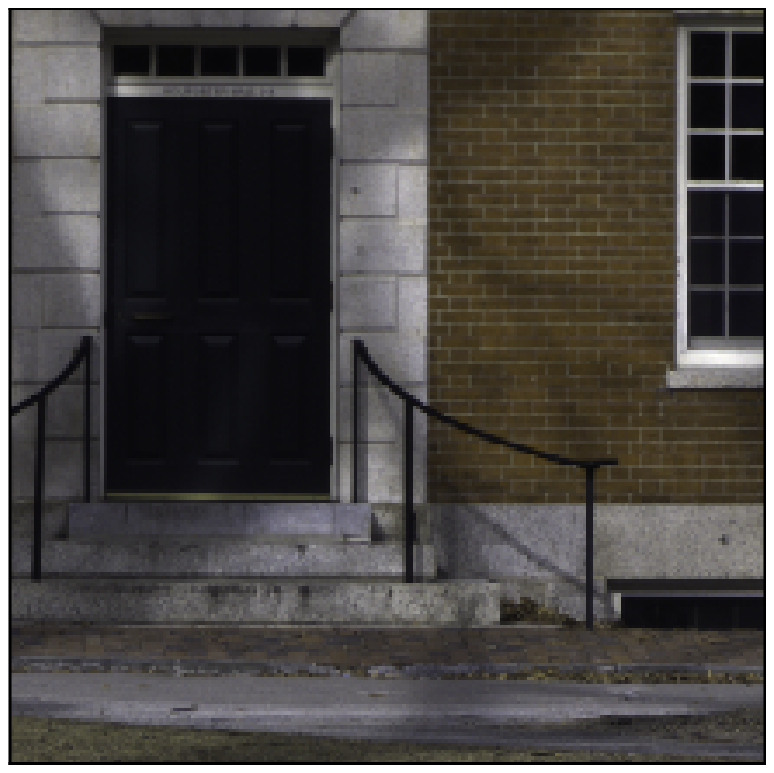}
    \end{minipage}
    &
    \hspace{-15pt}
    \begin{minipage}{\setwideb\linewidth}
    \includegraphics[width=\linewidth]{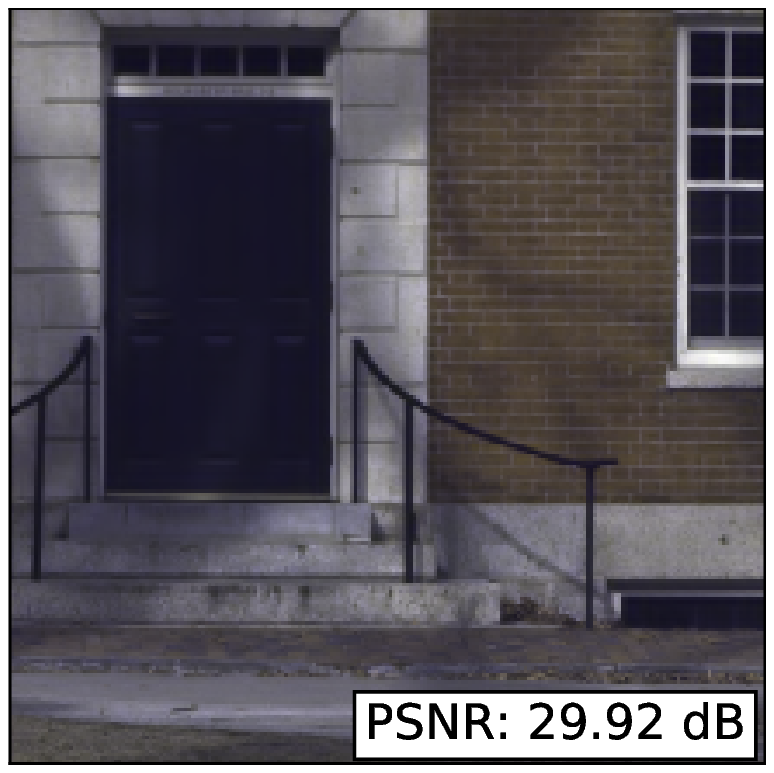}
    \end{minipage}
    &
    \hspace{-15pt}
    \begin{minipage}{\setwideb\linewidth}
    \includegraphics[width=\linewidth]{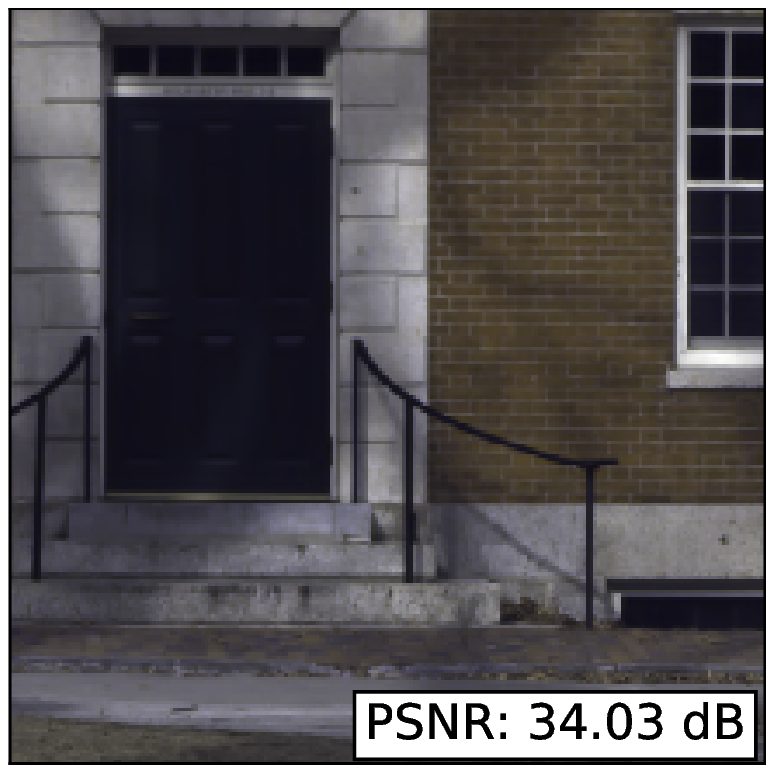}
    \end{minipage}
    &
    \hspace{-15pt}
    \begin{minipage}{\setwideb\linewidth}
    \includegraphics[width=\linewidth]{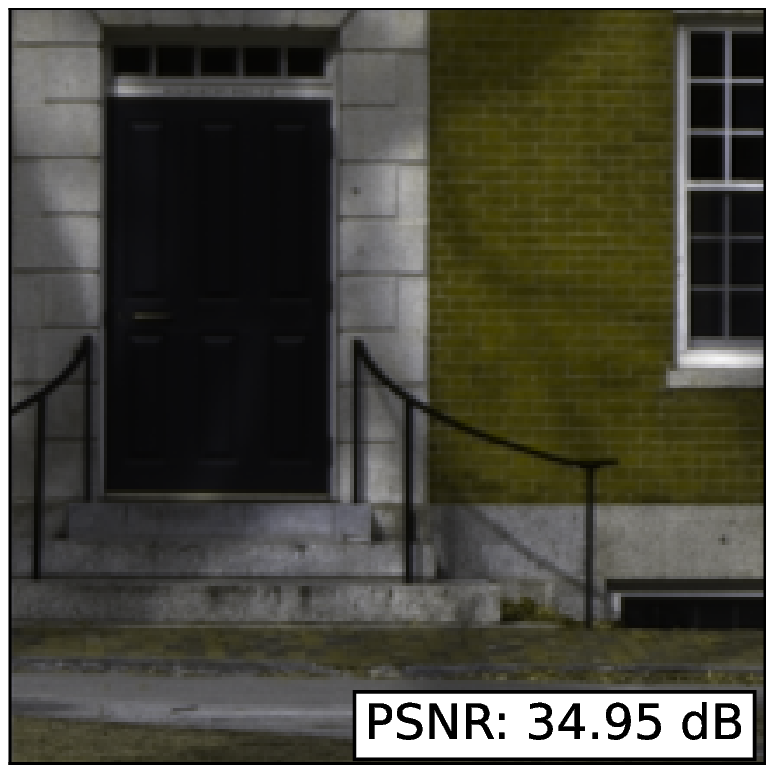}
    \end{minipage}
    &
    \hspace{-15pt}
    \begin{minipage}{\setwideb\linewidth}
    \includegraphics[width=\linewidth]{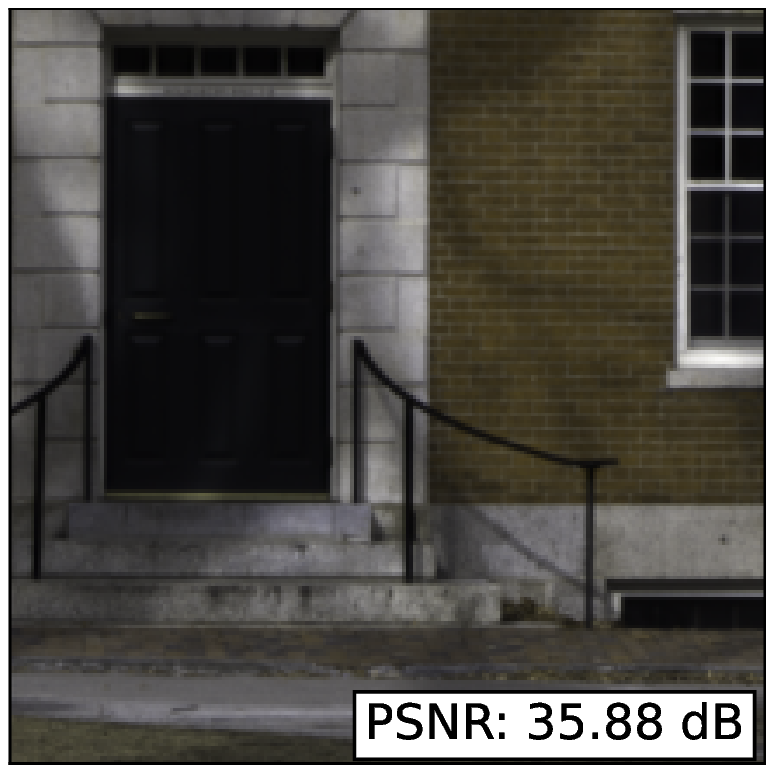}
    \end{minipage}
    &
    \hspace{-15pt}
    \begin{minipage}{\setwideb\linewidth}
    \includegraphics[width=\linewidth]{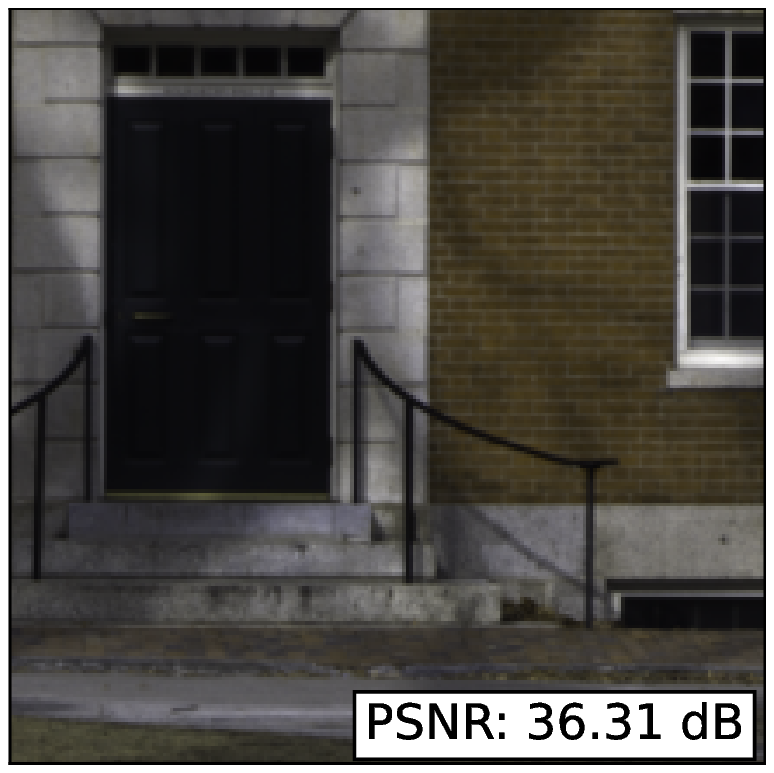}
    \end{minipage}
	\\
	\hspace{-15pt}
	\footnotesize Original
	&
	\hspace{-15pt}
	\footnotesize Rec + FUSE
	&
	\hspace{-15pt}
	\footnotesize SIFCM
	&
	\hspace{-15pt}
	\footnotesize ISTA-Net
	&
	\hspace{-15pt}
	\footnotesize OPINE-Net
	&
	\footnotesize \hspace{-15pt}
	Proposed
\end{tabular}
\end{center}
\caption{Harvard database. RGB composite of the reconstructions obtained by the state-of-the-art fusion methods that recover the high-resolution spectral image from 3D-CASSI compressive measurements.}
    \label{fig:image_comparison}
\end{figure*}

\begin{figure}
\begin{center}
    \begin{tabular}{c c c}
  \hspace{-15pt}     
    \begin{minipage}[c]{0.33\linewidth}
    \resizebox{1.05\textwidth}{!}{
    \begin{tikzpicture}
		\begin{axis}[
		legend cell align = left,
		legend pos = south east,
		scale = 0.90,
		grid = major,
		ylabel style = {yshift=-0.25cm},
		ymin = 0,
		ymax = 0.90,
		xmin = 420,
		xmax = 720,
		xlabel = \texttt{\Large  Wavelength[nm]},
		]
        \addplot[color=mycolor5, smooth, line width = 2.00pt] table[x ={x},y = y1]{FigureData/building_signatures1.dat};
        \addlegendentry{\large Rec. + Fusion}
		\addplot[color=mycolor1, smooth, line width = 2.00pt] table[x ={x},y = y2]{FigureData/building_signatures1.dat};
		\addlegendentry{\large DFCM}
		\addplot[color=mycolor3, smooth, line width = 2.00pt] table[x ={x},y = y5]{FigureData/building_signatures1.dat};
		\addlegendentry{\large ISTA-Net}
		\addplot[color=mycolor4, smooth, line width = 2.00pt] table[x ={x},y = y3]{FigureData/building_signatures1.dat};
		\addlegendentry{\large Proposed}
		\addplot[color=mycolor2, smooth, line width = 2.00pt] table[x ={x},y = y4]{FigureData/building_signatures1.dat};
		\addlegendentry{\large Original}
		\end{axis}
	\end{tikzpicture}
	}
    \end{minipage}

    &
    \hspace{-10pt}
        \begin{minipage}[c]{0.33\linewidth}
    \resizebox{1.05\textwidth}{!}{
    \begin{tikzpicture}
		\begin{axis}[
		legend cell align = left,
		legend pos = south east,
		scale = 0.90,
		grid = major,
		ylabel style = {yshift=-0.25cm},
		ymin = 0,
		ymax = 0.27,
		xmin = 420,
		xmax = 720,
		ytick={0,0.1,0.2},
		xlabel = \texttt{\Large Wavelength[nm]},
		]
        \addplot[color=mycolor5, smooth, line width = 2.00pt] table[x ={x},y = y1]{FigureData/building_signatures2.dat};
        \addlegendentry{\large Rec. + Fusion}
		\addplot[color=mycolor1, smooth, line width = 2.00pt] table[x ={x},y = y2]{FigureData/building_signatures2.dat};
		\addlegendentry{\large DFCM}
		\addplot[color=mycolor3, smooth, line width = 2.00pt] table[x ={x},y = y5]{FigureData/building_signatures2.dat};
		\addlegendentry{\large ISTA-Net}
		\addplot[color=mycolor4, smooth, line width = 2.00pt] table[x ={x},y = y3]{FigureData/building_signatures2.dat};
		\addlegendentry{\large Proposed}
		\addplot[color=mycolor2, smooth, line width = 2.00pt] table[x ={x},y = y4]{FigureData/building_signatures2.dat};
		\addlegendentry{\large Original}
		\end{axis}
	\end{tikzpicture}
	}
    \end{minipage}
    
    &
    \hspace{-10pt}
        \begin{minipage}[c]{0.33\linewidth}
    \resizebox{1.05\textwidth}{!}{
    \begin{tikzpicture}
		\begin{axis}[
		legend cell align = left,
		legend pos = south east,
		scale = 0.90,
		grid = major,
		ylabel style = {yshift=-0.25cm},
		ymin = 0,
		ymax = 0.70,
		xmin = 420,
		xmax = 720,
		xlabel = \texttt{\Large Wavelength[nm]},
		]
        \addplot[color=mycolor5, smooth, line width = 2.00pt] table[x ={x},y = y1]{FigureData/building_signatures3.dat};
        \addlegendentry{\large Rec. + Fusion}
		\addplot[color=mycolor1, smooth, line width = 2.00pt] table[x ={x},y = y2]{FigureData/building_signatures3.dat};
		\addlegendentry{\large DFCM}
		\addplot[color=mycolor3, smooth, line width = 2.00pt] table[x ={x},y = y5]{FigureData/building_signatures3.dat};
		\addlegendentry{\large ISTA-Net}
		\addplot[color=mycolor4, smooth, line width = 2.00pt] table[x ={x},y = y3]{FigureData/building_signatures3.dat};
		\addlegendentry{\large Proposed}
		\addplot[color=mycolor2, smooth, line width = 2.00pt] table[x ={x},y = y4]{FigureData/building_signatures3.dat};
		\addlegendentry{\large Original}
		\end{axis}
	\end{tikzpicture}
	}
    \end{minipage}
    
    \end{tabular}
    \end{center}
    \caption{Original and reconstructed of selected spectral signatures from images in Fig. \ref{fig:image_comparison}.}
    \label{fig:signatures}
\end{figure}
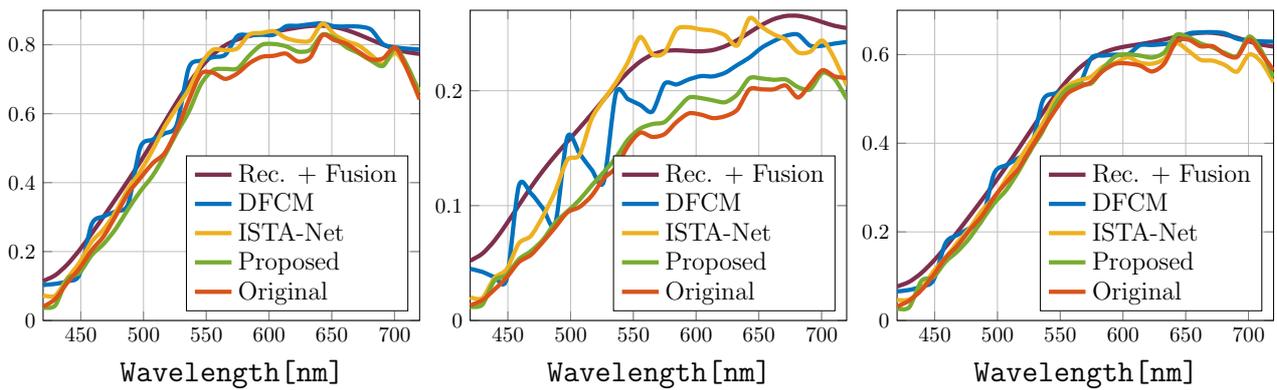

Table \ref{tab:comparison_compressive} quantitatively shows the performance yielded by the various fusion methods from compressive measurements in terms of PSNR, SSIM, SAM, and computation time. In particular, this table contains the ensemble averages obtained from fusing 24 multi-sensor compressive observations for various compression rates. The best values are written in bold font. As can be seen in this table, the proposed fusion approach provides outstanding image quality results compared to the other methods, while achieving competitive computational times.

\setlength{\tabcolsep}{2.5pt}
\begin{table}[]\footnotesize
    \caption{Comparison of the average reconstruction metrics obtained from 12 testing image of the Harvard database for different compression rates.}
    \label{tab:comparison_compressive}
    \centering
    \begin{tabular}{c|c|c c c c c}
    \hline
    \hline
    Comp. & Quality & \multirow{2}{*}{Rec + FUSE} & \multirow{2}{*}{SIFCM} & \multirow{2}{*}{ISTA-Net} & \multirow{2}{*}{OPINE-Net} & \multirow{2}{*}{Proposed} \\
    Rate & Metric \\
    \hline
    \hline
    \multirow{5}{*}{$25.0\%$} & PSNR [dB] & 31.40 $\pm$ 3.11 & 32.19 $\pm$ 2.73 & 31.10 $\pm$ 2.68 & 35.44 $\pm$ 4.26 & \textbf{36.15} $\pm$ \textbf{3.84} \\
    & SSIM & 0.904 $\pm$ 0.026 & 0.910 $\pm$ 0.027 & 0.912 $\pm$ 0.031 & 0.935 $\pm$ 0.066 & \textbf{0.955} $\pm$ \textbf{0.029} \\
    & SAM  & 0.097 $\pm$ 0.033 & 0.118 $\pm$ 0.040 & 0.143 $\pm$ 0.055 & 0.124 $\pm$ 0.099 & \textbf{0.095} $\pm$ \textbf{0.051}\\
    & Running time [s]& 66.08 $\pm$ 1.59 & 224.83 $\pm$ 4.11 & 0.27$\pm$ 0.02 & \textbf{0.26} $\pm$ \textbf{0.02} & 0.30 $\pm$ 0.02\\
    & Training time & - & - & \textbf{2h34min} & 2h45min & 2h54min \\
    \hline
    \hline
    \multirow{5}{*}{$37.5\%$} & PSNR [dB] & 30.27 $\pm$ 3.11 & 30.29 $\pm$ 2.55 & 33.55 $\pm$ 2.90 & 35.05 $\pm$ 2.94 & \textbf{37.06} $\pm$ \textbf{3.35} \\
    & SSIM & 0.888 $\pm$ 0.025 & 0.894 $\pm$ 0.025 & 0.938 $\pm$ 0.031 & 0.947 $\pm$ 0.025 & \textbf{0.965} $\pm$ \textbf{0.018} \\
    & SAM  & 0.102 $\pm$ 0.025 & 0.105 $\pm$ 0.031 & 0.109 $\pm$ 0.050 & 0.099 $\pm$ 0.044 & \textbf{0.083} $\pm$ \textbf{0.038} \\
    & Running time[s] & 67.56 $\pm$ 1.08 & 226.71 $\pm$ 3.25 & 0.028 $\pm$ 0.05  & \textbf{0.28} $\pm$ \textbf{0.02} &  0.30 $\pm$ 0.02 \\
    & Training time & - & - & \textbf{2h43min} & 2h46min & 3h1min \\
    \hline
    \hline
    \multirow{5}{*}{$50.0\%$} & PSNR [dB] & 31.89 $\pm$ 3.58 & 32.25 $\pm$ 3.10 & 36.93 $\pm$ 3.46 & 35.93 $\pm$ 2.20 & \textbf{39.47} $\pm$ \textbf{3.12} \\
    & SSIM & 0.908 $\pm$ 0.024 & 0.916 $\pm$ 0.025 & 0.956 $\pm$ 0.028 & 0.941 $\pm$ 0.019 & \textbf{0.976} $\pm$ \textbf{0.013} \\
    & SAM  & 0.087 $\pm$ 0.024 & 0.087 $\pm$ 0.024 & 0.088 $\pm$ 0.044 & 0.101 $\pm$ 0.034 & \textbf{0.064} $\pm$ \textbf{0.027} \\
    & Running time[s] & 69.54 $\pm$ 1.80 & 235.50 $\pm$ 5.11 & \textbf{0.30} $\pm$ \textbf{0.01} & 0.30 $\pm$ 0.01 & 0.32 $\pm$ 0.01 \\
    & Training time & - & - & \textbf{2h53min} & 2h55min & 3h9min \\
    \hline
    \hline
    \end{tabular}
\end{table}

\subsection{CAVE database}

In this experiment, we use the Stuff section of the CAVE spectral image database. This database contains fifteen high-resolution images of previously constructed scenes under controlled illumination \cite{YasumaGeneralized2010}. Furthermore, the spectral images exhibit dimensions of $512 \times 512 \times 31$ that include reflectance data from $400$ nm to $700$ nm. The first column of Fig. \ref{fig:image_comparison_CAVE} displays the RGB composite of three testing images. The training set comprises 12 spectral images and their respective compressive measurements for a compression ratio of $25\%$. For this database, we use 256 epochs. For comparative purposes, we obtain the fused images estimated by the Rec + FUSE approach, the SIFCM method, the ISTA-Net technique, and the OPINE-Net architecture. The last five columns of Fig. \ref{fig:image_comparison_CAVE} shows the RGB composite of spectral images recovered by the various methods. In this case, LADMM-Net, ISTA-Net, and OPINE-Net are implemented with 10 processing layers. As can be observed, the proposed image fusion method outperforms the remaining methods. Table \ref{tab:comparison_cave} displays the performance results obtained by the different spectral fusion methods in terms of PSNR, SSIM, SAM, running time, and training times. As can be seen in this table, the proposed approach outperforms the other methods in terms of image quality metrics, while it also exhibits competitive computation times.

\newcommand\setwidecave{0.162}
\begin{figure*}
\begin{center}
\begin{tabular}{c c c c c c}
    \hspace{-8pt}
    \begin{minipage}{\setwidecave\linewidth}
    \includegraphics[width=\linewidth]{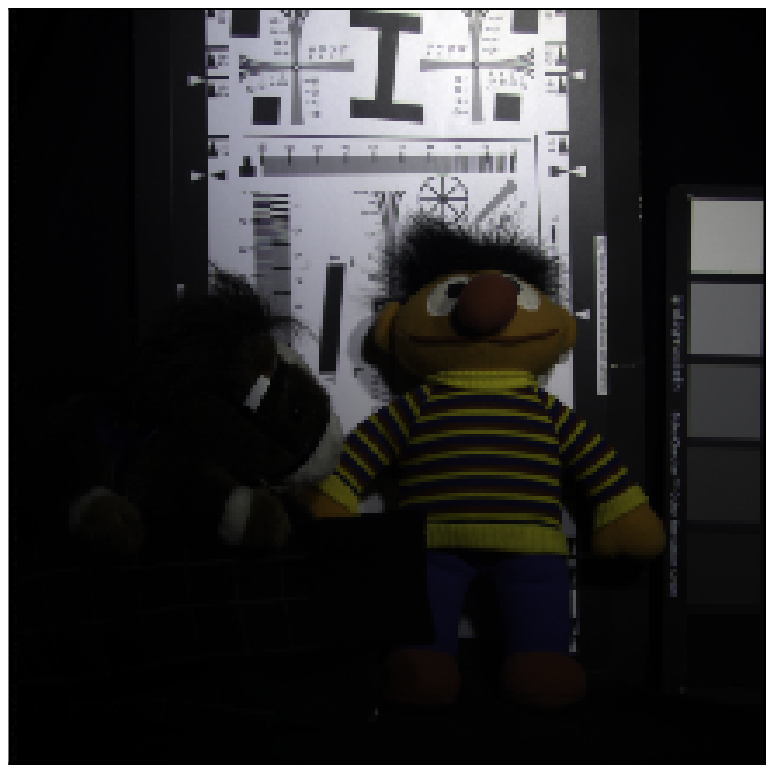}
    \end{minipage}
    &
    \hspace{-8pt}
    \begin{minipage}{\setwidecave\linewidth}
    \includegraphics[width=\linewidth]{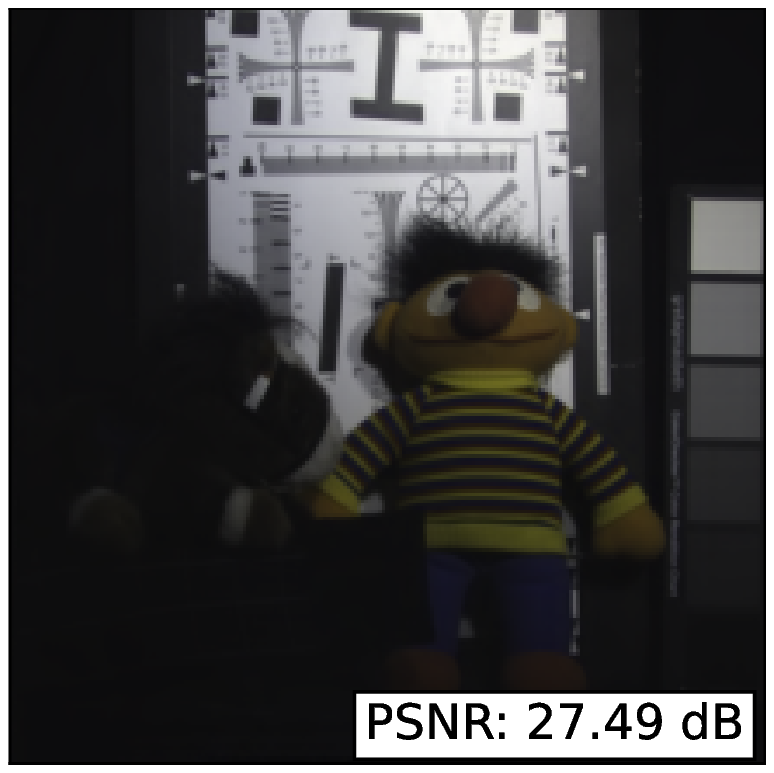}
    \end{minipage}
    &
    \hspace{-8pt}
    \begin{minipage}{\setwidecave\linewidth}
    \includegraphics[width=\linewidth]{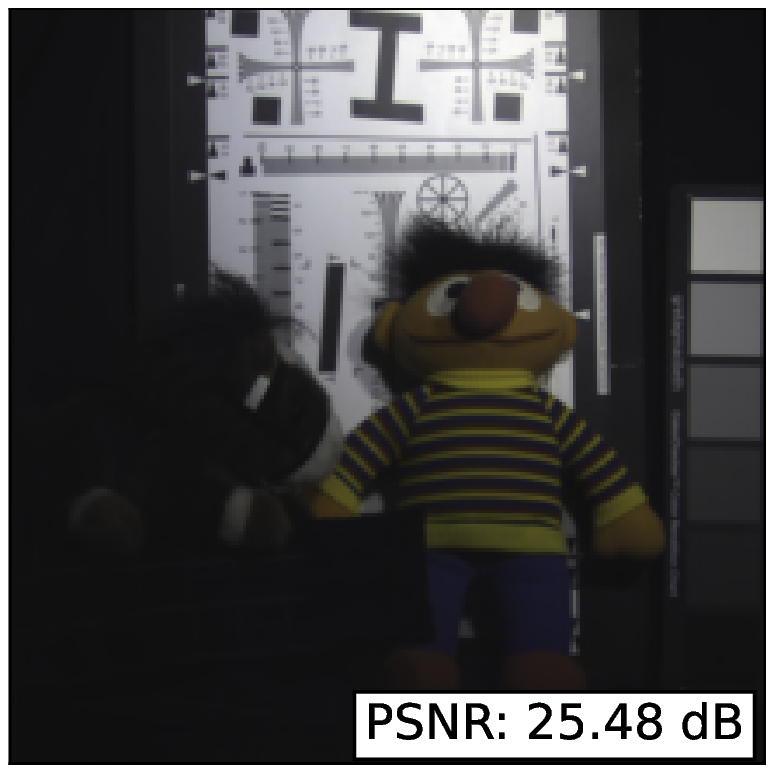}
    \end{minipage}
    &
    \hspace{-8pt}
    \begin{minipage}{\setwidecave\linewidth}
    \includegraphics[width=\linewidth]{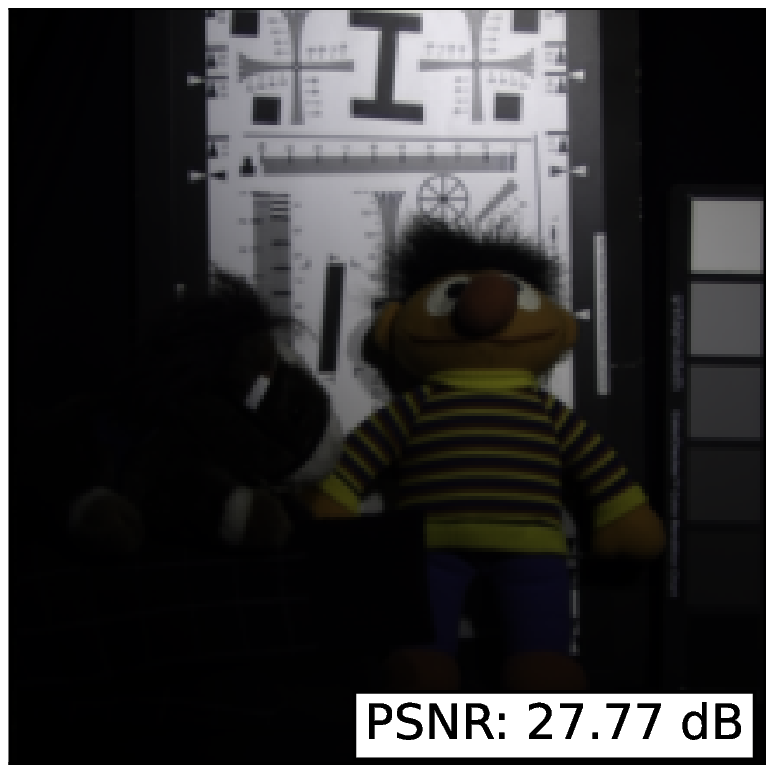}
    \end{minipage}
    &
    \hspace{-8pt}
    \begin{minipage}{\setwidecave\linewidth}
    \includegraphics[width=\linewidth]{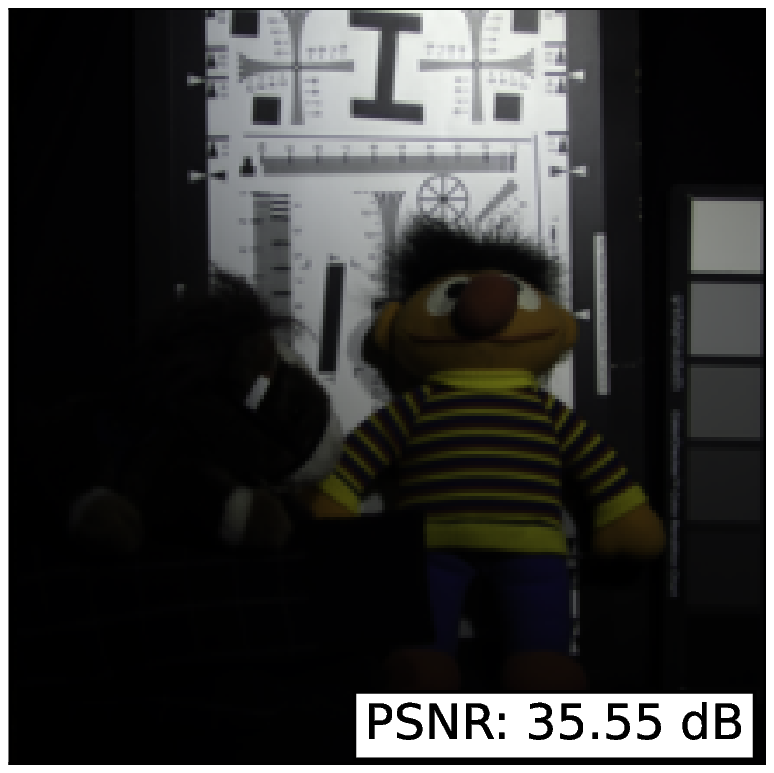}
    \end{minipage}
    &
    \hspace{-8pt}
    \begin{minipage}{\setwidecave\linewidth}
    \includegraphics[width=\linewidth]{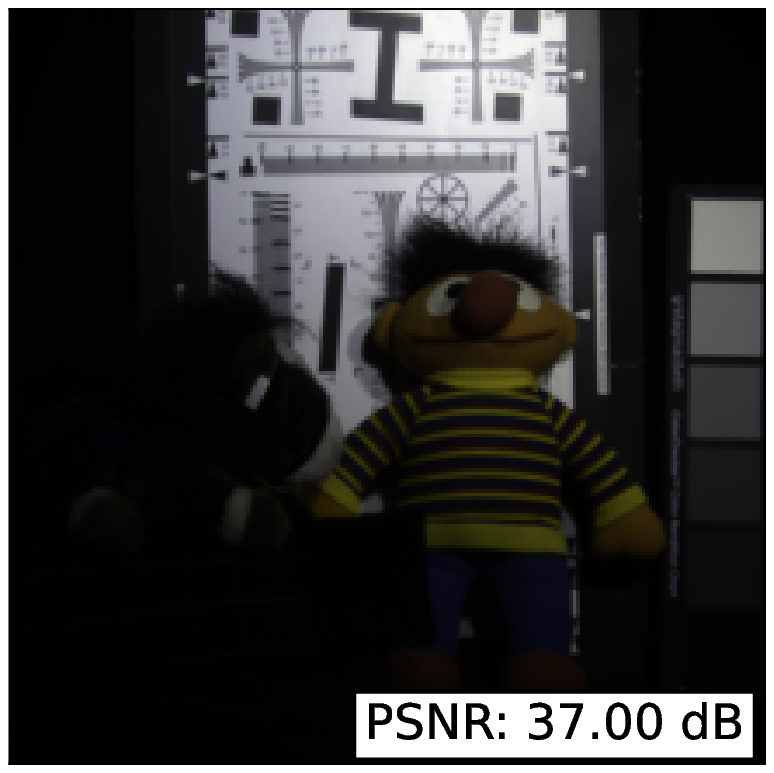}
    \end{minipage}
    \\
    \hspace{-8pt}
    \begin{minipage}{\setwidecave\linewidth}
    \includegraphics[width=\linewidth]{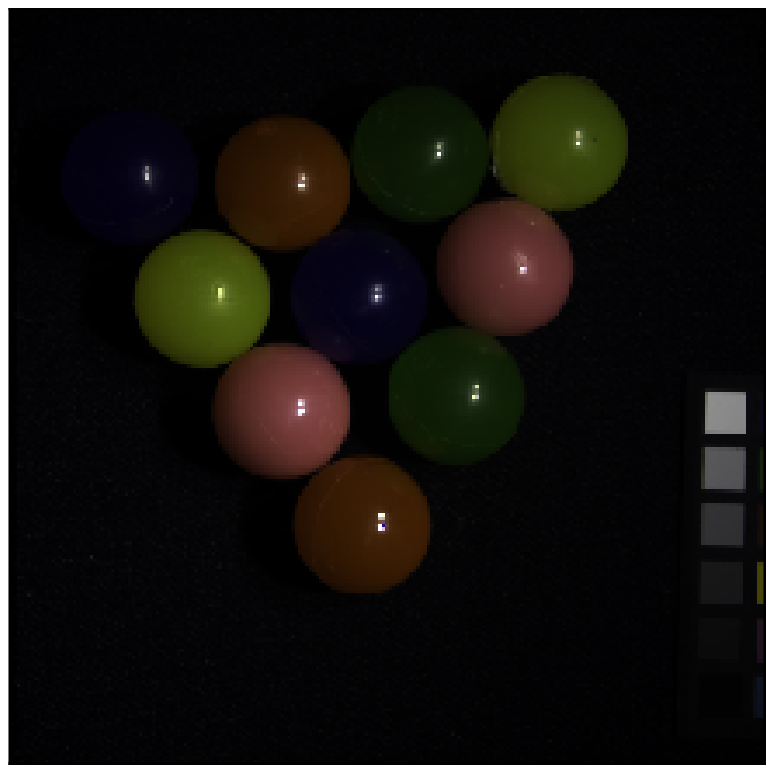}
    \end{minipage}
    &
    \hspace{-8pt}
    \begin{minipage}{\setwidecave\linewidth}
    \includegraphics[width=\linewidth]{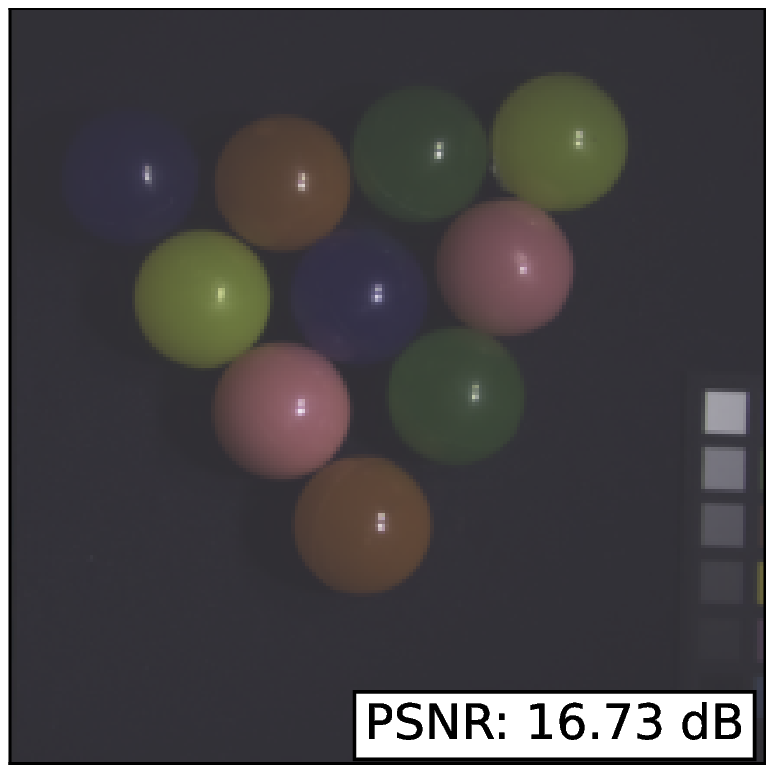}
    \end{minipage}
    &
    \hspace{-8pt}
    \begin{minipage}{\setwidecave\linewidth}
    \includegraphics[width=\linewidth]{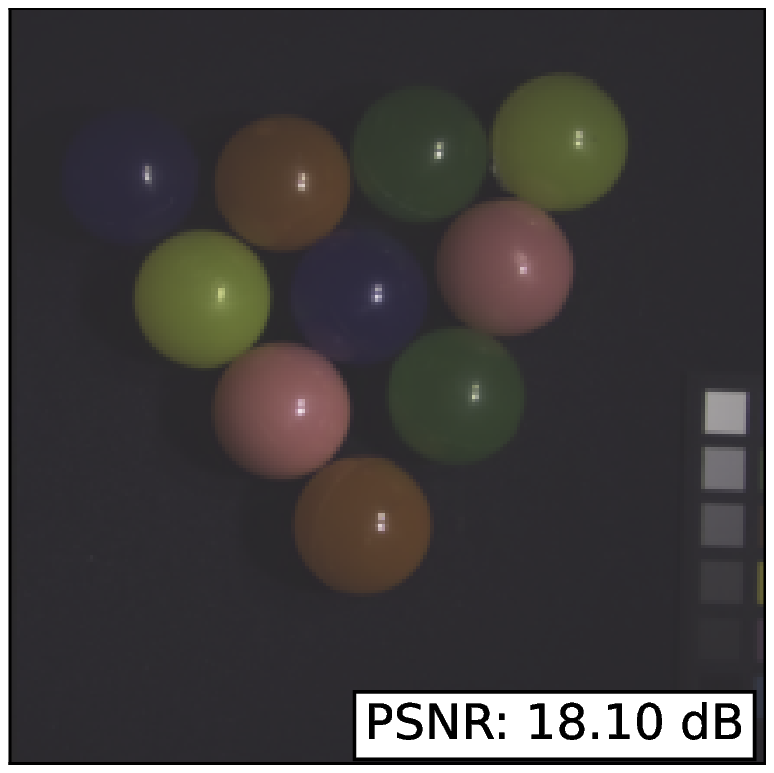}
    \end{minipage}
    &
    \hspace{-8pt}
    \begin{minipage}{\setwidecave\linewidth}
    \includegraphics[width=\linewidth]{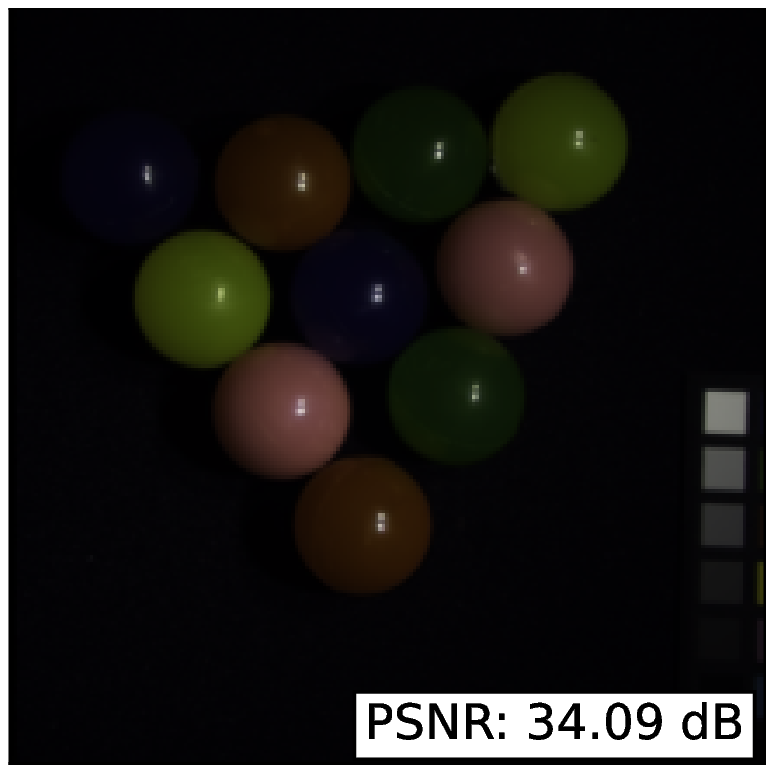}
    \end{minipage}
    &
    \hspace{-8pt}
    \begin{minipage}{\setwidecave\linewidth}
    \includegraphics[width=\linewidth]{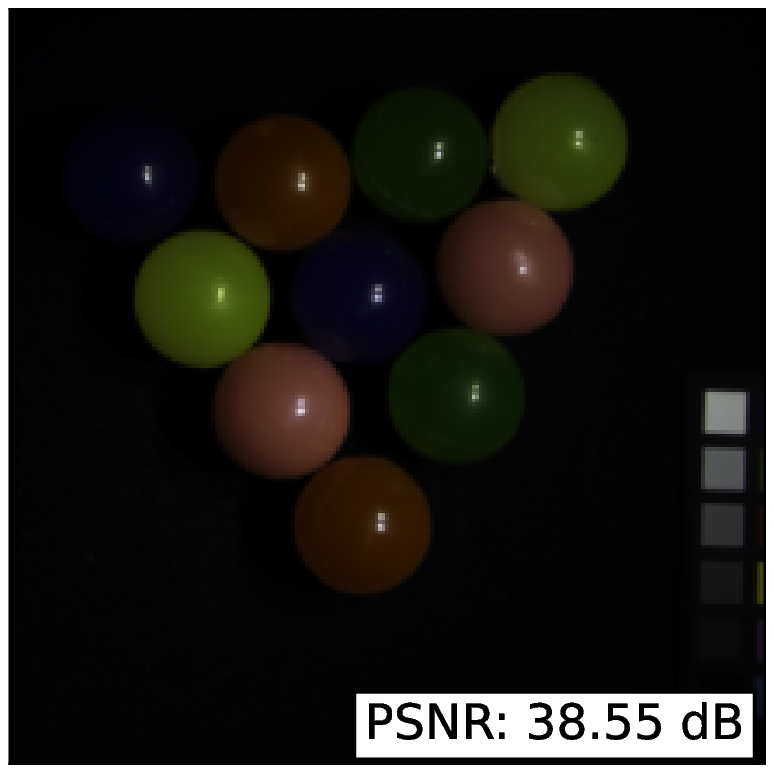}
    \end{minipage}
    &
    \hspace{-8pt}
    \begin{minipage}{\setwidecave\linewidth}
    \includegraphics[width=\linewidth]{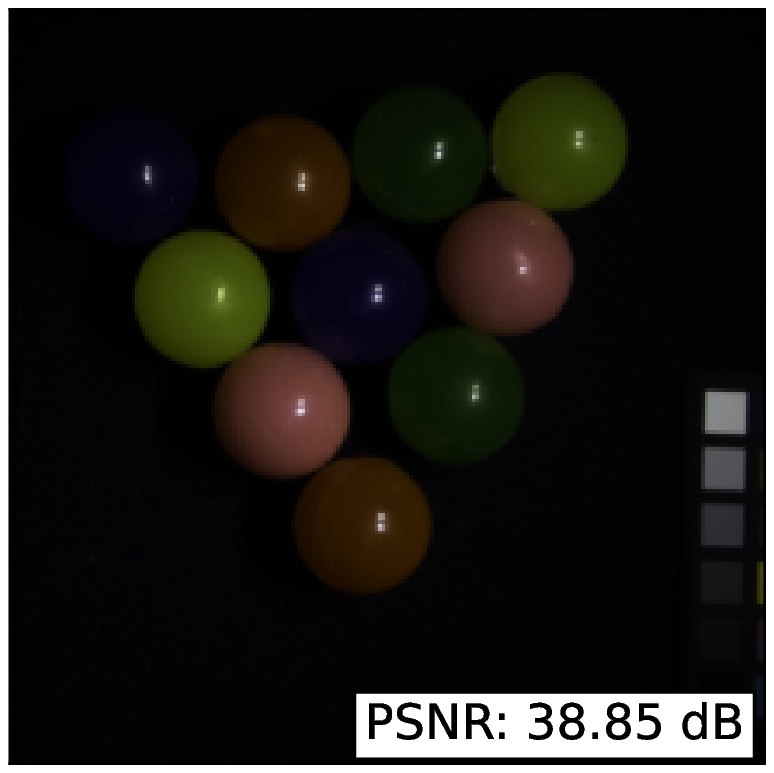}
    \end{minipage}
    \\
    \hspace{-8pt}
    \begin{minipage}{\setwidecave\linewidth}
    \includegraphics[width=\linewidth]{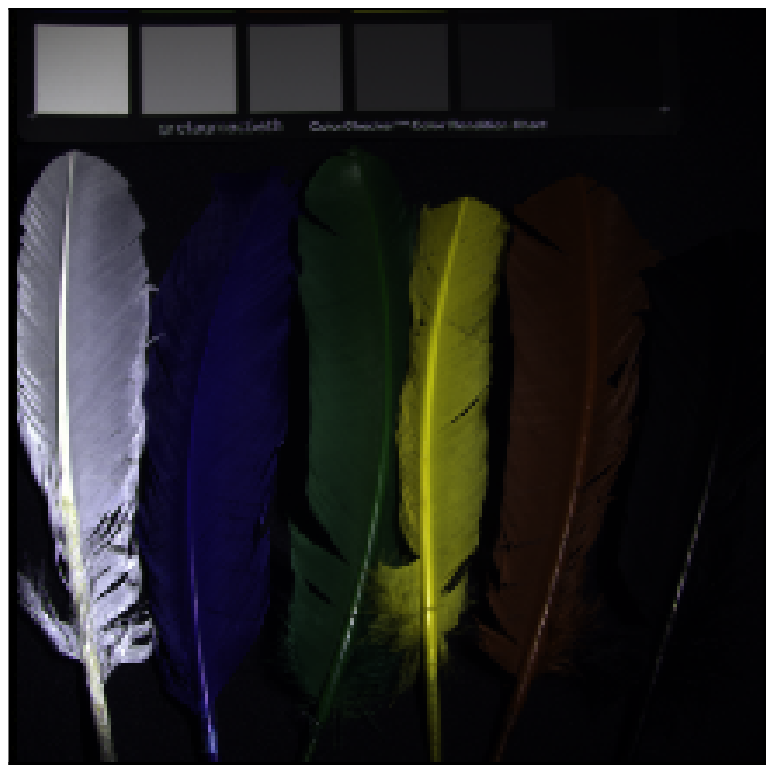}
    \end{minipage}
    &
    \hspace{-8pt}
    \begin{minipage}{\setwidecave\linewidth}
    \includegraphics[width=\linewidth]{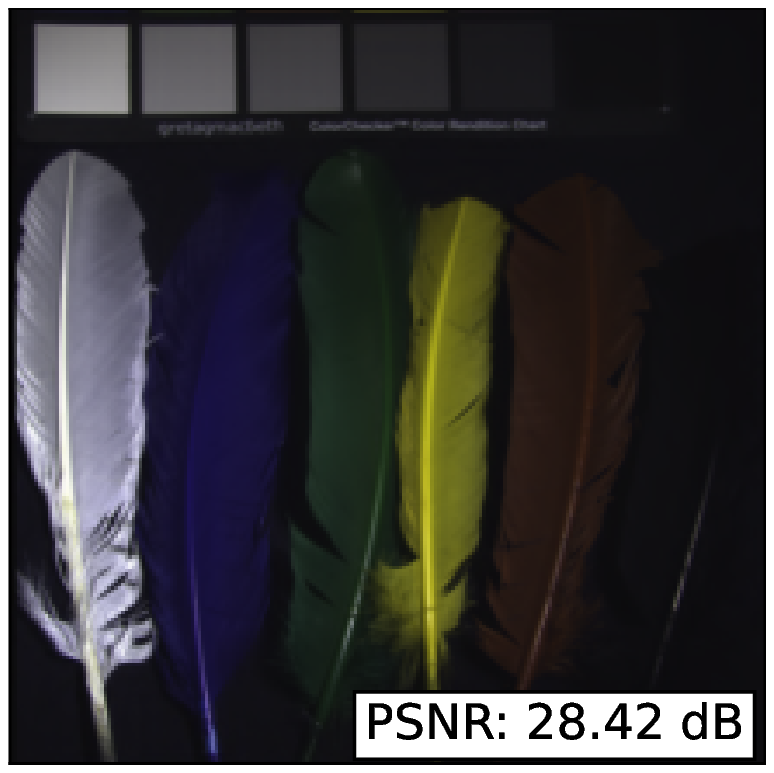}
    \end{minipage}
    &
    \hspace{-8pt}
    \begin{minipage}{\setwidecave\linewidth}
    \includegraphics[width=\linewidth]{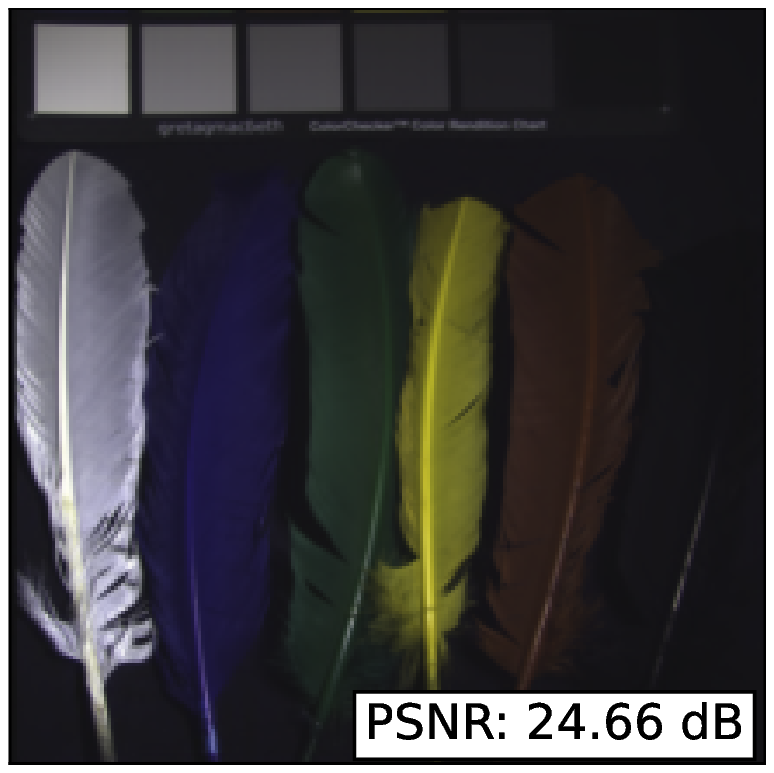}
    \end{minipage}
    &
    \hspace{-8pt}
    \begin{minipage}{\setwidecave\linewidth}
    \includegraphics[width=\linewidth]{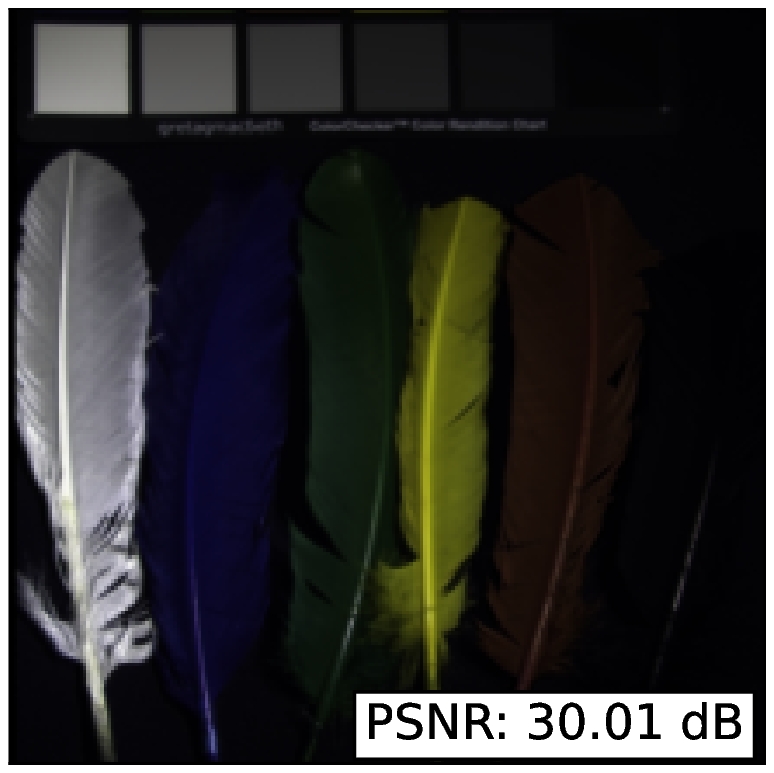}
    \end{minipage}
    &
    \hspace{-8pt}
    \begin{minipage}{\setwidecave\linewidth}
    \includegraphics[width=\linewidth]{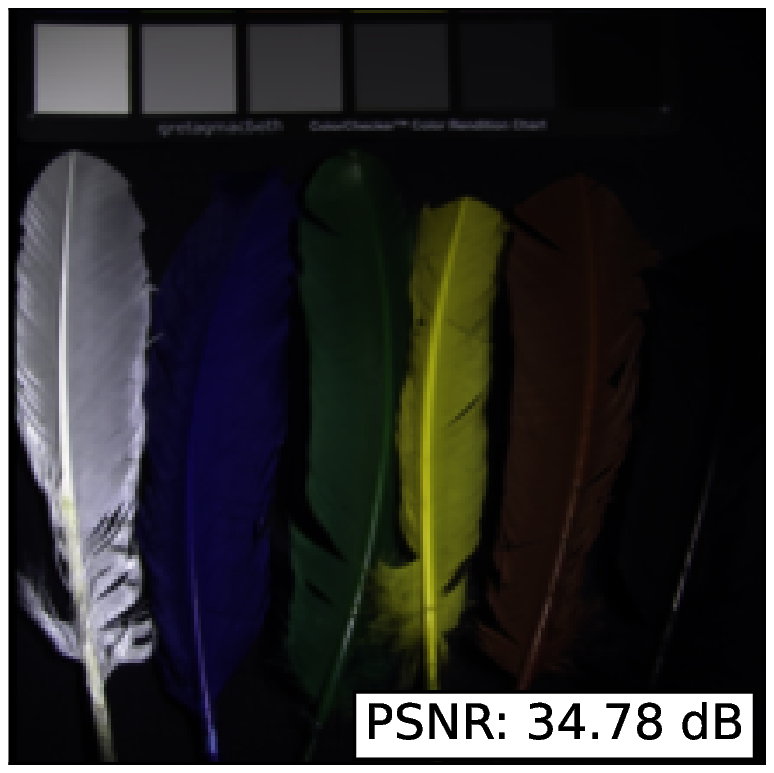}
    \end{minipage}
    &
    \hspace{-8pt}
    \begin{minipage}{\setwidecave\linewidth}
    \includegraphics[width=\linewidth]{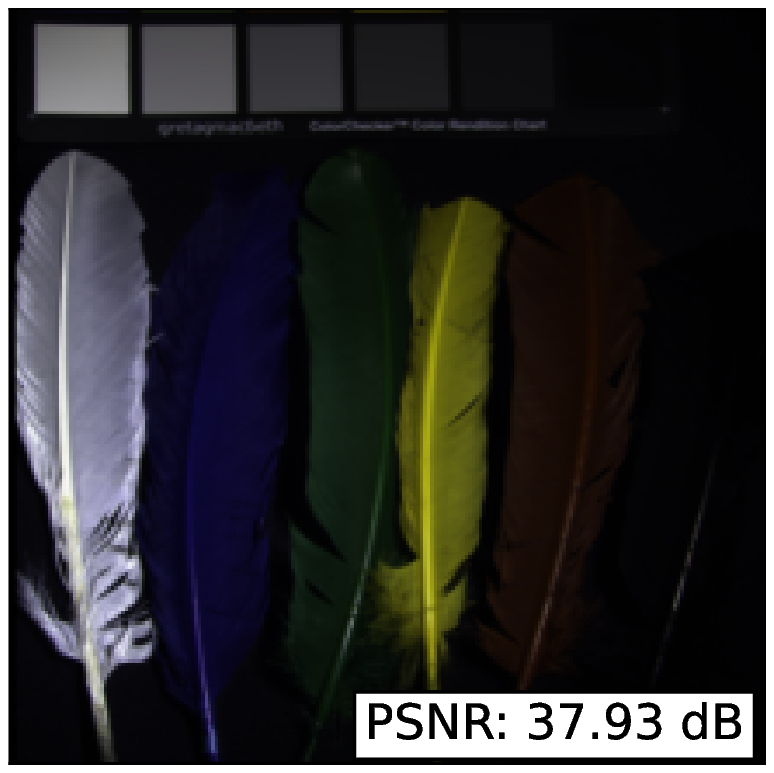}
    \end{minipage}
	\\
	\hspace{-15pt}
	\footnotesize Original
	&
	\hspace{-15pt}
	\footnotesize Rec + FUSE
	&
	\hspace{-15pt}
	\footnotesize SIFCM
	&
	\hspace{-15pt}
	\footnotesize ISTA-Net
	&
	\hspace{-15pt}
	\footnotesize OPINE-Net
	&
	\hspace{-15pt}
	\footnotesize Proposed
\end{tabular}
\end{center}
\caption{CAVE database. RGB composite of the reconstructions obtained by the state-of-the-art fusion methods that recover the high-resolution spectral image from 3D-CASSI compressive measurements.}
    \label{fig:image_comparison_CAVE}
\end{figure*}

\setlength{\tabcolsep}{2.5pt}
\begin{table}[]\footnotesize
    \caption{Comparison of the average reconstruction metrics obtained from testing images of the CAVE database for two compression rates.}
    \label{tab:comparison_cave}
    \centering
    \begin{tabular}{c|c|c c c c c}
    \hline
    \hline
    Comp. & Quality & \multirow{2}{*}{Rec + FUSE} & \multirow{2}{*}{SIFCM} & \multirow{2}{*}{ISTA-Net} & \multirow{2}{*}{OPINE-Net} & \multirow{2}{*}{Proposed} \\
    Rate & Metric \\
    \hline
    \hline
    \multirow{5}{*}{$25.0\%$} & PSNR [dB] & 24.22 $\pm$ 6.50 & 22.75 $\pm$ 4.05 & 30.63 $\pm$ 2.61 & 36.29 $\pm$ 1.63 & \textbf{37.92} $\pm$ \textbf{0.75} \\
    & SSIM & 0.571 $\pm$ 0.274 & 0.534 $\pm$ 0.176 & 0.944 $\pm$ 0.010 & 0.964 $\pm$ 0.003 & \textbf{0.973} $\pm$ \textbf{0.006} \\
    & SAM  & 0.362 $\pm$ 0.167 & 0.385 $\pm$ 0.148 & 0.115 $\pm$ 0.034 & 0.081 $\pm$ 0.020 & \textbf{0.071} $\pm$ \textbf{0.027}\\
    & Running time [s]& 67.57 $\pm$ 0.64 & 223.95 $\pm$ 0.49 & \textbf{0.26} $\pm$ \textbf{0.01} & 0.27 $\pm$ 0.01 & 0.33 $\pm$ 0.03\\
    & Training time & - & - & \textbf{45min} & 47min & 49min \\
    \hline
    \hline
    \multirow{5}{*}{$37.5\%$} & PSNR [dB] & 21.62 $\pm$ 5.06 & 22.07 $\pm$ 3.50 & 35.28 $\pm$ 2.38 & 35.93 $\pm$ 1.98 & \textbf{41.10} $\pm$ \textbf{1.21} \\
    & SSIM & 0.505 $\pm$ 0.193 & 0.516 $\pm$ 0.164 & 0.967 $\pm$ 0.006 & 0.955 $\pm$ 0.005 & \textbf{0.988} $\pm$ \textbf{0.001} \\
    & SAM  & 0.373 $\pm$ 0.155 & 0.371 $\pm$ 0.142 & 0.080 $\pm$ 0.021 & 0.089 $\pm$ 0.022 & \textbf{0.051} $\pm$ \textbf{0.018}\\
    & Running time [s]& 69.59 $\pm$ 1.16 & 229.25 $\pm$ 0.73 & 0.30 $\pm$ 0.03 & \textbf{0.27} $\pm$ \textbf{0.01} & 0.32 $\pm$ 0.01\\
    & Training time & - & - & \textbf{49min} & 49min & 52min \\
    \hline
    \hline
    \end{tabular}
\end{table}

\begin{figure*}
\begin{center}
\begin{tabular}{c c c c c c}
    \hspace{-8pt}
    \begin{minipage}{\setwidecave\linewidth}
    \includegraphics[width=\linewidth]{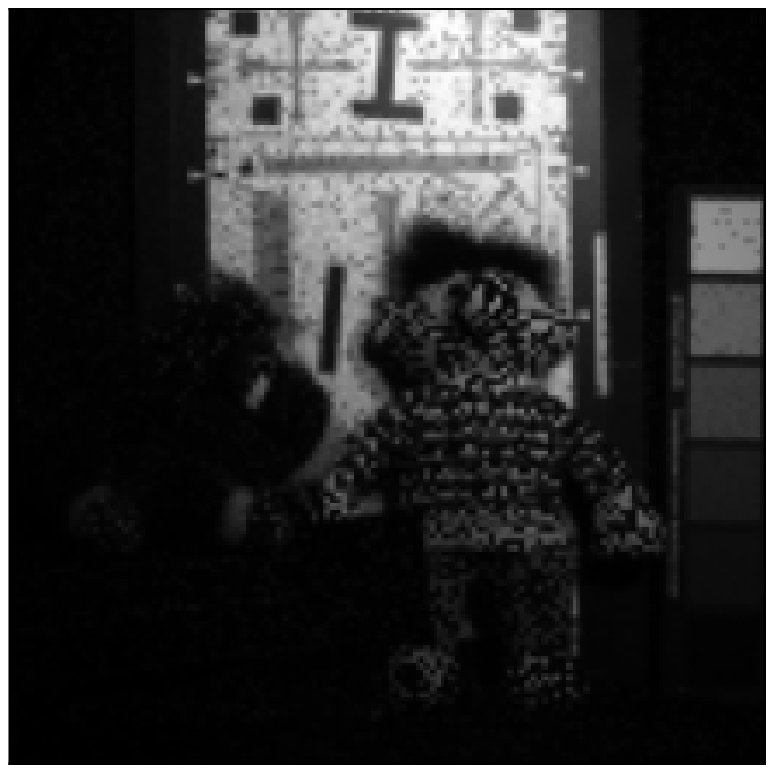}
    \end{minipage}
    &
    \hspace{-8pt}
    \begin{minipage}{\setwidecave\linewidth}
    \includegraphics[width=\linewidth]{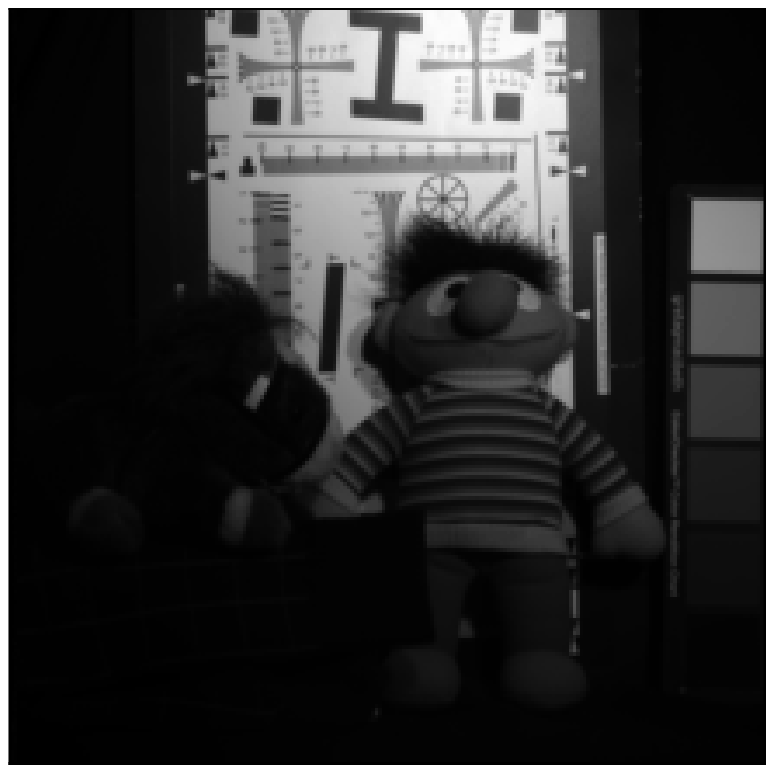}
    \end{minipage}
    &
    \hspace{-8pt}
    \begin{minipage}{\setwidecave\linewidth}
    \includegraphics[width=\linewidth]{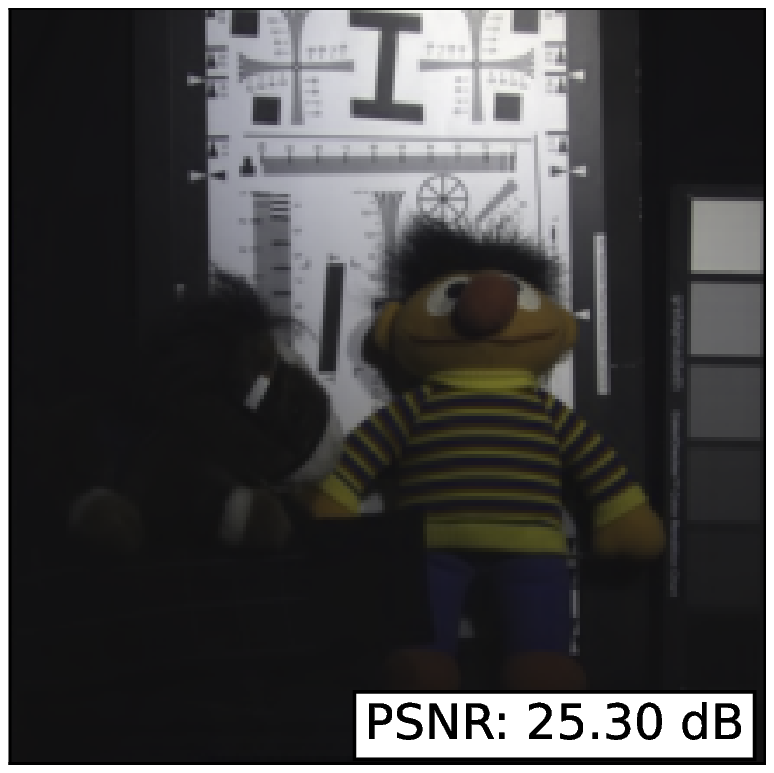}
    \end{minipage}
    &
    \hspace{-8pt}
    \begin{minipage}{\setwidecave\linewidth}
    \includegraphics[width=\linewidth]{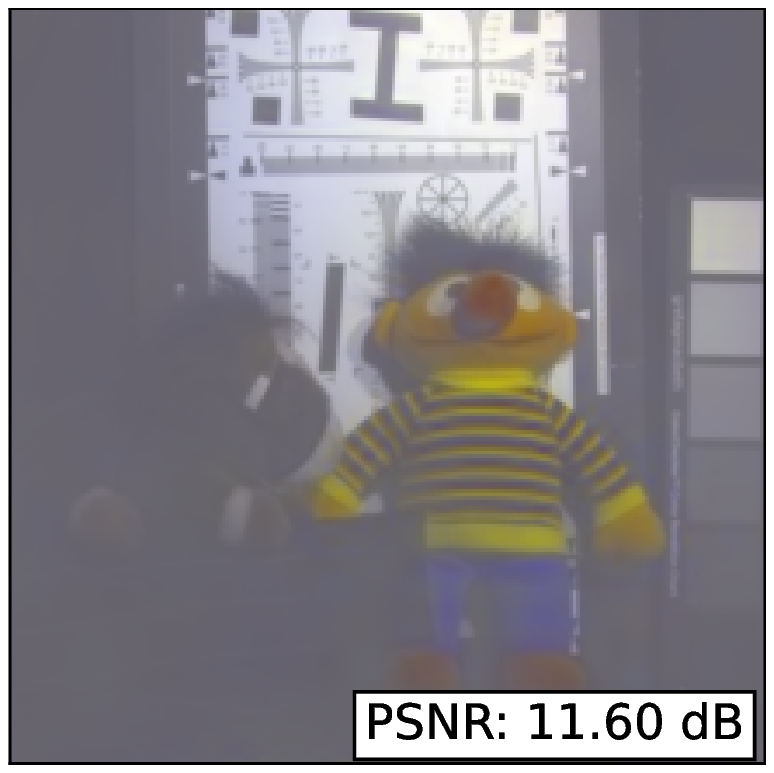}
    \end{minipage}
    &
    \hspace{-8pt}
    \begin{minipage}{\setwidecave\linewidth}
    \includegraphics[width=\linewidth]{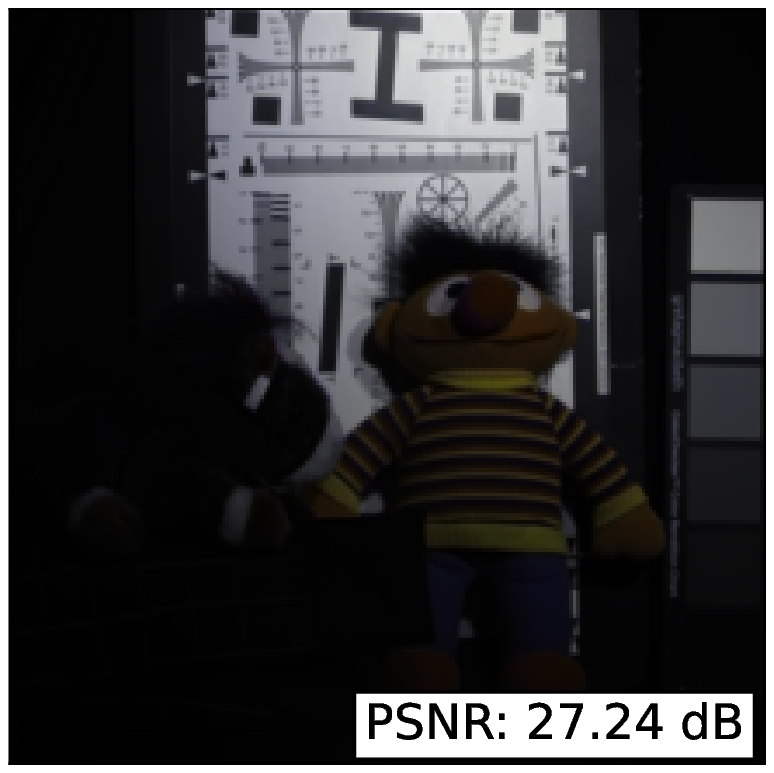}
    \end{minipage}
    &
    \hspace{-8pt}
    \begin{minipage}{\setwidecave\linewidth}
    \includegraphics[width=\linewidth]{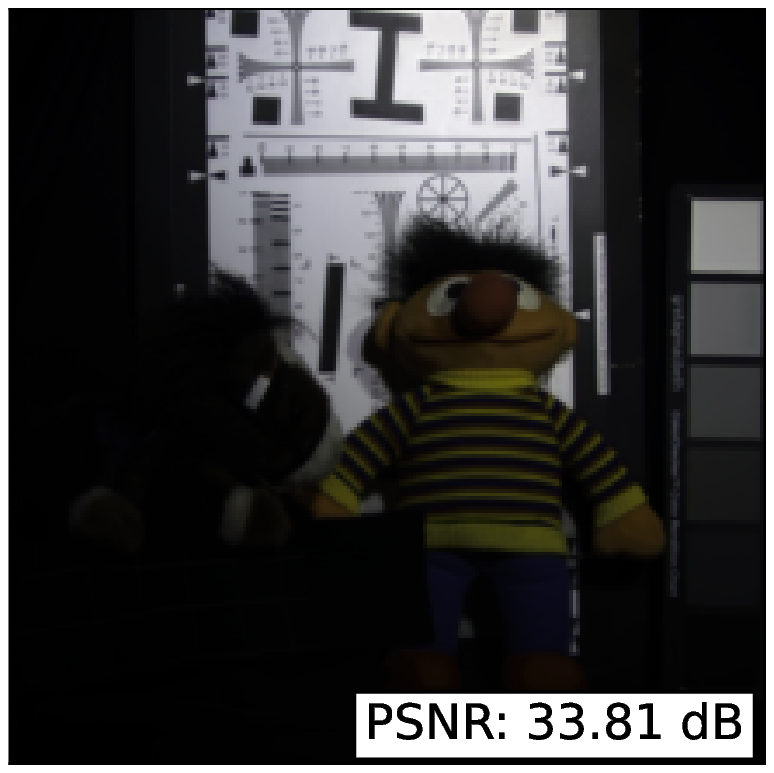}
    \end{minipage}
    \\
    \hspace{-8pt}
    \begin{minipage}{\setwidecave\linewidth}
    \includegraphics[width=\linewidth]{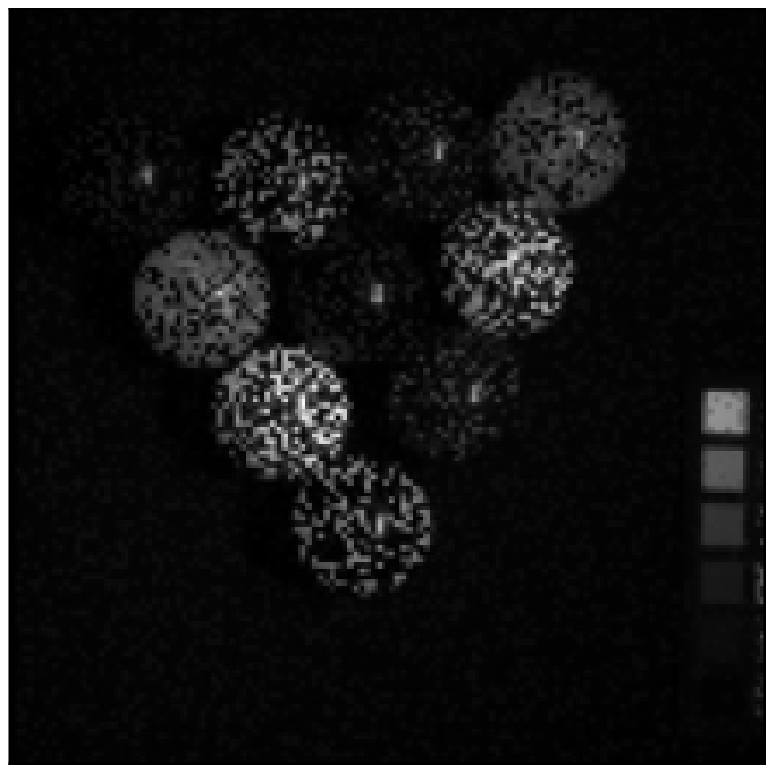}
    \end{minipage}
    &
    \hspace{-8pt}
    \begin{minipage}{\setwidecave\linewidth}
    \includegraphics[width=\linewidth]{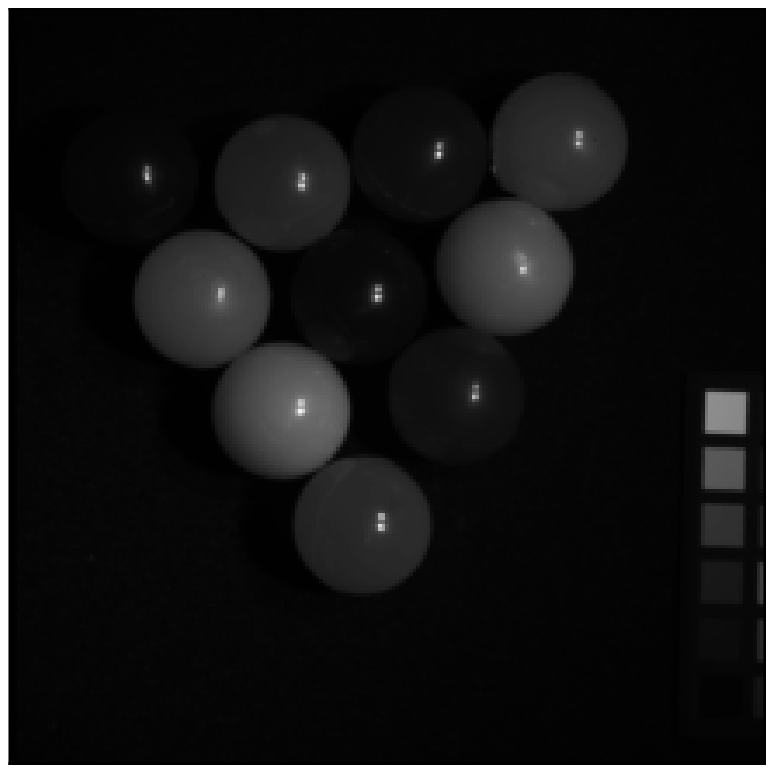}
    \end{minipage}
    &
    \hspace{-8pt}
    \begin{minipage}{\setwidecave\linewidth}
    \includegraphics[width=\linewidth]{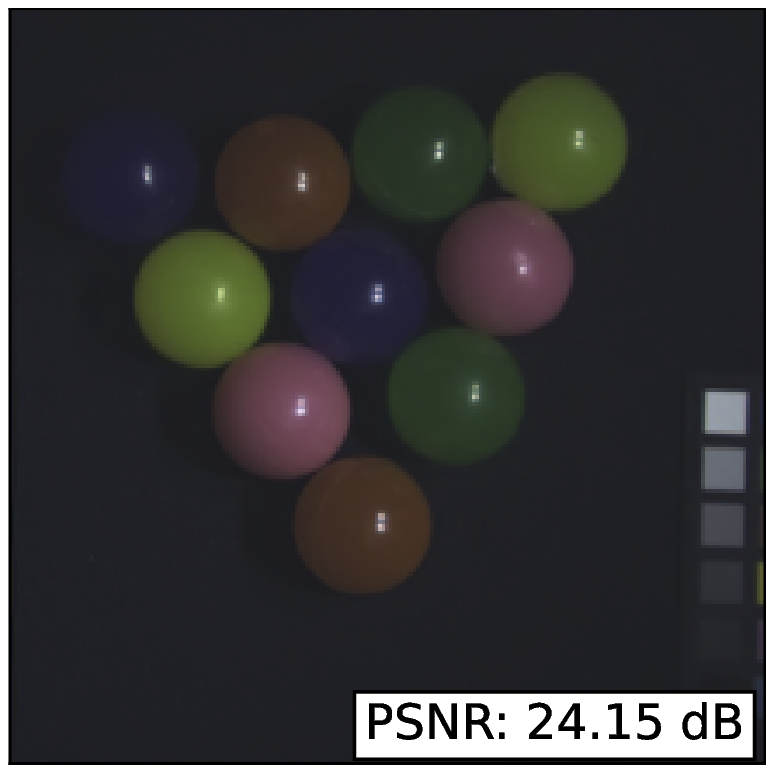}
    \end{minipage}
    &
    \hspace{-8pt}
    \begin{minipage}{\setwidecave\linewidth}
    \includegraphics[width=\linewidth]{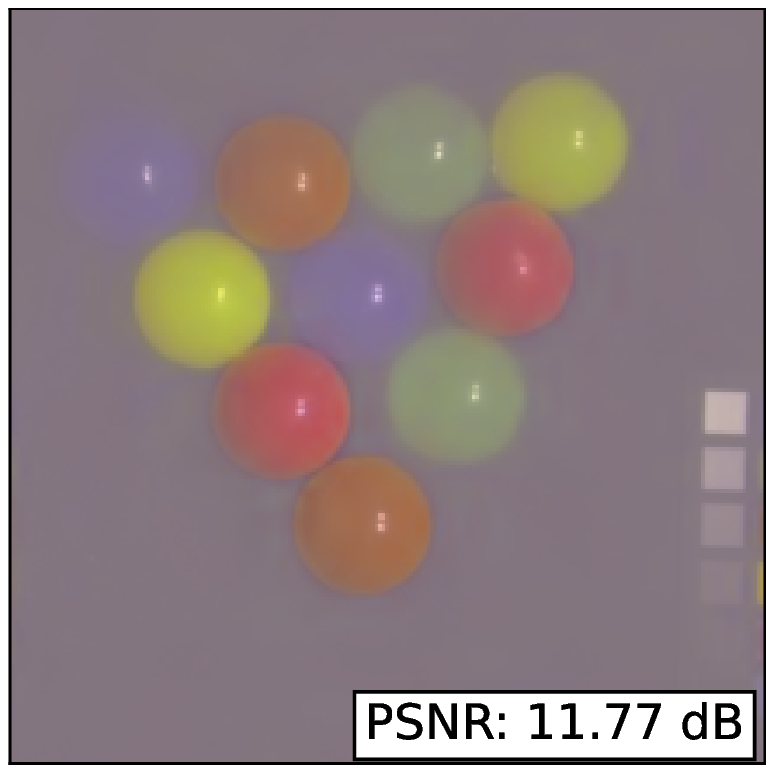}
    \end{minipage}
    &
    \hspace{-8pt}
    \begin{minipage}{\setwidecave\linewidth}
    \includegraphics[width=\linewidth]{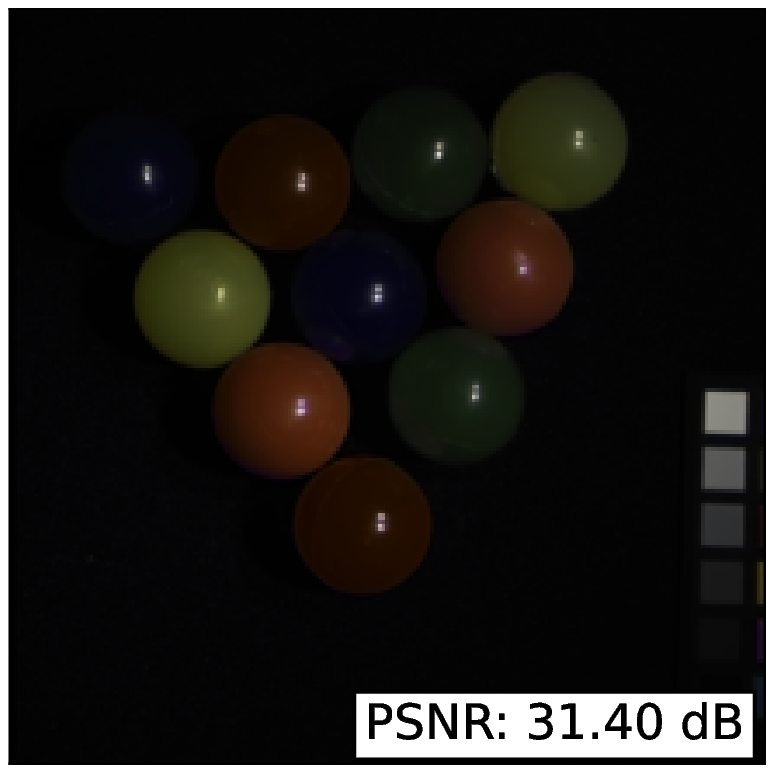}
    \end{minipage}
    &
    \hspace{-8pt}
    \begin{minipage}{\setwidecave\linewidth}
    \includegraphics[width=\linewidth]{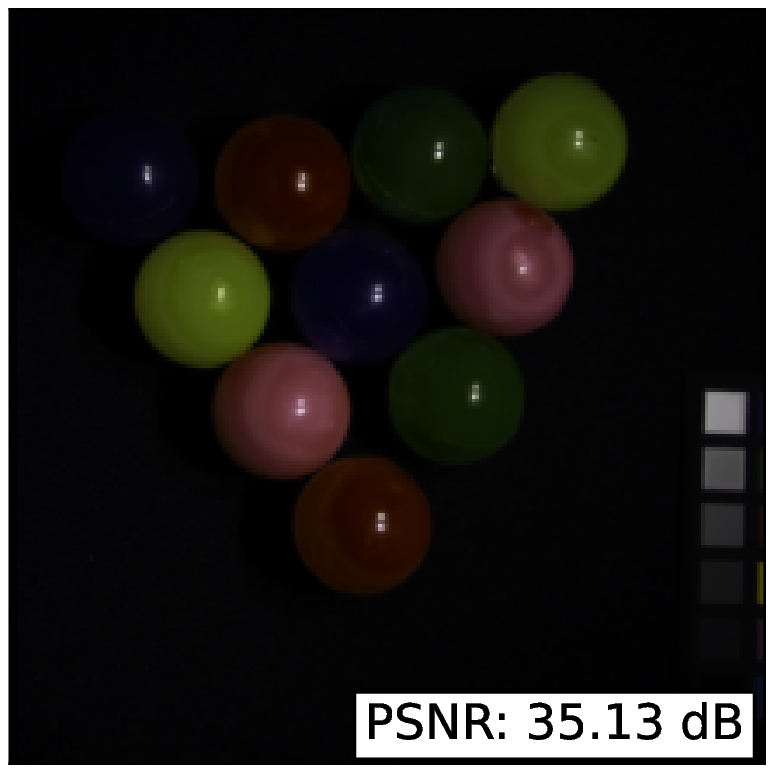}
    \end{minipage}
    \\
    \hspace{-8pt}
    \begin{minipage}{\setwidecave\linewidth}
    \includegraphics[width=\linewidth]{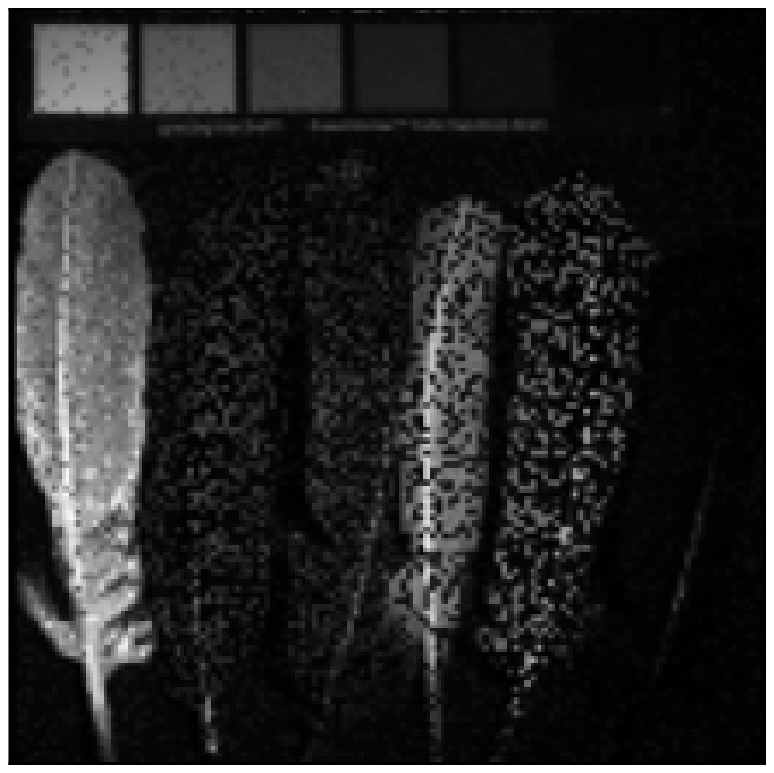}
    \end{minipage}
    &
    \hspace{-8pt}
    \begin{minipage}{\setwidecave\linewidth}
    \includegraphics[width=\linewidth]{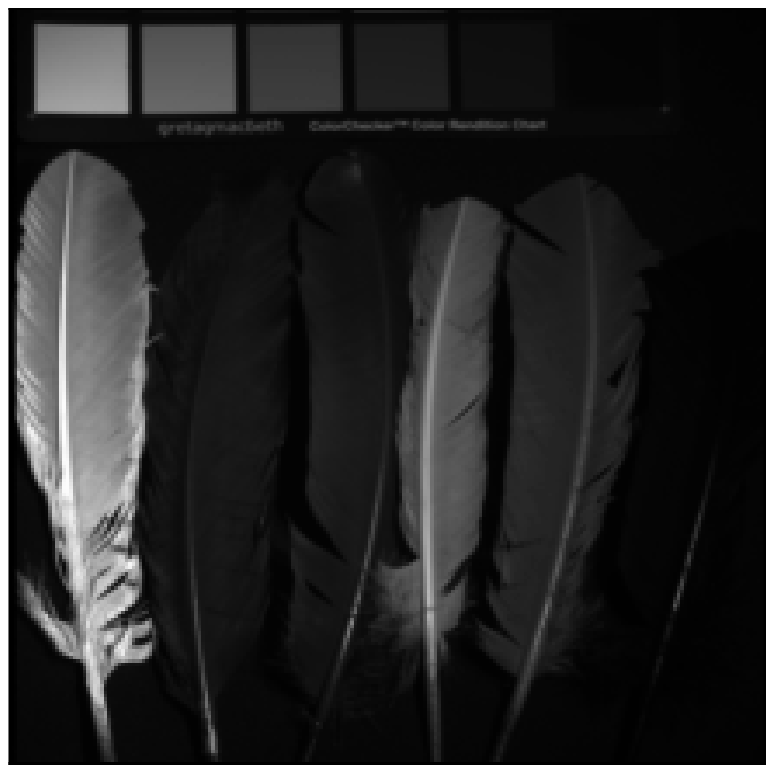}
    \end{minipage}
    &
    \hspace{-8pt}
    \begin{minipage}{\setwidecave\linewidth}
    \includegraphics[width=\linewidth]{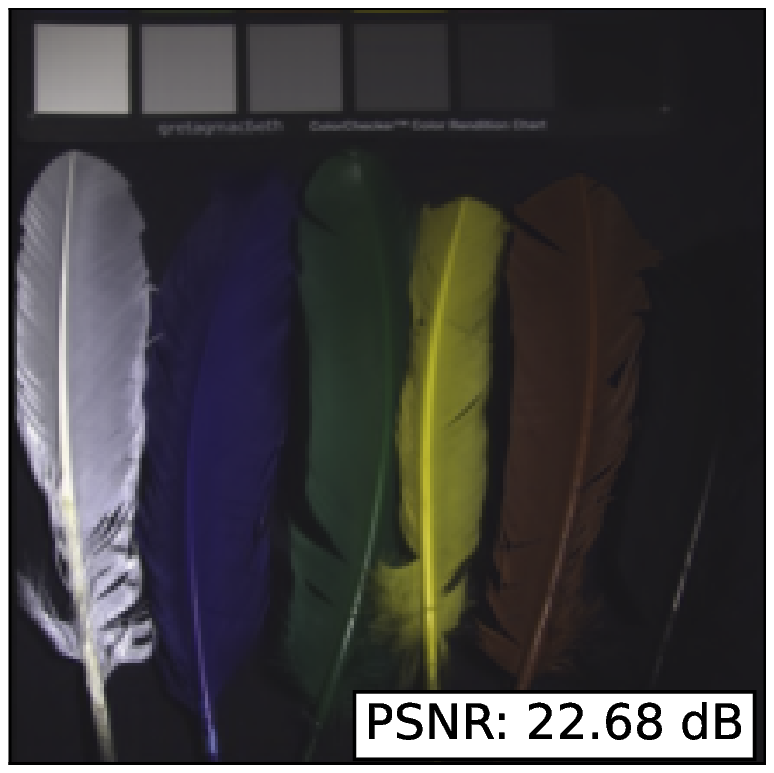}
    \end{minipage}
    &
    \hspace{-8pt}
    \begin{minipage}{\setwidecave\linewidth}
    \includegraphics[width=\linewidth]{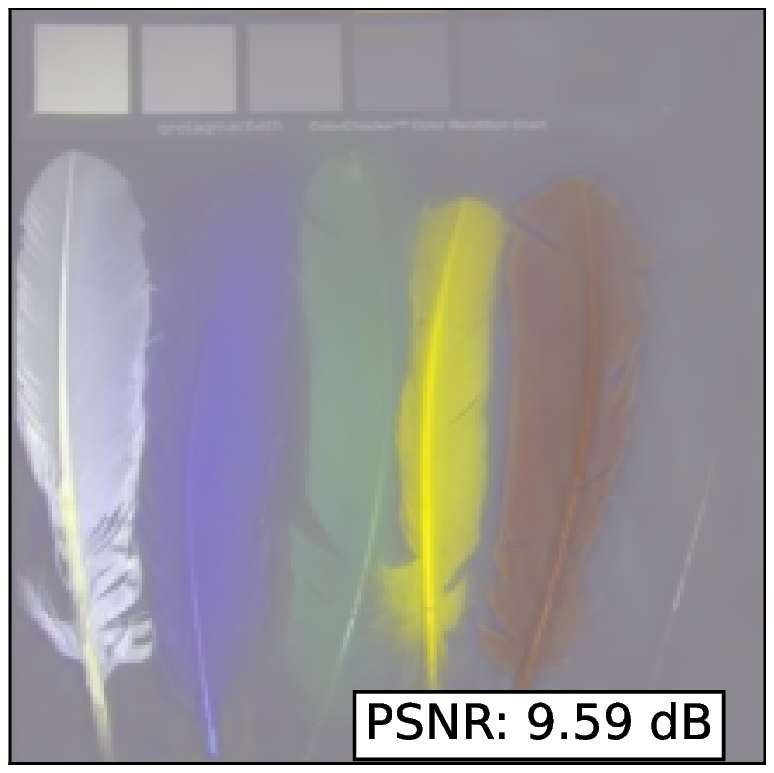}
    \end{minipage}
    &
    \hspace{-8pt}
    \begin{minipage}{\setwidecave\linewidth}
    \includegraphics[width=\linewidth]{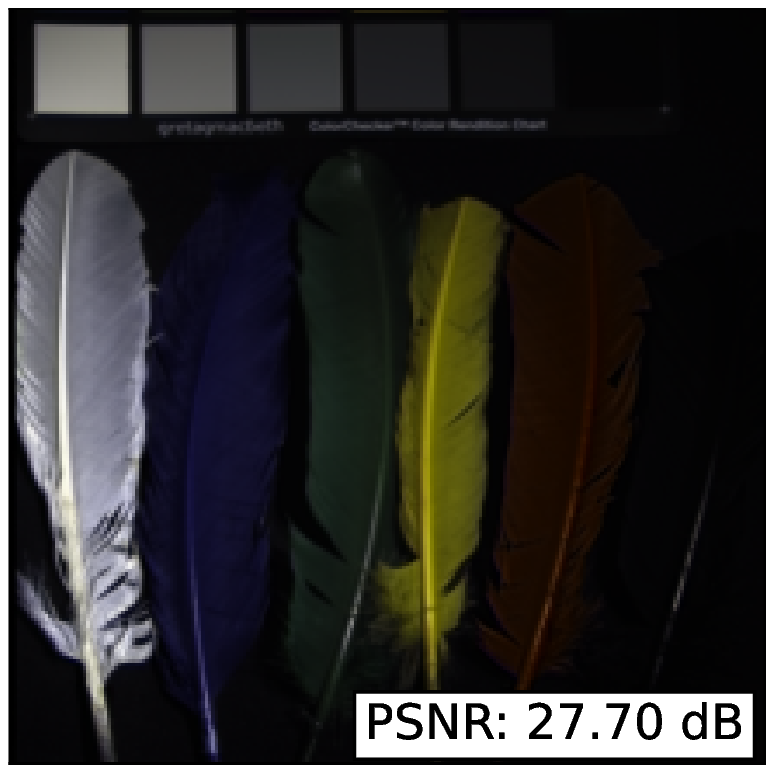}
    \end{minipage}
    &
    \hspace{-8pt}
    \begin{minipage}{\setwidecave\linewidth}
    \includegraphics[width=\linewidth]{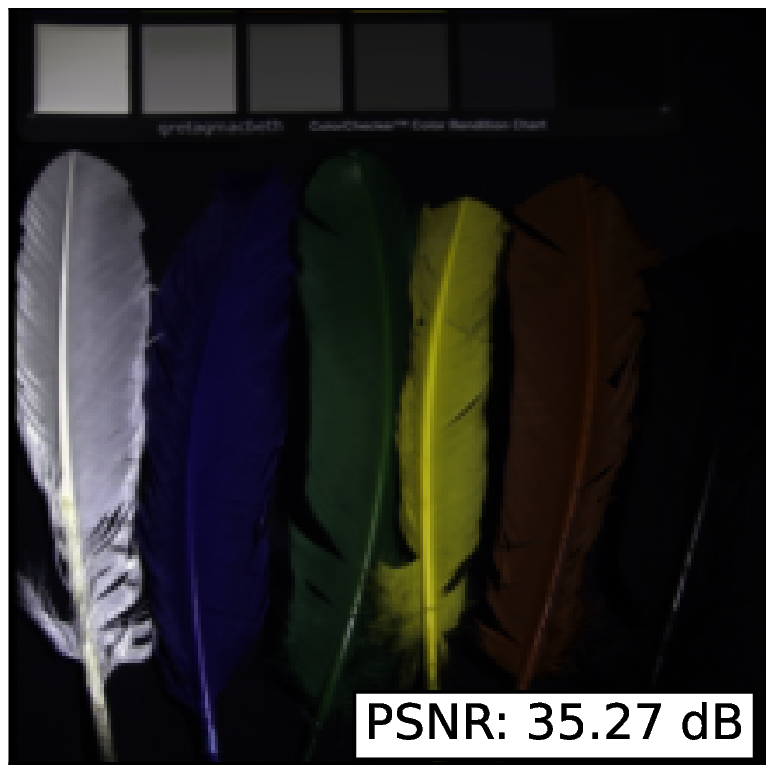}
    \end{minipage}
	\\
	\hspace{-15pt}
	\footnotesize HS shot
	&
	\hspace{-15pt}
	\footnotesize Grayscale image
	&
	\hspace{-15pt}
	\footnotesize Rec + FUSE
	&
	\hspace{-15pt}
	\footnotesize SIFCM
	&
	\hspace{-15pt}
	\footnotesize OPINE-Net
	&
	\hspace{-15pt}
	\footnotesize Proposed
\end{tabular}
\end{center}
\caption{CAVE database. RGB composite of the reconstructions obtained by the various fusion methods from hyperspectral compressive measurements and a high-spatial-resolution grayscale image.}
    \label{fig:image_comparison_SIDE}
\end{figure*}

We also evaluate the performance of the proposed fusion approach using a different acquisition setting. More precisely, the dual-arm acquisition architecture is simulated to obtain a single monochrome snapshot instead of multiple MS 3D-CASSI snapshots. In other words, the MS 3D-CASSI arm captures a single high-spatial-resolution grayscale image with a 31:1 spectral decimation factor ($q=31$). On the other hand, the HS 3D-CASSI acquires multi-frame compressive snapshots with a 16:1 spatial downsampling factor ($p=4$). This case can be considered as a compressive hyperspectral pansharpening that attempts to obtain a high-resolution spectral image from hyperspectral compressive measurements and a high-spatial-resolution panchromatic image. The first column of Fig. \ref{fig:image_comparison_SIDE} displays HS 3D-CASSI snapshots, while the second column shows high-spatial-resolution grayscale images. The last four columns in Fig. \ref{fig:image_comparison_SIDE} show the RGB composites obtained by various fusion methods for this particular acquisition setting. As can be observed in this figure, the proposed architecture outperforms the other approaches achieving an improvement gain of at least 3 dB.

\subsection{Laboratory data set}

\begin{figure}
\begin{center}
    \begin{tabular}{c c c c c c}
    \hspace{-12pt}
    \begin{minipage}{\setwideb\linewidth}
    \includegraphics[width=\linewidth]{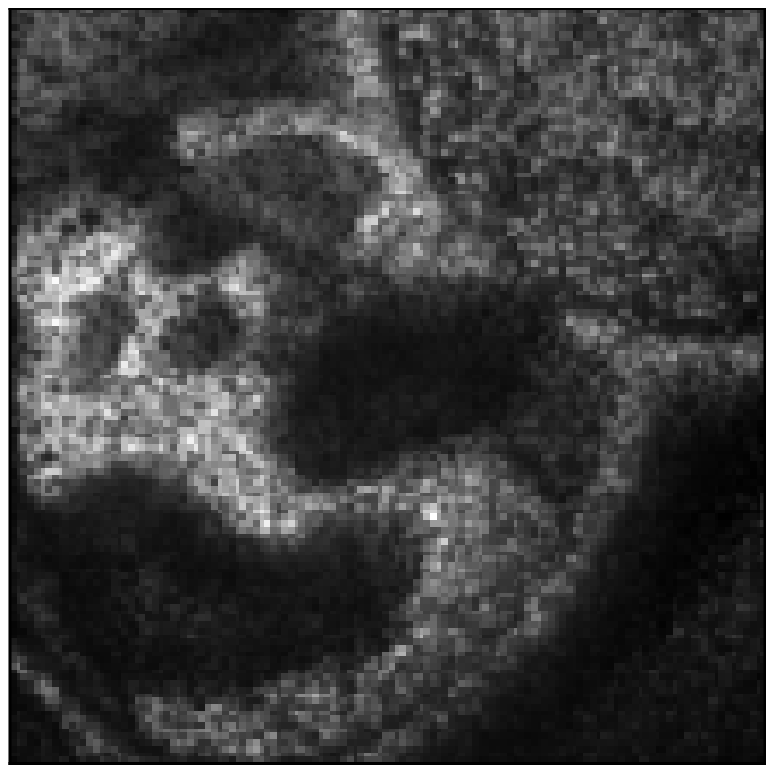}
    \end{minipage}
    &
	\hspace{-12pt}
    \begin{minipage}{\setwideb\linewidth}
    \includegraphics[width=\linewidth]{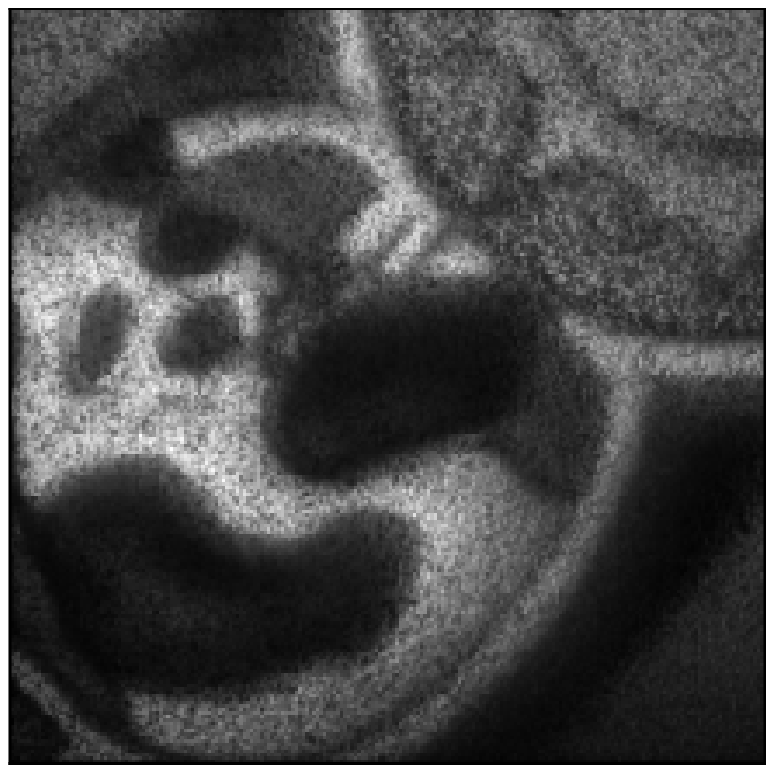}
    \end{minipage}
    &
	\hspace{-12pt}
    \begin{minipage}{\setwideb\linewidth}
    \includegraphics[width=\linewidth]{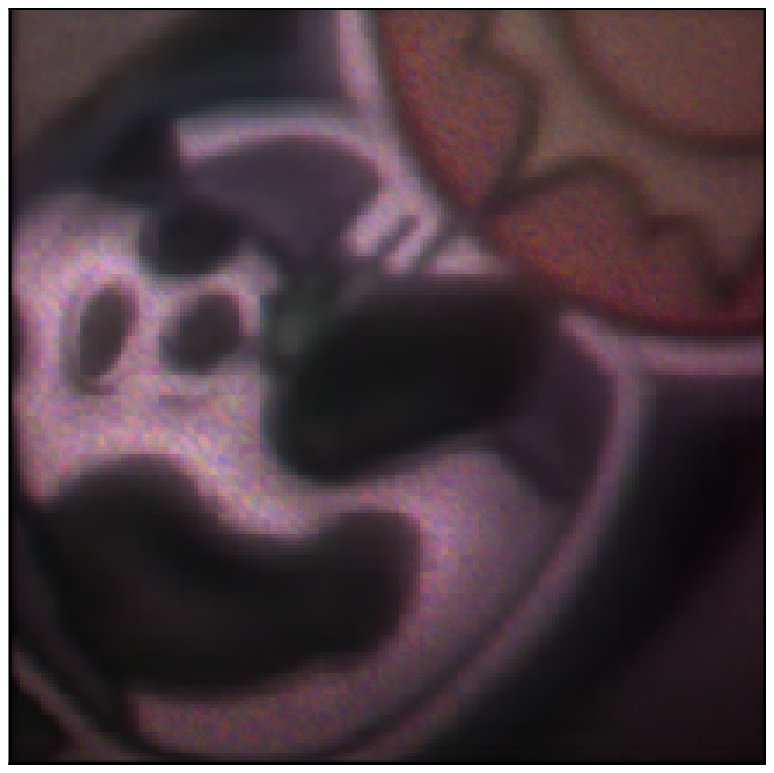}
    \end{minipage}
    &
	\hspace{-12pt}
    \begin{minipage}{\setwideb\linewidth}
    \includegraphics[width=\linewidth]{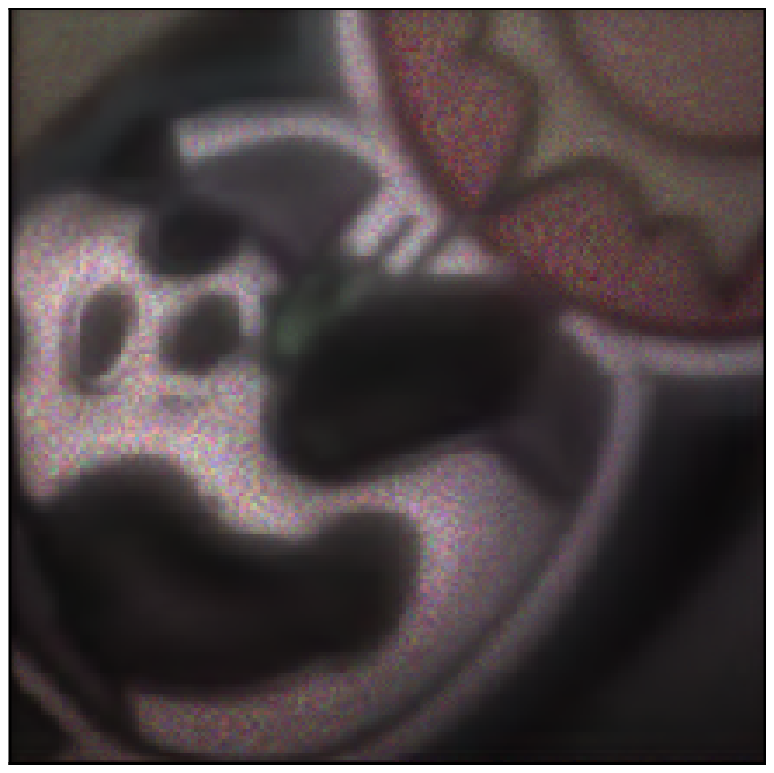}
    \end{minipage}
    &
	\hspace{-12pt}
    \begin{minipage}{\setwideb\linewidth}
    \includegraphics[width=\linewidth]{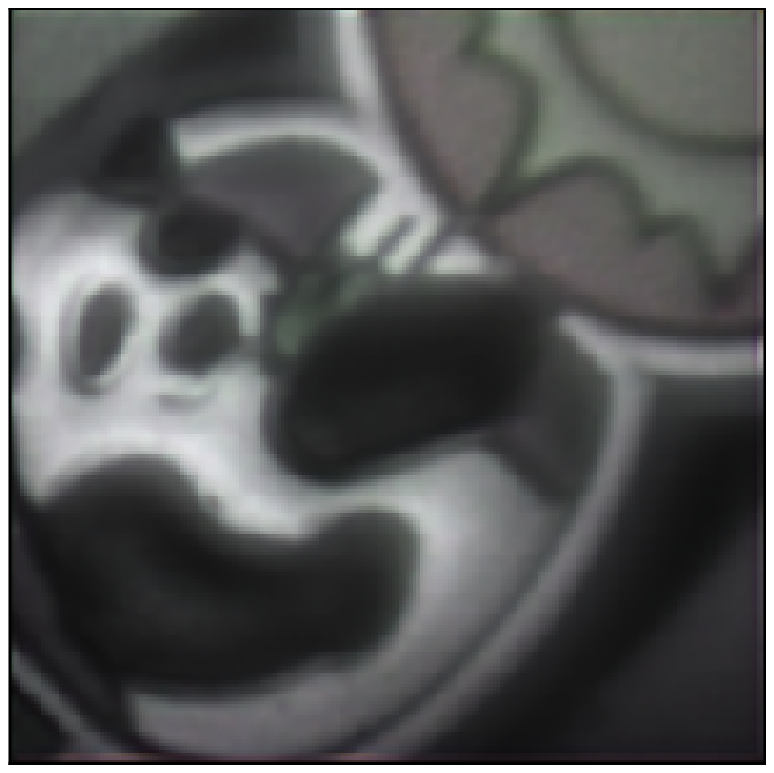}
    \end{minipage}
    &
	\hspace{-12pt}
    \begin{minipage}{\setwideb\linewidth}
    \includegraphics[width=\linewidth]{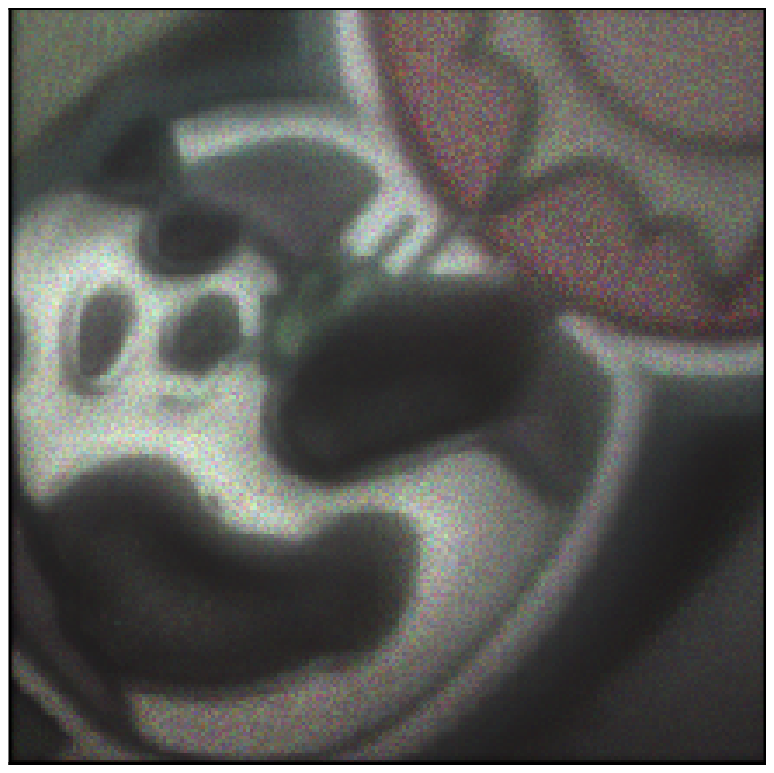}
    \end{minipage}
    \\
    \scriptsize
    \hspace{-12pt}
    \scriptsize (a) HS shot
    &
    \hspace{-12pt}
    \scriptsize (b) MS shot
    &
     \hspace{-12pt}
     \scriptsize (c) Rec $+$ FUSE
    &
    \scriptsize
     \hspace{-12pt}
    (d) SIFCM
     &
     \hspace{-12pt}
    \scriptsize (e) ISTA-Net
     & 
     \hspace{-12pt}
    \scriptsize (f) Proposed
    \end{tabular}
\end{center}
    \caption{Experimental data set. (a) shot captured by the hyperspectral arm, (b) shot captured by the multispectral arm. RGB composite of fused images obtained by (c) Rec $+$ Fuse, (d) SIFCM, (e) ISTA-Net, and (f) the proposed LADMM-Net.}
    \label{fig:real_data}
\end{figure}

We also evaluate the proposed reconstruction approach from compressive measurements captured by an optical setup that implements the dual-arm system based on the 3D-CASSI architecture. To be more precise, the experimental setup comprises a 100 mm objective lens, a 100 mm relay lens, a digital micromirror device (DMD), and a camera detector based on a charged coupled device (CCD). To obtain the dual-resolution compressive measurements the patterns of the coded aperture were separately emulated using the DMD. In particular, we fixed the spatial resolution of the DMD to $256 \times 256$ to capture the compressive measurements of the multispectral arm, while the DMD spatial resolution was fixed to $128 \times 128$ to obtain the compressive samples of the hyperspectral arm. Figures \ref{fig:real_data}(a) and (b) show, respectively, a snapshot captured by the hyperspectral arm and a projection obtained by the multispectral arm. For comparative purposes, Figs. \ref{fig:real_data}(c)-(e) show the RGB composites of the spectral images recovered by the methods Rec $+$ Fuse, SIFCM, and ISTA-Net, respectively. An RGB composite of the fused image obtained by the proposed approach is shown in Fig \ref{fig:real_data}(f).

\subsection{CS reconstruction of natural images}

The proposed algorithm unrolling technique can be adapted to solve a particular compressive sensing problem. In this section, the performance of the proposed architecture is tested to recover grayscale images from compressive random projections. In this regard, the target variable update is given by
\begin{equation}\label{eq:au_eq2}
    \pmb{f}^{(k)} = \pmb{f}^{(k-1)} - \frac{1}{\alpha^{(k)}} \left[ \pmb{H}^{\top}\left(\pmb{H}\pmb{f}^{(k-1)} - \pmb{y}\right) + \rho^{(k)} \pmb{r}^{(k-1)} \right]
\end{equation}
where the CS-based acquisition process can be described as $\pmb{y} = \pmb{H}\pmb{f}  + \pmb{\eta}$, where $\pmb{y} \in \mathbb{R}^{m}$ is the vector containing the compressive measurements; $\pmb{H} \in \mathbb{R}^{m \times n}$ denotes the measurement matrix with $m \ll n$; $\pmb{f} \in \mathbb{R}^{n}$ is the image of interest in vector form; and $\pmb{\eta} \in \mathbb{R}^{M}$ is the noise vector whose entries are frequently characterized as iid random samples following a zero-mean Gaussian distribution \cite{CandesIntroduction2008, DonohoCompressed2006}. Notice that this network architecture can be used in other compressive sampling applications including sparse channel estimation \cite{vargas2020admm}, smart grids \cite{sun2016data}, among others.

In this sense, we obtain the training data by extracting $B = 5120$ image blocks with dimensions $33 \times 33$ from the Berkeley BSD dataset \cite{MartinFTM01}. Furthermore, the testing data set includes $12$ standard images with dimensions  $256 \times 256$. For a given compression ratio, the measurement matrix is generated whose elements are random samples following a zero-mean Gaussian distribution. The sampling matrix columns $\{\mathbf{h}_{\jmath}\}_{\jmath = 1}^B$ are normalized such that $\|\mathbf{h}_{\jmath}\|_2^2 =1$ for $\jmath=1,\ldots,B$.

We compare the performance of the proposed deep network with respect to other reconstruction techniques such as the gradient projection for sparse reconstruction (GPSR) \cite{FiguereidoGradient2007}, the linearized version of the alternating direction method of multipliers (LADMM) \cite{esser2010general}, the denoising-based approximated message passing (D-AMP) method \cite{MetzlerFrom2016}, the ReconNet architecture \cite{kulkarni2016reconnet}, the LISTA approach \cite{gregor2010learning}, the CSNet technique \cite{ShiDeep2017}, the ISTA-Net network \cite{zhang2018ISTA}, and the OPINE-Net architecture \cite{zhang2020optimization}. Notice that GPSR, LADMM, and D-AMP are model-based CS reconstruction methods. Furthermore, ReconNet and CSNet architectures belong to the class of deep learning techniques. Additionally, LISTA, ISTA-Net, OPINE-Net, and LADMM-Net are algorithm unrolling techniques based on deep networks. The algorithm unrolling techniques are tested using a similar number of learnable parameters. For this application, different models were trained for distinct compression ratios $\{5 \%, 10\%, 15\%, 20\%, 25\% \}$ with $512$ training epochs.

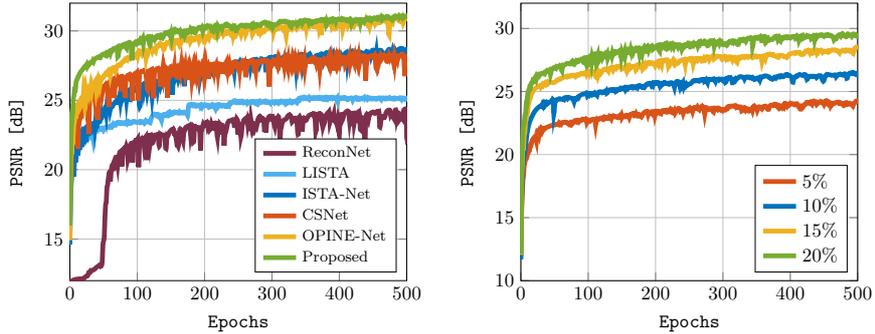
\begin{figure}
\begin{center}
    \begin{tabular}{c c}
  \hspace{-10pt}     
    \begin{minipage}[c]{0.35\linewidth}
    \resizebox{1.00\textwidth}{!}{
    \begin{tikzpicture}
		\begin{axis}[
		legend cell align = left,
		legend pos = south east,
		scale = 0.90,
		grid = major,
		ylabel = \texttt{PSNR [dB]},
		ylabel style = {yshift=-0.25cm},
		ymin = 12,
		ymax = 32,
		xmin = 0,
		xmax = 500,
		xlabel = \texttt{Epochs},
		]
        \addplot[color=mycolor5, smooth, line width = 2.50pt] table[x ={x},y = y3]{FigureData/learning_methods.dat};
        \addlegendentry{\footnotesize ReconNet}
        \addplot[color=mycolor6, smooth, line width = 2.50pt] table[x ={x},y = y]{FigureData/learning_LISTA.dat};
        \addlegendentry{\footnotesize LISTA}
       	\addplot[color=mycolor1, smooth, line width = 2.50pt] table[x ={x},y = y2]{FigureData/learning_methods.dat};
       	\addlegendentry{\footnotesize ISTA-Net}
        \addplot[color=mycolor2, smooth, line width = 2.50pt] table[x ={x},y = y]{FigureData/learning_CSNET.dat};
        \addlegendentry{\footnotesize CSNet}
        \addplot[color=mycolor3, smooth, line width = 2.50pt] table[x ={x},y = y]{FigureData/learning_OPINE.dat};
        \addlegendentry{\footnotesize OPINE-Net}
        \addplot[color=mycolor4, smooth, line width = 2.50pt] table[x ={x},y = y]{FigureData/learning_LADMM2.dat};
        \addlegendentry{\footnotesize Proposed}
		\end{axis}
	\end{tikzpicture}
	}
    \end{minipage} 
    
    &

    \begin{minipage}[c]{0.35\linewidth}
    \resizebox{1.00\textwidth}{!}{
    \begin{tikzpicture}
		\begin{axis}[
		legend cell align = left,
		legend pos = south east,
		scale = 0.90,
		grid = major,
		ylabel = \texttt{PSNR [dB]},
		ylabel style = {yshift=-0.25cm},
		ymin = 10,
		ymax = 32,
		xmin = 0,
		xmax = 500,
		xlabel = \texttt{Epochs},
		]
        \addplot[color=mycolor2, smooth, line width = 2.50pt] table[x ={x},y = y4]{FigureData/learning_compression.dat};
        \addlegendentry{$5\%$}
        \addplot[color=mycolor1, smooth, line width = 2.50pt] table[x ={x},y = y3]{FigureData/learning_compression.dat};
        \addlegendentry{$10\%$}
        \addplot[color=mycolor3, smooth, line width = 2.50pt] table[x ={x},y = y2]{FigureData/learning_compression.dat};
        \addlegendentry{$15\%$}
        \addplot[color=mycolor4, smooth, line width = 2.50pt] table[x ={x},y = y1]{FigureData/learning_compression.dat};
        \addlegendentry{$20\%$}
		\end{axis}
	\end{tikzpicture}
	}
    \end{minipage}

    \end{tabular}
    \end{center}
    \caption{(Left) PSNR obtained by the different deep learning approaches versus the epochs for a compression ratio of $25\%$. (Right) PSNR obtained by the proposed approach versus the epochs for different compression rates.}
    \label{fig:learning}
\end{figure}

\begin{table}\footnotesize
\caption{The ensemble average of the PSNR in DB obtained by the various CS reconstruction methods from compressive measurements for different compression ratios.}
    \centering
    \begin{tabular}{c|c|c|c|c|c}
    \hline
    \hline
    & \multicolumn{5}{c}{Compression rate} \\
    \hline
    Method   & $5\%$ & $10\%$ & $15\%$ & $20\%$ & $25\%$  \\
    \hline
    \hline
    GPSR   & $15.04 \pm 1.25$ & $18.16 \pm 2.37$ & $19.28 \pm 2.94$ & $20.49 \pm 3.08$ & $21.25 \pm 3.05$  \\
    LADMM  & $15.33 \pm 1.57$ & $18.10 \pm 2.13$ & $19.48 \pm 3.26$ & $20.65 \pm 3.24$ & $21.30 \pm 2.98$  \\
    D-AMP  & $11.23 \pm 1.10$ & $18.24 \pm 2.00$ & $21.82 \pm 2.97$ & $24.64 \pm 2.98$ & $27.58 \pm 3.84$  \\
    \hline
    ReconNet & $12.88 \pm 1.66$ & $21.94 \pm 3.25$ & $22.90 \pm 3.27$ & $23.51 \pm 3.23$ & $23.68 \pm 3.16$  \\
    LISTA    & $20.96 \pm 3.18$ & $22.51 \pm 3.10$ & $23.84 \pm 3.02$ & $24.81 \pm 3.10$ & $24.61 \pm 2.76$  \\
    CSNet    & $21.23 \pm 3.35$ & $23.33 \pm 3.41$ & $24.55 \pm 3.63$ & $25.53 \pm 3.59$ & $26.62 \pm 3.66$  \\
    ISTA-Net & $21.45 \pm 3.01$ & $22.59 \pm 3.03$ & $24.04 \pm 3.13$ & $25.26 \pm 3.25$ & $26.94 \pm 3.44$  \\
    OPINE-Net & 21.71 $\pm$ 3.12 & $\mathbf{24.46 \pm 3.50}$ & 26.12 $\pm$ 3.58 & 27.50 $\pm$3.66 & 28.31 $\pm$ 3.74  \\
    Proposed & $\mathbf{22.08 \pm 3.26}$ & $24.30 \pm 3.54$ & $\mathbf{26.15 \pm 3.91}$ & $\mathbf{27.76 \pm 3.95}$ & $\mathbf{29.24 \pm 3.96}$  \\
    \hline
    \hline
    \end{tabular}
    \label{tab:CS_image_psnr}
\end{table}

\begin{figure*}
\begin{center}
\begin{tabular}{c c c c c}
    \includegraphics[width=0.13\linewidth]{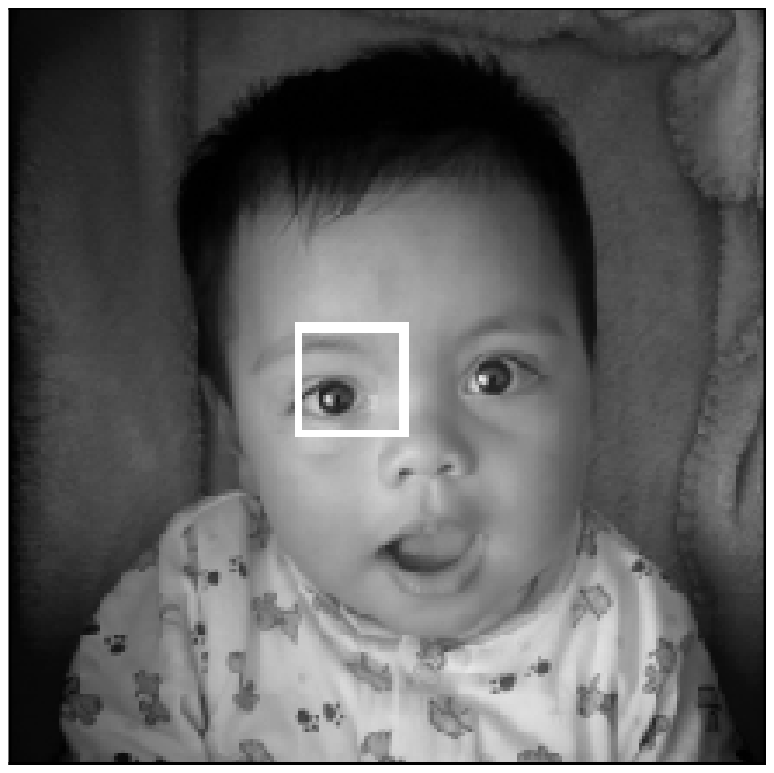}
    &
    \includegraphics[width=0.13\linewidth]{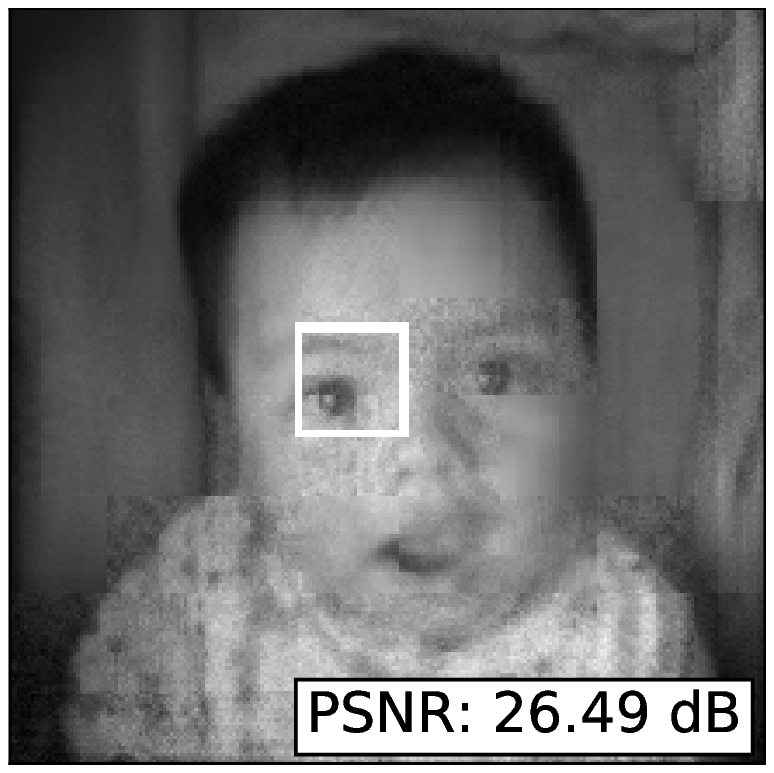}
    &
    \includegraphics[width=0.13\linewidth]{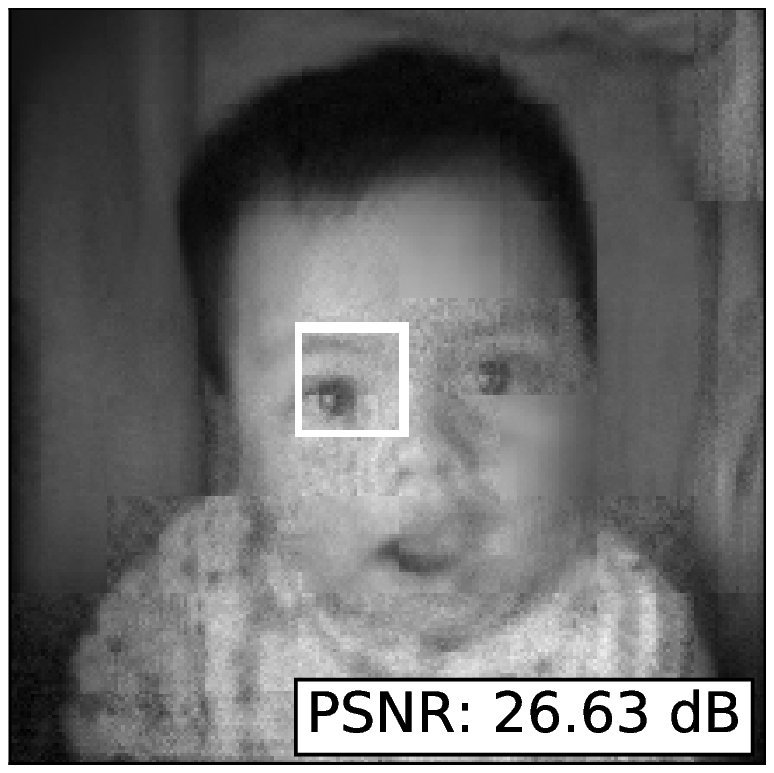}
    &
    \includegraphics[width=0.13\linewidth]{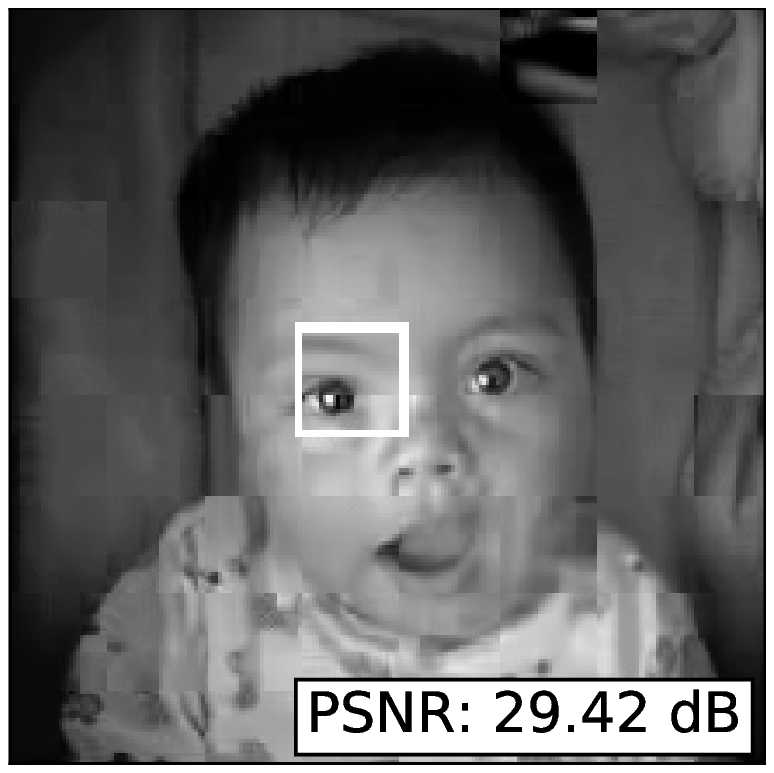}\vspace{-2pt}
    &
    \includegraphics[width=0.13\linewidth]{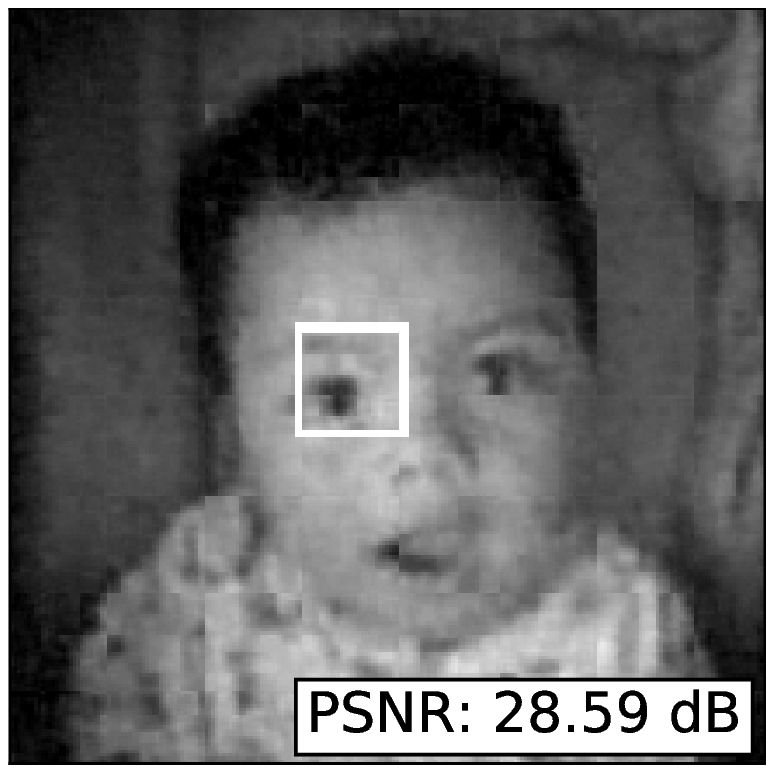}
	\\
    \includegraphics[width=0.13\linewidth]{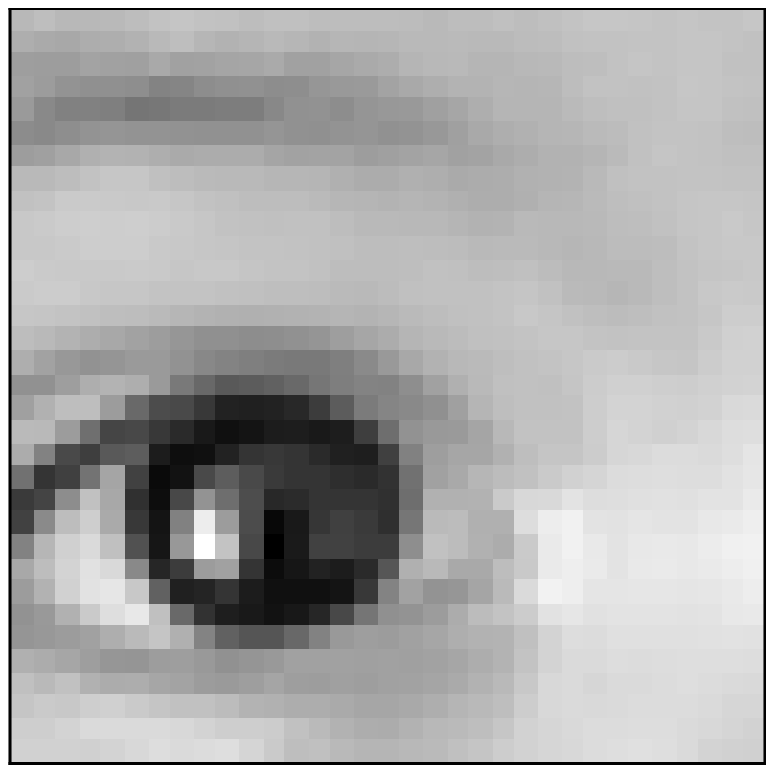}
    &
    \includegraphics[width=0.13\linewidth]{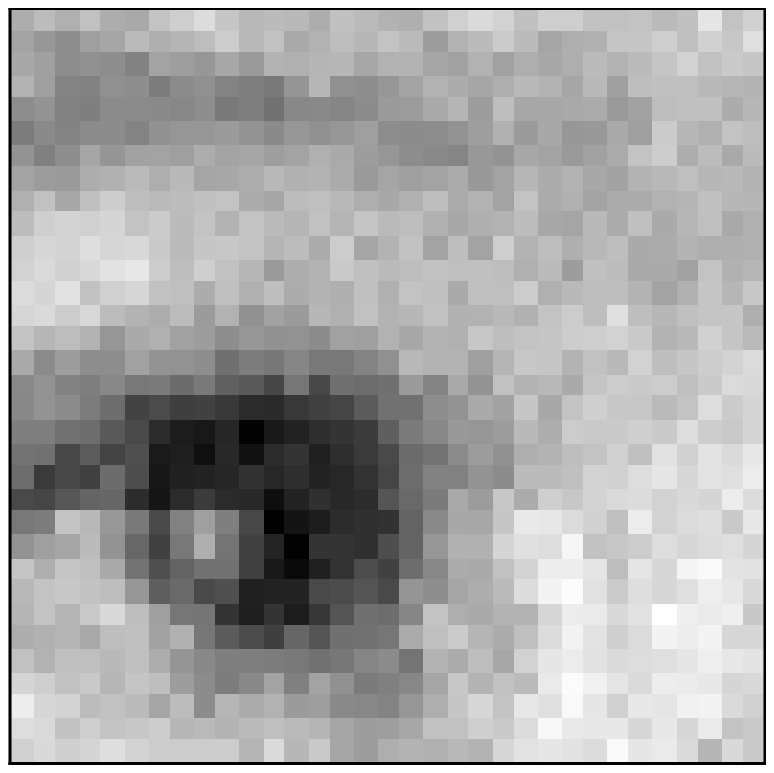}
    &
    \includegraphics[width=0.13\linewidth]{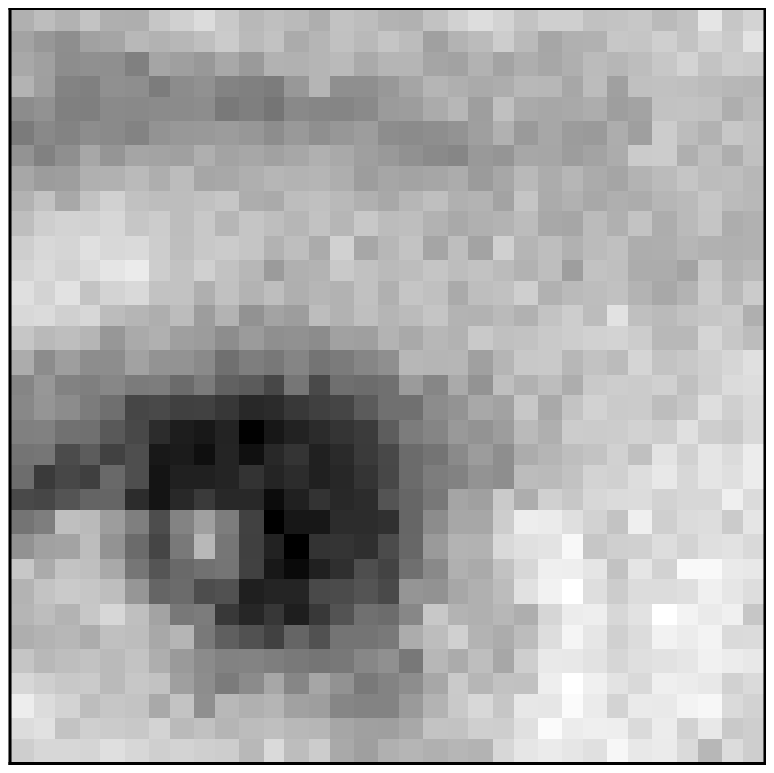}
    &
    \includegraphics[width=0.13\linewidth]{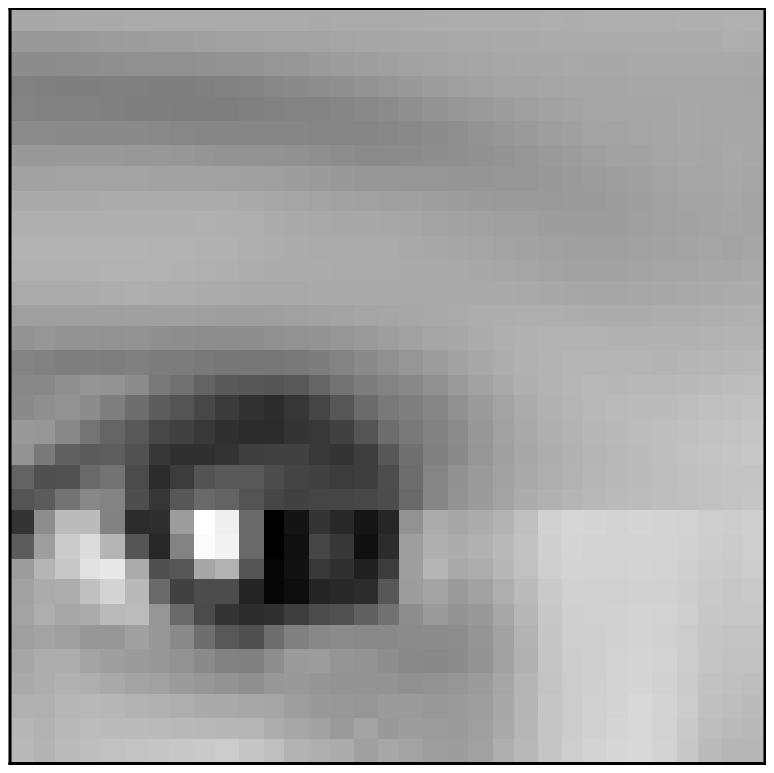}\vspace{-4pt}
    &
    \includegraphics[width=0.13\linewidth]{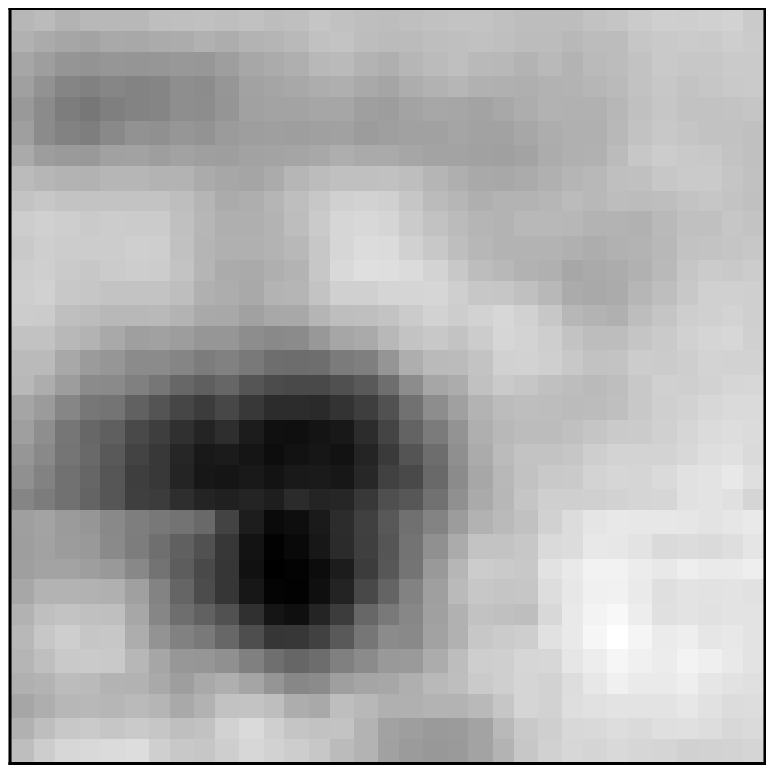}
    \\
    \scriptsize
    (a) Original
    &
    \scriptsize
    (b) GPSR
    &
    \scriptsize
    (c) LADMM
    &
    \scriptsize
    (d) D-AMP \vspace{2pt}
    &
    \scriptsize
    (e) ReconNet 
	\\
    \includegraphics[width=0.13\linewidth]{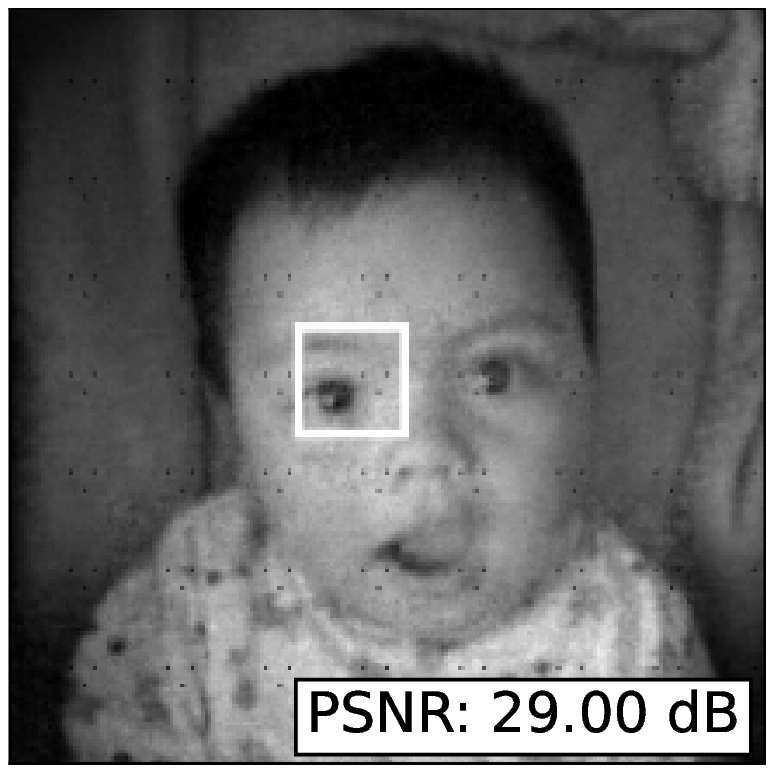}
    &
	\includegraphics[width=0.13\linewidth]{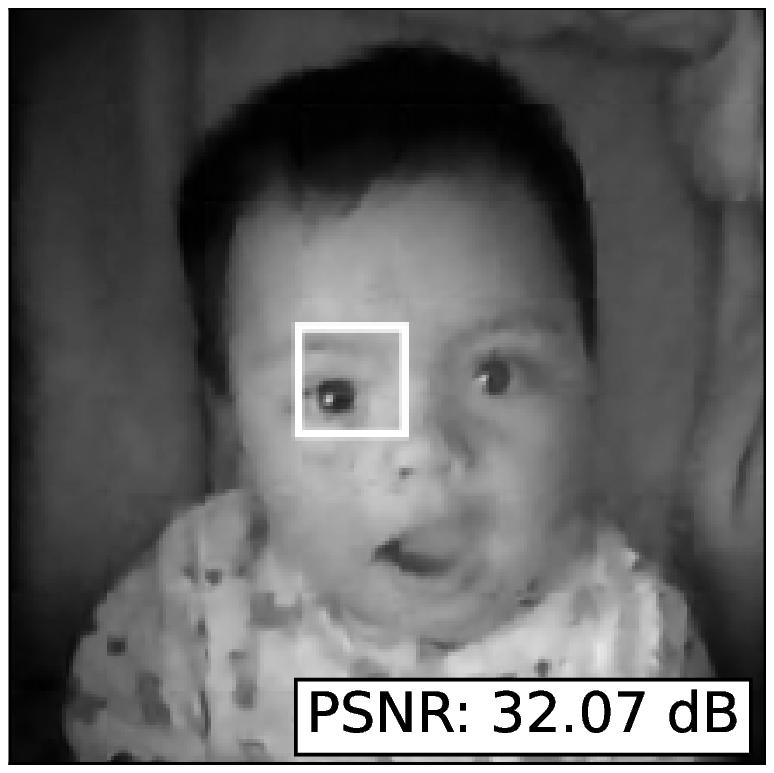}
	&
	\includegraphics[width=0.13\linewidth]{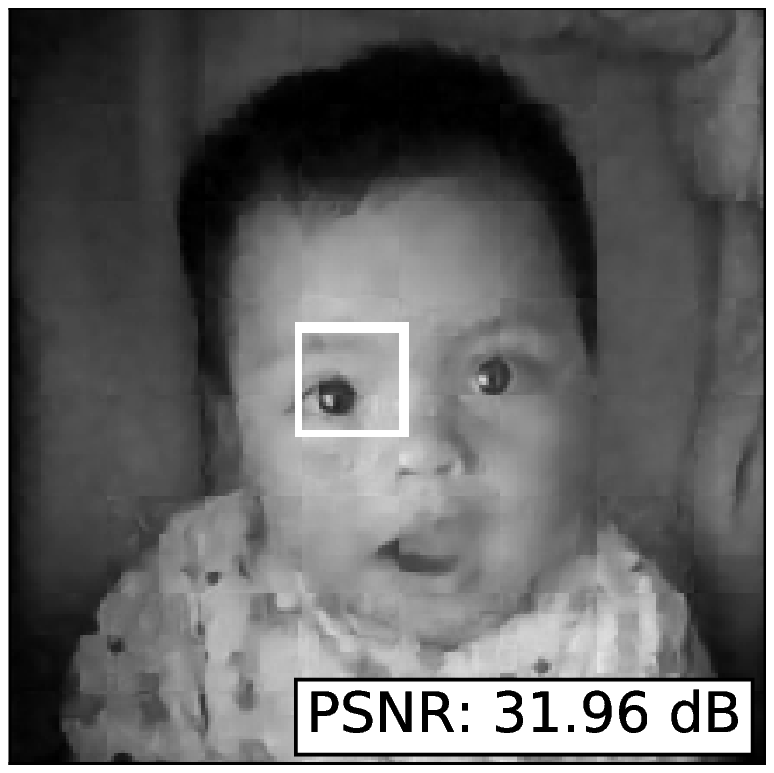}
	&
	\includegraphics[width=0.13\linewidth]{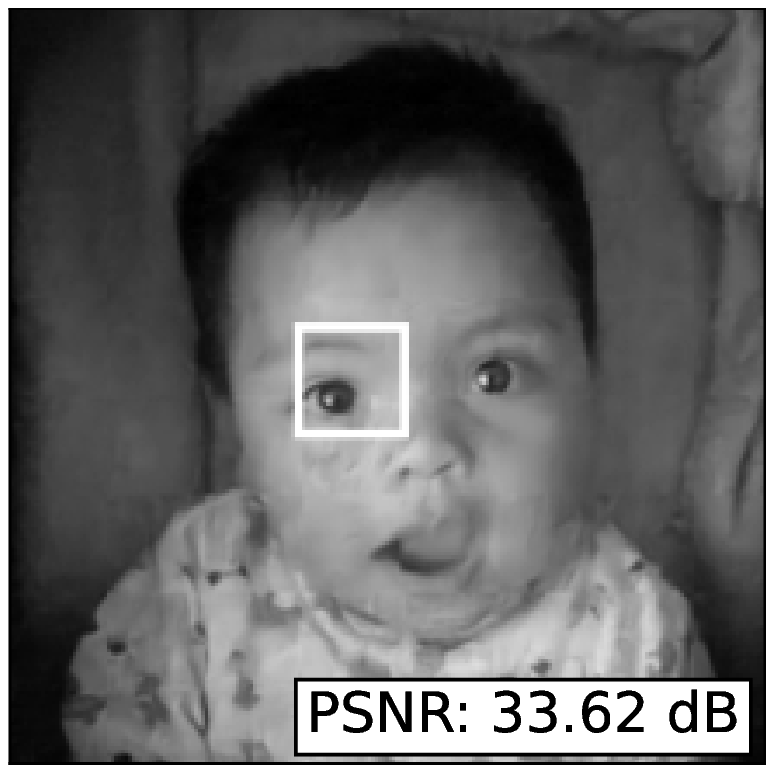}
	&
    \includegraphics[width=0.13\linewidth]{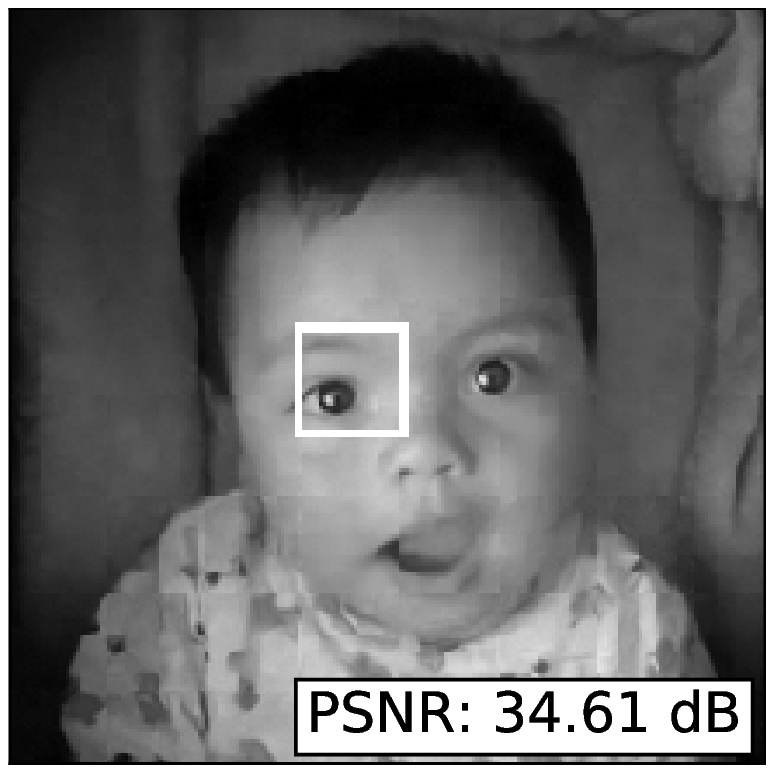}\vspace{-1pt}
    \\
    \includegraphics[width=0.13\linewidth]{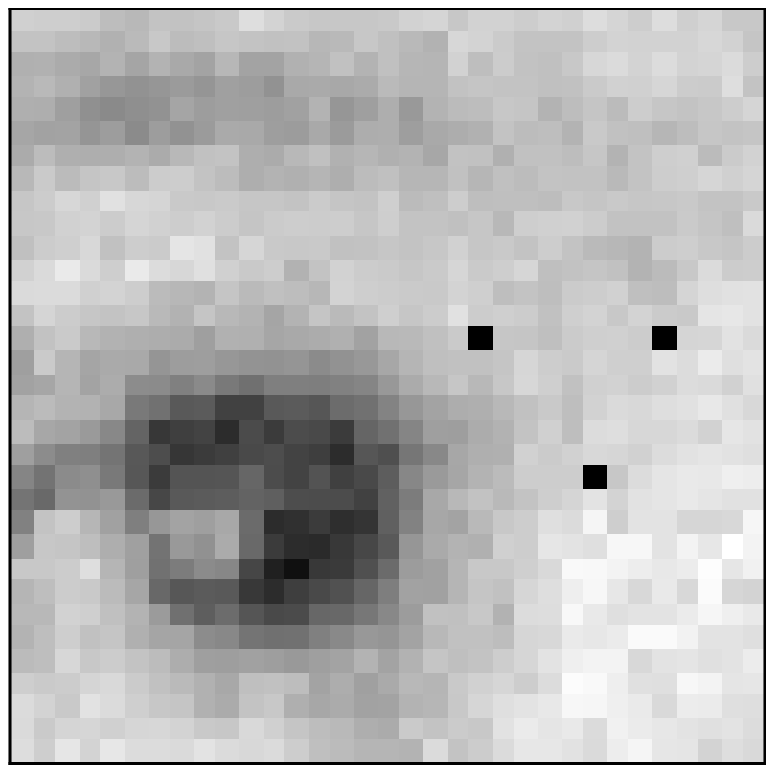}
	&
	\includegraphics[width=0.13\linewidth]{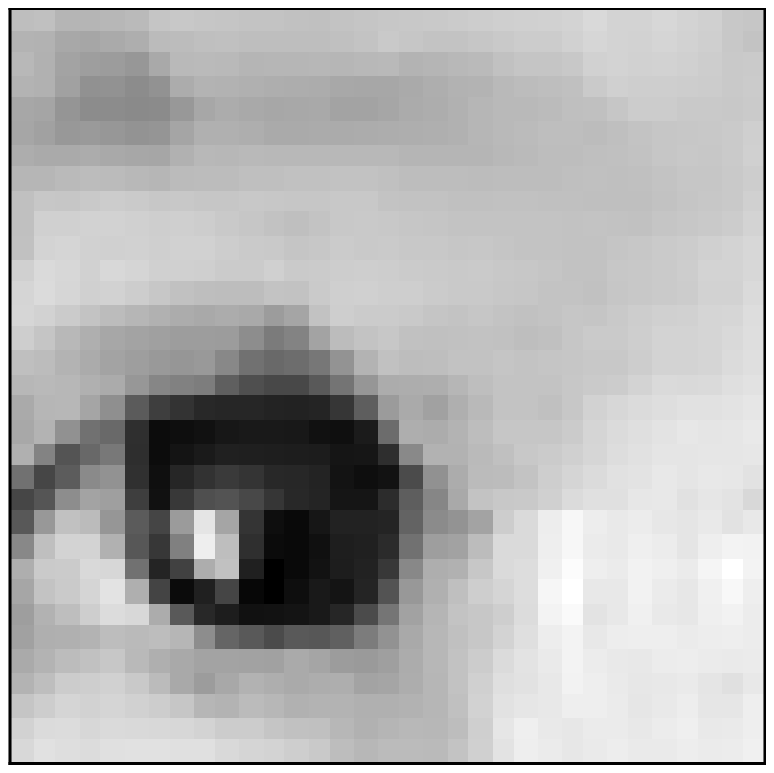}
	&
	\includegraphics[width=0.13\linewidth]{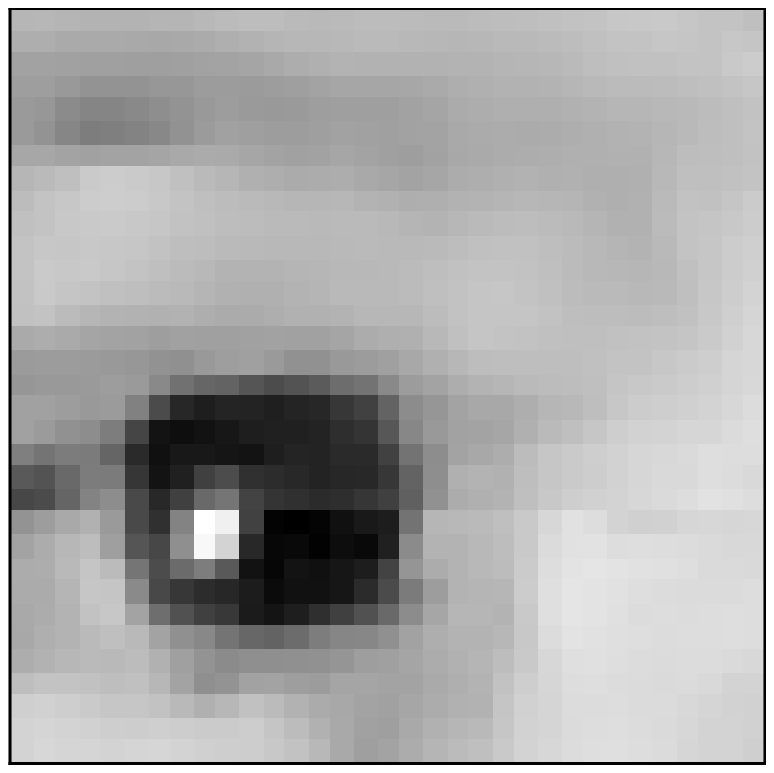}
	&
	\includegraphics[width=0.13\linewidth]{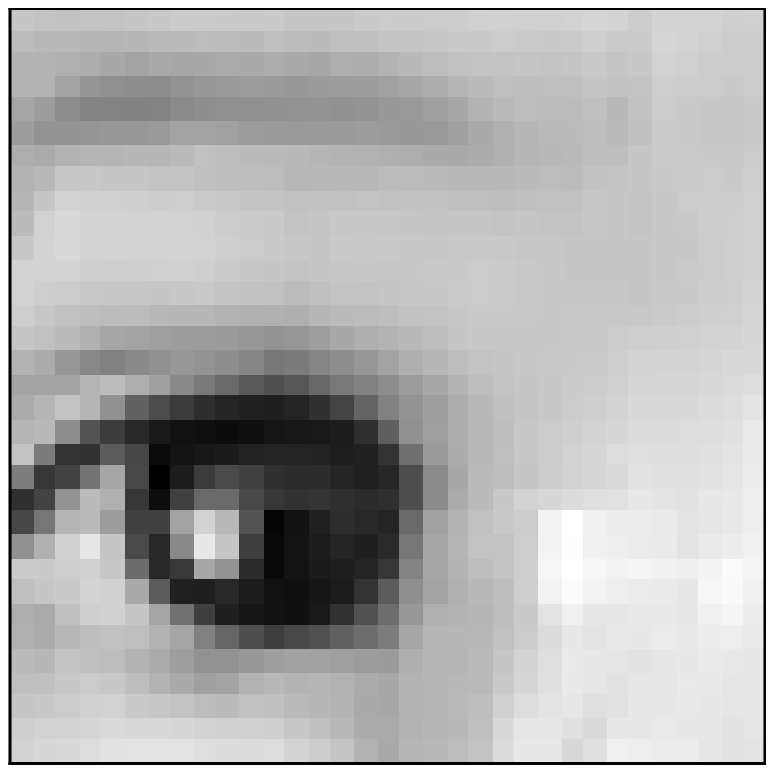}
	&
    \includegraphics[width=0.13\linewidth]{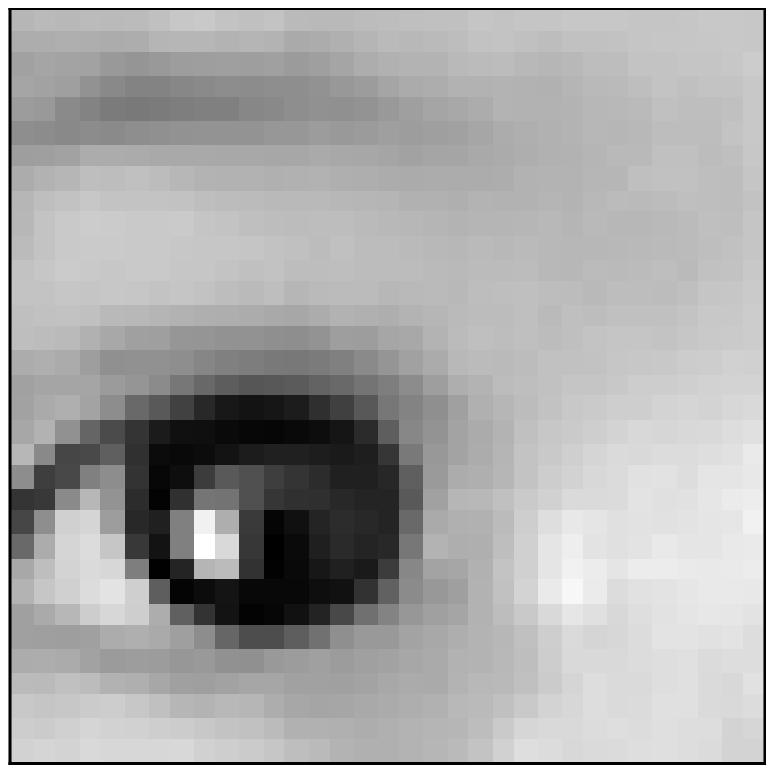}\vspace{5pt}
    \\
    \scriptsize
    (f) LISTA
    &
    \scriptsize
    (g) CSNet
    &
    \scriptsize
    (h) ISTA-Net
    &
    \scriptsize
    (i) OPINE-Net \vspace{2pt}
    &
    \scriptsize
    (j) Proposed
\end{tabular}
\end{center}
\caption{Baby (upper) and Boy (bottom). Recovered images by the various CS reconstruction methods from compressive measurements with a compression ratio of $25\%$.}\label{fig:CS_images_baby}
\end{figure*}

\begin{figure*}
\begin{center}
\begin{tabular}{c c c c c c c}
    \includegraphics[width=0.13\linewidth]{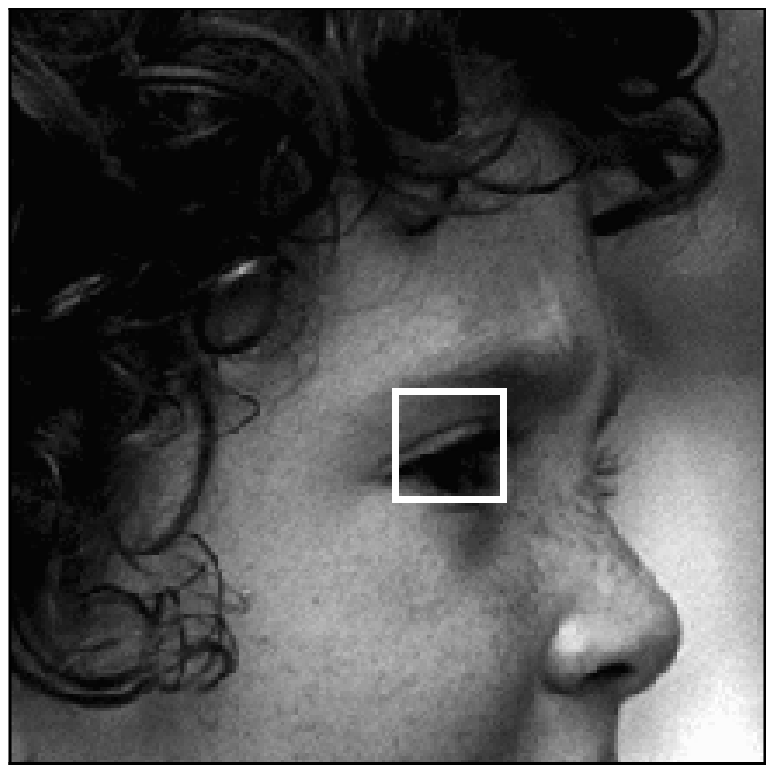}
    &
    \includegraphics[width=0.13\linewidth]{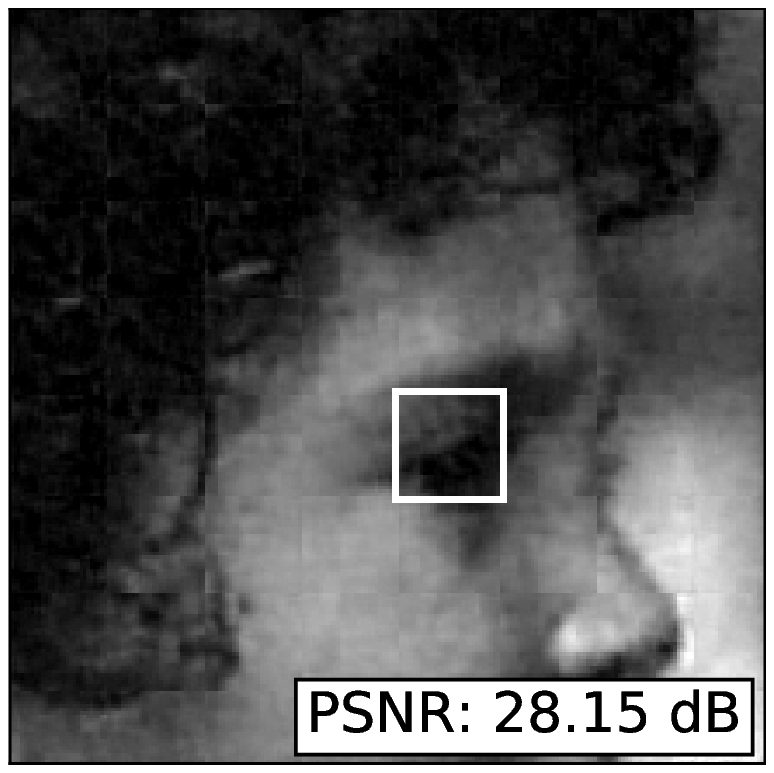}
    &
    \includegraphics[width=0.13\linewidth]{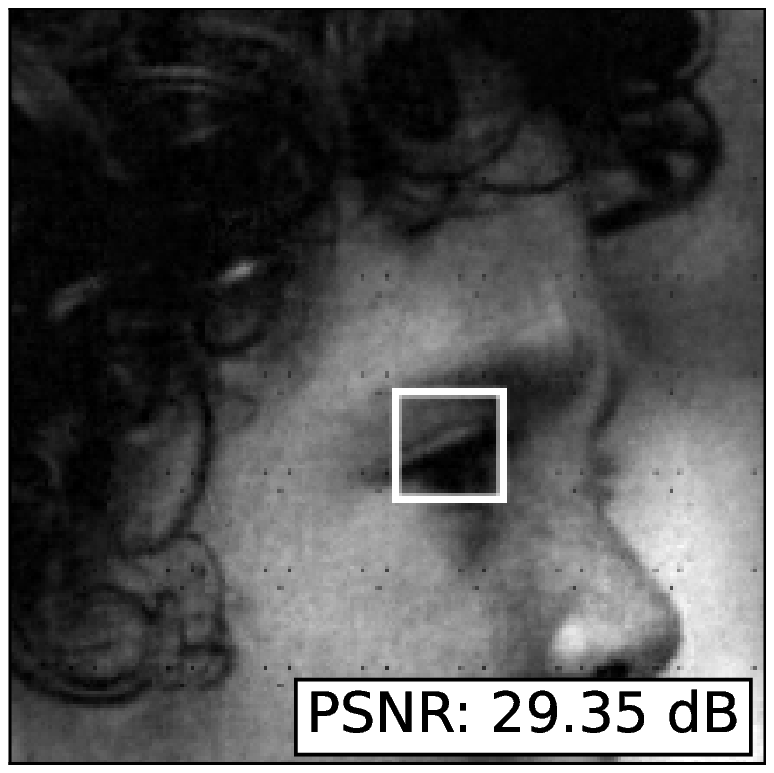}
    &
    \includegraphics[width=0.13\linewidth]{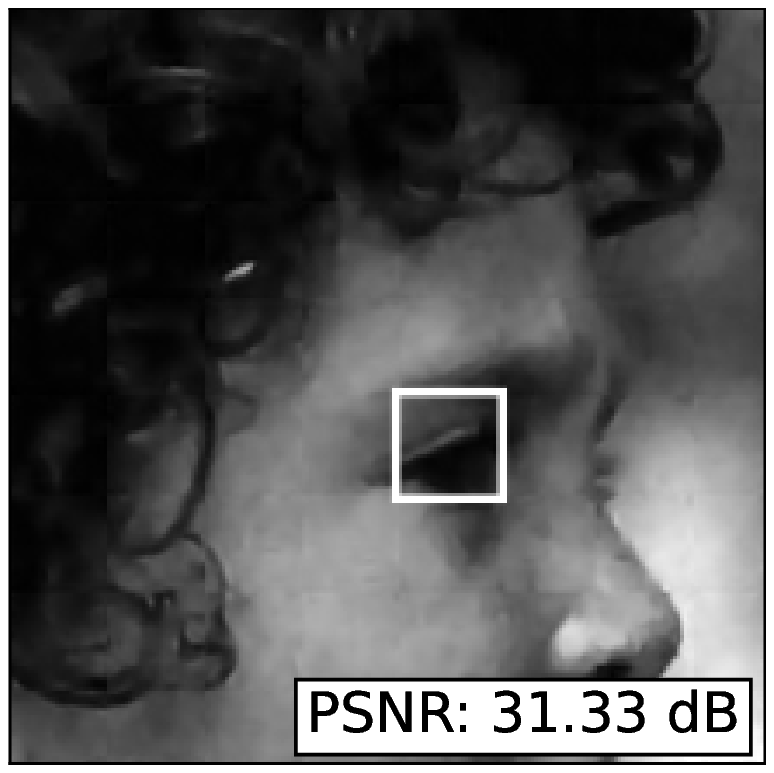}\vspace{-2pt}
    &
    \includegraphics[width=0.13\linewidth]{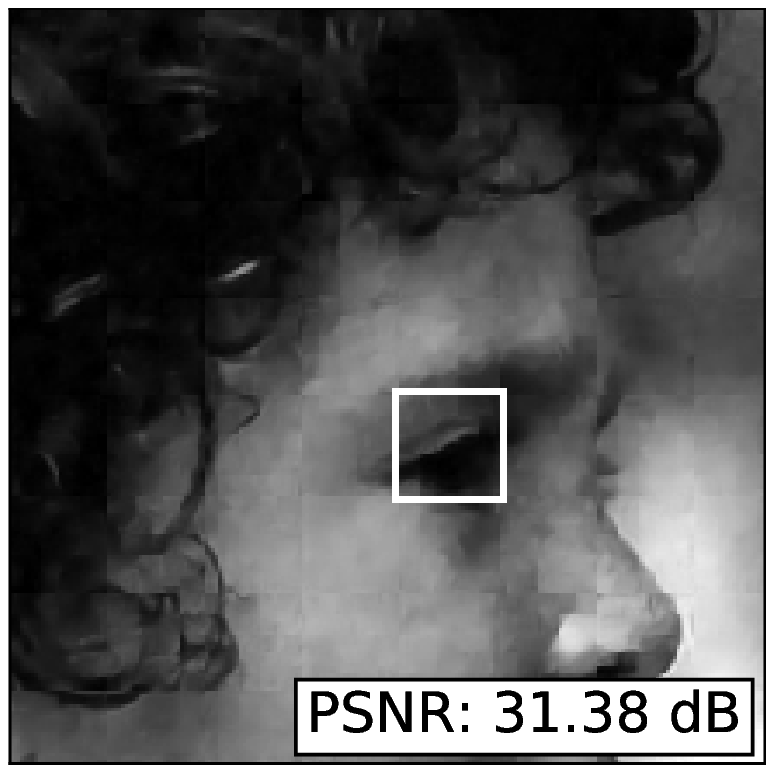}
    &
    \includegraphics[width=0.13\linewidth]{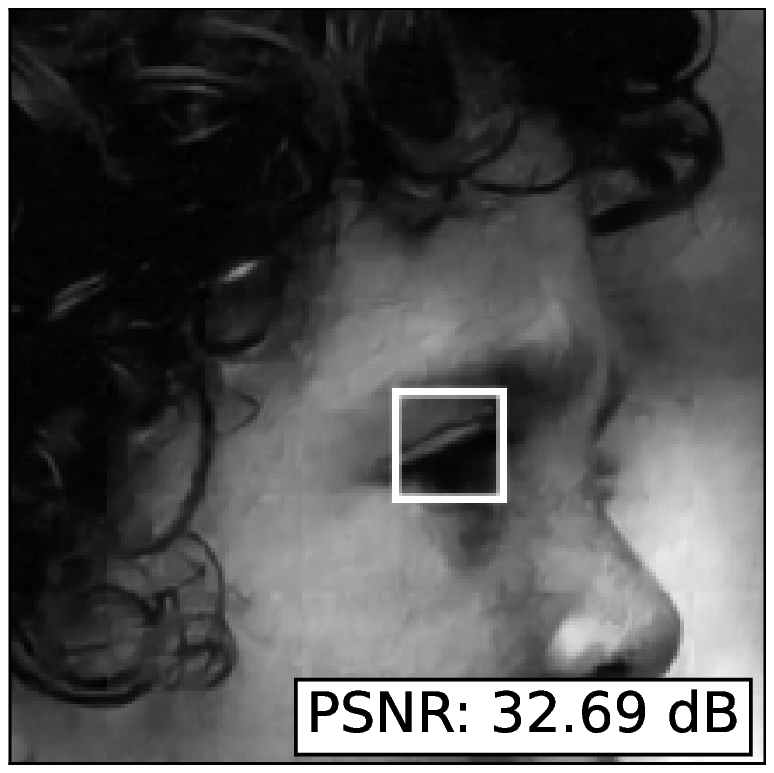}
    &
    \includegraphics[width=0.13\linewidth]{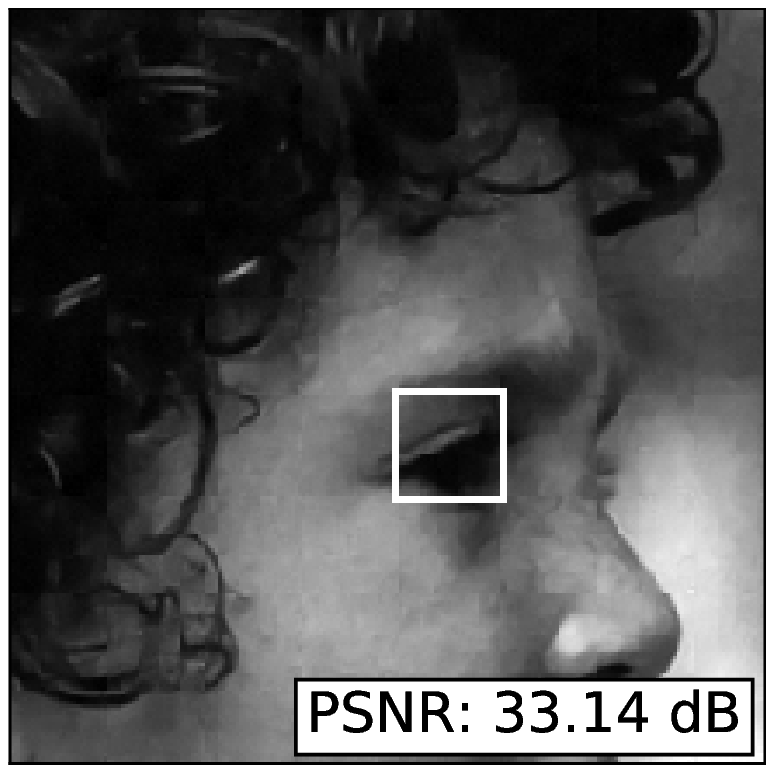}\vspace{-1pt}
    \\
    \includegraphics[width=0.13\linewidth]{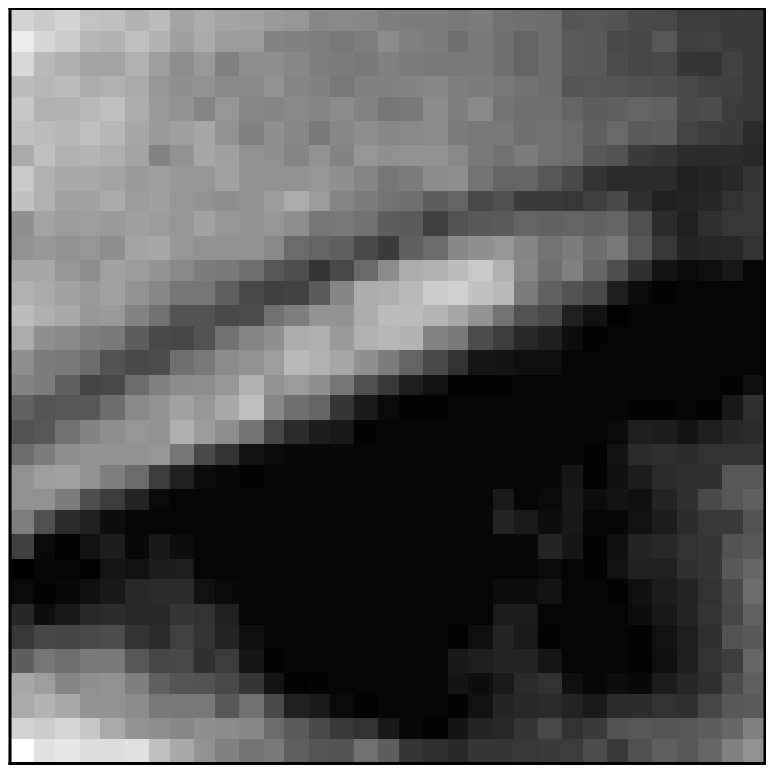}
    &
    \includegraphics[width=0.13\linewidth]{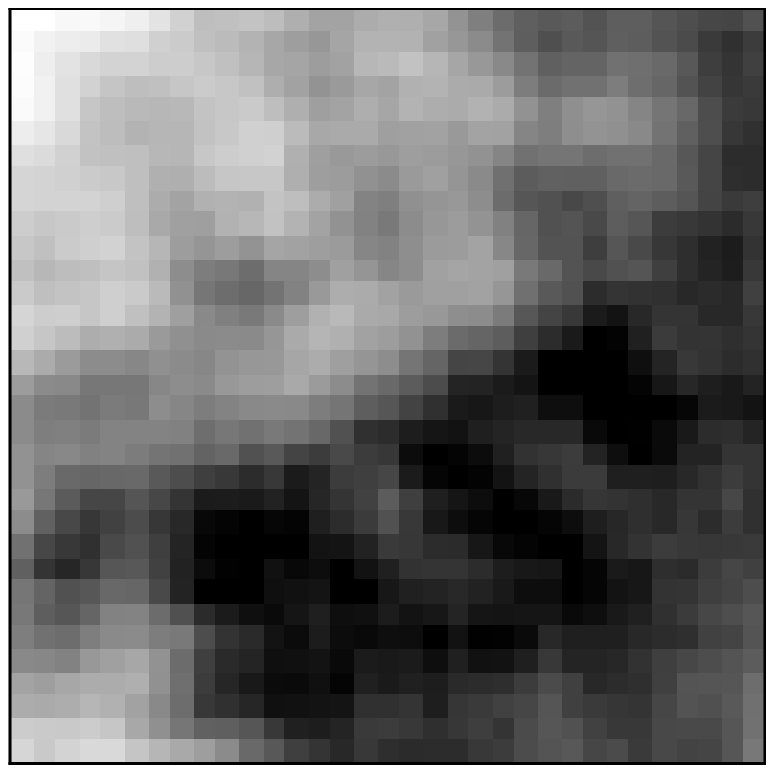}
    &
    \includegraphics[width=0.13\linewidth]{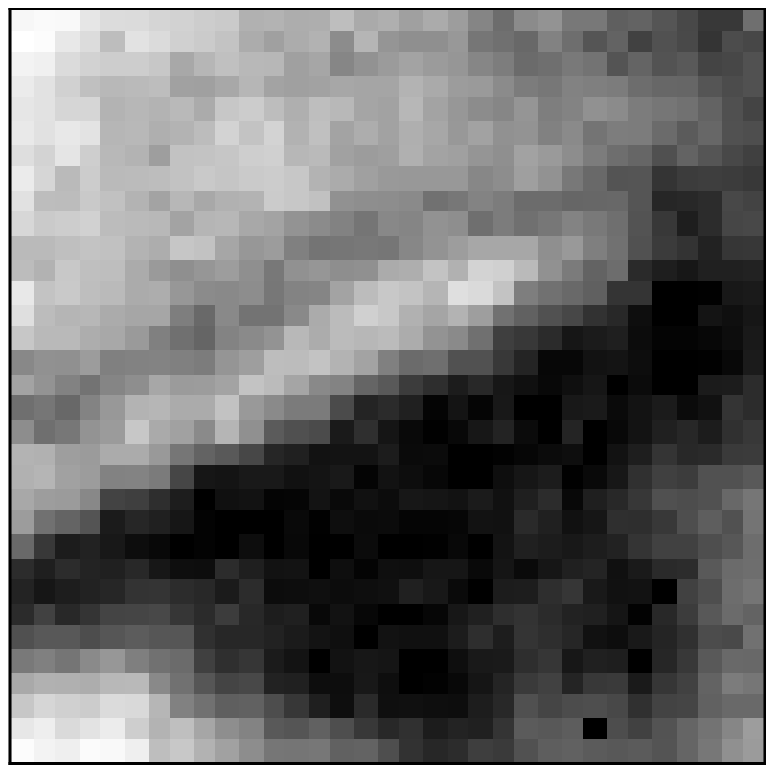}
    &
    \includegraphics[width=0.13\linewidth]{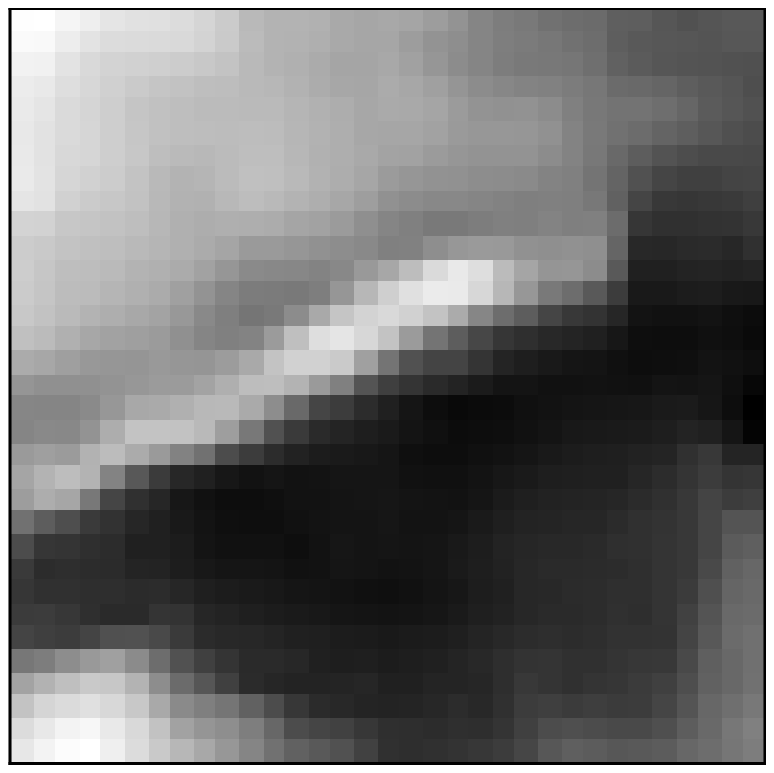}\vspace{-
    4pt}
    &
    \includegraphics[width=0.13\linewidth]{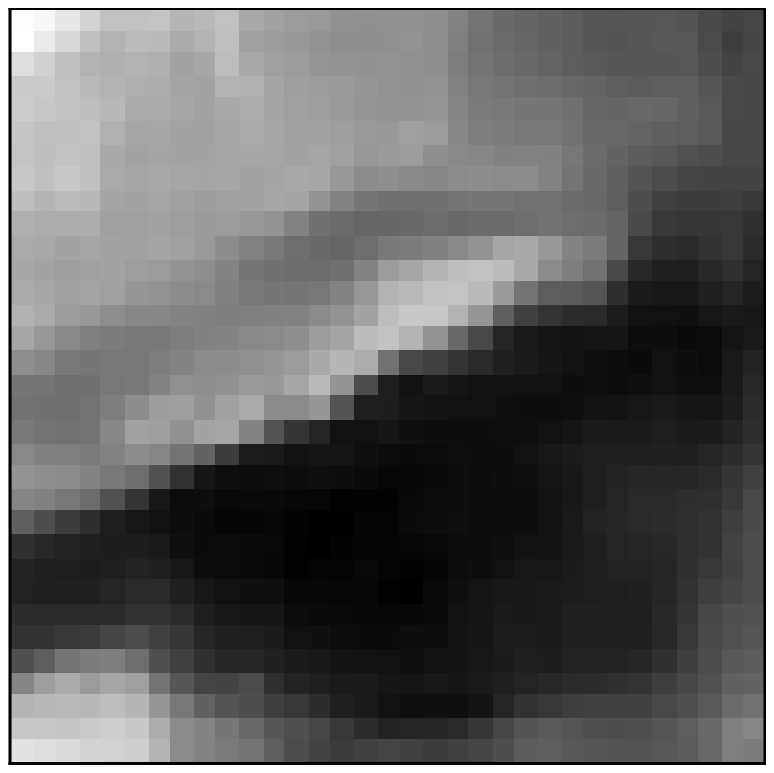}
    &
    \includegraphics[width=0.13\linewidth]{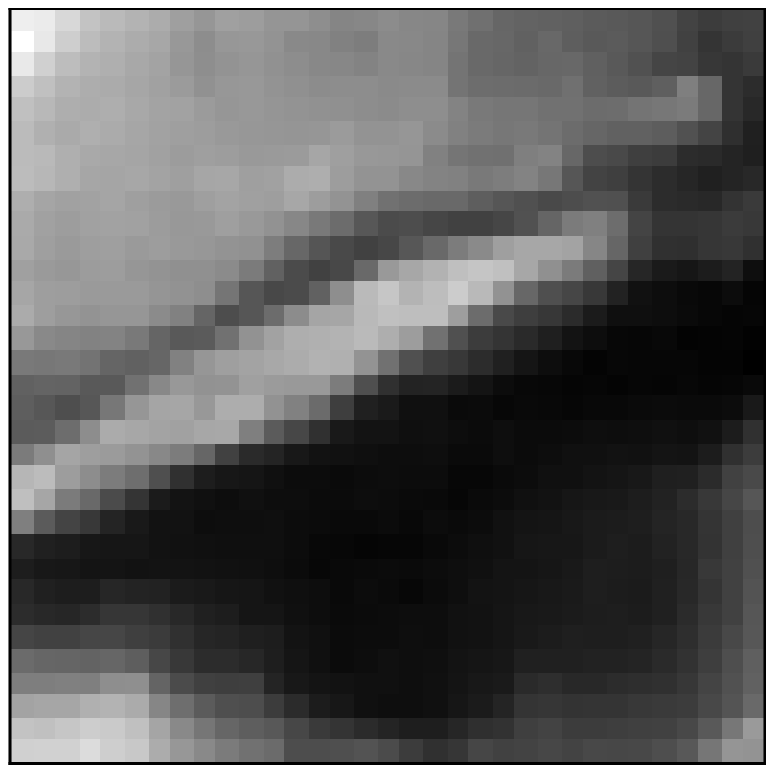}
    &
    \includegraphics[width=0.13\linewidth]{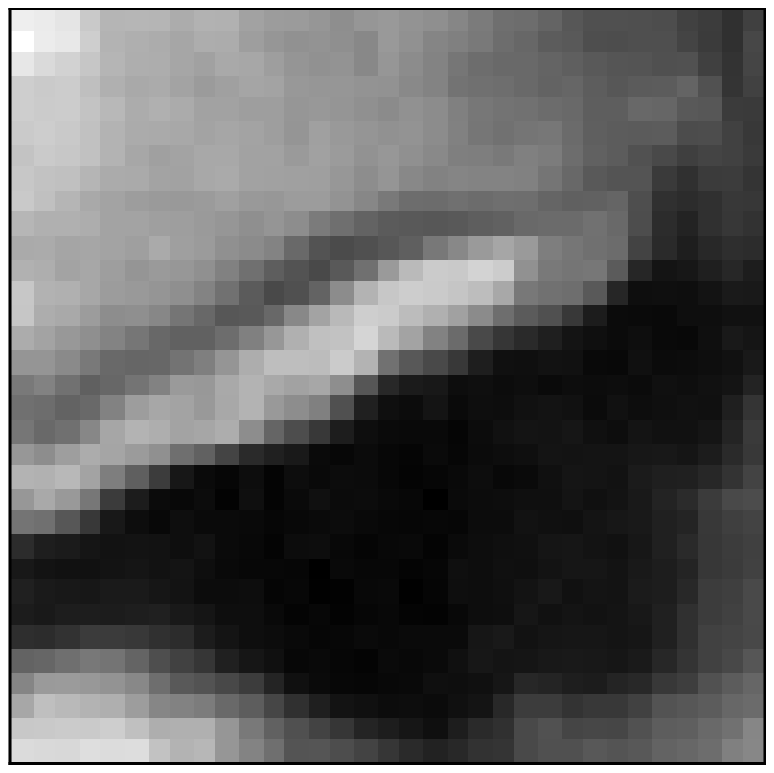}\vspace{4pt}
    \\
    \scriptsize
    (a) Original
    &
    \scriptsize
    (b) ReconNet
    &
    \scriptsize
    (c) LISTA
    &
    \scriptsize
    (d) CSNet
    &
    \scriptsize
    (e) ISTA-Net 
    &
    \scriptsize
    (f) OPINE-Net
    &
    \scriptsize
    (g) Proposed
\end{tabular}
\end{center}
\caption{Baby (upper) and Boy (bottom). Recovered images by the various CS reconstruction methods from compressive measurements with a compression ratio of $25\%$.}\label{fig:CS_images_boy}
\end{figure*}

To evaluate the performance of the proposed algorithm unrolling approach at the training stage, Fig. \ref{fig:learning}(left) displays the average PSNR of the images recovered by the various network-based methods as the number of learning epochs increases. As can be seen in this figure, the proposed approach exhibits an outstanding performance for the entire evaluation interval. Moreover, Fig. \ref{fig:learning}(right) shows the average PSNR of the images recovered by the proposed method as the number of epochs increases for different compression ratios. Higher PSNR values are reached in the learning curves as the compression ratio increases. To quantitatively evaluate the proposed technique, Table \ref{tab:CS_image_psnr} shows the ensemble average of the PSNR in dB obtained by the various CS reconstruction methods from compressive random projections for different compression ratios. The higher values are in bold font. As can be seen in this table, the proposed approach exhibits a remarkable outcomes across the compression ratios under test. In addition, Fig. \ref{fig:CS_images_baby} illustrates the images recovered by the various CS reconstruction methods from compressive measurements with a compression ratio of $25\%$. As can be observed in this figure, the proposed approach is able to properly recover smooth-wise regions preserving, in turn, the structure of the edges. Finally, Fig. \ref{fig:CS_images_boy} displays the reconstructed image obtained by various methods relied on deep netwoks. Fig. \ref{fig:CS_images_boy} displays the recovered images obtained by the different reconstruction methods relied on deep networks. Outstanding performance is exhibited by the proposed approach.

\section{Conclusion}\label{sec:conclusions}

In this work, a deep learning architecture under the algorithm unrolling approach was proposed for estimating high-resolution spectral images from multi-sensor compressive measurements. More precisely, the proposed image fusion method cast each iteration of a linearized version of the ADMM algorithm into a CNN-based structure whose concatenation of deploys a deep network. The linearized approach allows estimating the target variable without resorting to computationally costly matrix inversions. Therefore, the proposed image fusion architecture exhibits competitive running times. On the other hand, the network-based structure learns the relevant information embedded in both the auxiliary variable and the Lagrange multiplier that improves the estimation performance. Notice that the network-based architecture learns its parameters using an end-to-end training scheme. In addition, we evaluated the performance of the proposed approach on two spectral image databases and one real data set. The results yielded by the proposed architecture outperformed those obtained by state-of-the-art methods. To since the proposed approach can be seen as a deep unrolling method for solving inverse problems, the proposed architecture was tested to reconstruct natural images from compressive random projections. In future work, we are interested in evaluating other regularization functions and training different deep networks such as convolutional generators, variational autoencoders, and generative adversarial networks. Furthermore, we are also interested in implementing the proposed approach in further imaging applications including image denoising, inpainting, and deblurring.

\section{Acknowledgments}

This project has received funding from the European Union’s Horizon 2020 research and innovation programme under the Marie Skłodowska-Curie grant agreement No 754382, GOT ENERGY TALENT. The content of this article does not reflect the official opinion of the European Union. Responsibility for the information and views expressed herein lies entirely with the authors.

\section{References}
\bibliographystyle{elsarticle-num-names}
\bibliography{bibtex/bib/reference}

\end{document}